\documentclass[a4paper,UKenglish,cleveref,autoref]{lipics-v2019}

\title{UnLimited TRAnsfers for Multi-Modal Route Planning: An Efficient Solution}

\titlerunning{UnLimited TRAnsfers for Multi-Modal Route Planning: An Efficient Solution}

\newcommand{\affiliationKIT}{Karlsruhe Institute of Technology (KIT), Karlsruhe, Germany}

\author{Moritz Baum}{\affiliationKIT}{moritz@ira.uka.de}{}{}

\author{Valentin Buchhold}{\affiliationKIT}{buchhold@kit.edu}{}{}

\author{Jonas Sauer}{\affiliationKIT}{jonas.sauer2@kit.edu}{}{}

\author{Dorothea Wagner}{\affiliationKIT}{dorothea.wagner@kit.edu}{}{}

\author{Tobias Z\"undorf}{\affiliationKIT}{zuendorf@kit.edu}{}{}

\authorrunning{M. Baum and V. Buchhold and J. Sauer and D. Wagner and T. Z\"undorf}

\Copyright{Moritz Baum and Valentin Buchhold and Jonas Sauer and Dorothea Wagner and Tobias Z\"undorf}

\supplement{\\\indent{}Source code for ULTRA is available at \url{https://github.com/kit-algo/ULTRA}.}

\funding{\\\indent{}This work was supported by Deutsche Forschungsgemeinschaft under grant number WA 654/23-2.}

\acknowledgements{\\\indent{}This manuscript is partially based on extended abstracts which appeared in the proceedings of the 27th Annual European Symposium on Algorithms~\cite{Bau19b} and the 20th Symposium on Algorithmic Approaches for Transportation Modelling, Optimization, and Systems~\cite{Sau20b}, as well as the PhD thesis of one of its authors~\cite{Zuen20}.
We thank Laurent Viennot and Tim Zeitz for helpful suggestions, and Sascha Witt for fruitful discussions about Trip-Based Routing.}

\nolinenumbers

\EventEditors{Michael A. Bender, Ola Svensson, and Grzegorz Herman}
\EventNoEds{3}
\EventLongTitle{27th Annual European Symposium on Algorithms (ESA 2019)}
\EventShortTitle{ESA 2019}
\EventAcronym{ESA}
\EventYear{2019}
\EventDate{September 9--11, 2019}
\EventLocation{Munich/Garching, Germany}
\EventLogo{}
\SeriesVolume{144}
\ArticleNo{13}

\usepackage[                                                                                                          
ruled,                                                                                                            
vlined,                                                                                                           
linesnumbered,                                                                                                    
]{algorithm2e}

\DontPrintSemicolon

\SetArgSty{textnormal}

\newlength{\vlineOffset}
\SetInd{1.5pt}{13.5pt}                                                                                                
\SetVlineSkip{\vlineOffset}

\makeatletter
\renewcommand{\algocf@Vline}[1]{
	\strut\par\nointerlineskip
	\algocf@push{\skiprule}
	\hbox{\vrule\hspace{-0.4pt}%
		\vtop{\algocf@push{\skiptext}
			\vtop{\algocf@addskiptotal #1}\Hlne}}\vskip\skiphlne
	\algocf@pop{\skiprule}
	\nointerlineskip%
}                                                                                                                     
\makeatother

\makeatletter
\renewcommand{\SetKwInOut}[2]{%
	\sbox\algocf@inoutbox{\KwSty{#2\algocf@typo:}}%
	\expandafter\ifx\csname InOutSizeDefined\endcsname\relax
	\newcommand\InOutSizeDefined{}\setlength{\inoutsize}{\wd\algocf@inoutbox}%
	\sbox\algocf@inoutbox{\parbox[t]{\inoutsize}{\KwSty{#2\algocf@typo\hfill:}}\hspace{0.42em}}%
	\setlength{\inoutindent}{\wd\algocf@inoutbox}%
	\else
	\ifdim\wd\algocf@inoutbox>\inoutsize%
	\setlength{\inoutsize}{\wd\algocf@inoutbox}%
	\sbox\algocf@inoutbox{\parbox[t]{\inoutsize}{\KwSty{#2\algocf@typo\hfill:}}\hspace{0.42em}}%
	\setlength{\inoutindent}{\wd\algocf@inoutbox}%
	\fi%
	\fi
	\algocf@newcommand{#1}[1]{%
		\ifthenelse{\boolean{algocf@hanginginout}}{\relax}{\algocf@seteveryparhanging{\relax}}%
		\ifthenelse{\boolean{algocf@inoutnumbered}}{\relax}{\algocf@seteveryparnl{\relax}}{%
			\let\\\algocf@newinout\hangindent=\inoutindent\hangafter=1\parbox[t]{\inoutsize}{%
				\KwSty{#2\algocf@typo:}%
			}\,~{##1}\par%
		}%
		\algocf@linesnumbered
		\ifthenelse{\boolean{algocf@hanginginout}}{\relax}{\algocf@reseteveryparhanging}%
	}%
}
\SetKwInOut{Input}{Input}
\SetKwInOut{Output}{Output}
\makeatother

\newcommand{\Comment}[2][r]{\tcp*[#1]{#2\hspace{-13pt}}}
\newcommand{\BlockEnd}{\vspace{-\vlineOffset}\vspace{-0.4pt}}

\SetKwFor{FOR}{for}{do}{}
\renewcommand{\For}[2]{\FOR{#1}{#2\BlockEnd}}

\SetKwFor{FOREACH}{for each}{do}{}
\renewcommand{\ForEach}[2]{\FOREACH{#1}{#2\BlockEnd}}
\renewcommand{\lForEach}[2]{\lFOREACH{#1}{#2}}
\newcommand{\ForEachComment}[3]{\FOREACH(\Comment[f]{#1}){#2}{#3\BlockEnd}}

\SetKwFor{WHILE}{while}{do}{}
\renewcommand{\While}[2]{\WHILE{#1}{#2\BlockEnd}}

\SetKwIF{IF}{ELSEIF}{ELSE}{if}{then}{else if}{else}{}
\renewcommand{\If}[2]{\IF{#1}{#2\BlockEnd}}
\renewcommand{\lIf}[2]{\lIF{#1}{#2}}

\newcommand{\Break}{\ensuremath{\operatorname{\textbf{break}}}\xspace}
\newcommand{\Continue}{\ensuremath{\operatorname{\textbf{continue}}}\xspace}

\usepackage[dvipsnames]{xcolor}

\usepackage{xparse}
\ExplSyntaxOn
\cs_set_eq:NN \fpeval \fp_eval:n
\ExplSyntaxOff

\newcommand{\tcolor}[3]{\expandafter\newcommand\csname #1\endcsname[1]{#2\fpeval{min(max(#3+##1,0),6)}}}

\definecolor{tobiasblue0}         {rgb}{0.847, 0.922, 0.986}
\definecolor{tobiasblue1}         {rgb}{0.666, 0.822, 0.934}
\definecolor{tobiasblue2}         {rgb}{0.500, 0.707, 0.856}
\definecolor{tobiasblue3}         {rgb}{0.350, 0.582, 0.756}
\definecolor{tobiasblue4}         {rgb}{0.216, 0.453, 0.636}
\definecolor{tobiasblue5}         {rgb}{0.099, 0.324, 0.499}
\definecolor{tobiasblue6}         {rgb}{0.000, 0.200, 0.348}

\definecolor{tobiasred0}          {rgb}{1.000, 0.885, 0.903}
\definecolor{tobiasred1}          {rgb}{0.943, 0.714, 0.707}
\definecolor{tobiasred2}          {rgb}{0.857, 0.544, 0.539}
\definecolor{tobiasred3}          {rgb}{0.753, 0.383, 0.396}
\definecolor{tobiasred4}          {rgb}{0.637, 0.235, 0.273}
\definecolor{tobiasred5}          {rgb}{0.519, 0.106, 0.165}
\definecolor{tobiasred6}          {rgb}{0.406, 0.000, 0.069}

\definecolor{tobiasgreen0}        {rgb}{0.800, 0.987, 0.700}
\definecolor{tobiasgreen1}        {rgb}{0.585, 0.885, 0.487}
\definecolor{tobiasgreen2}        {rgb}{0.433, 0.785, 0.339}
\definecolor{tobiasgreen3}        {rgb}{0.324, 0.681, 0.242}
\definecolor{tobiasgreen4}        {rgb}{0.241, 0.571, 0.180}
\definecolor{tobiasgreen5}        {rgb}{0.167, 0.451, 0.137}
\definecolor{tobiasgreen6}        {rgb}{0.082, 0.317, 0.100}

\definecolor{tobiasyellow0}       {rgb}{0.971, 0.900, 0.655}
\definecolor{tobiasyellow1}       {rgb}{0.942, 0.811, 0.400}
\definecolor{tobiasyellow2}       {rgb}{0.908, 0.720, 0.226}
\definecolor{tobiasyellow3}       {rgb}{0.863, 0.626, 0.117}
\definecolor{tobiasyellow4}       {rgb}{0.800, 0.524, 0.054}
\definecolor{tobiasyellow5}       {rgb}{0.713, 0.411, 0.021}
\definecolor{tobiasyellow6}       {rgb}{0.596, 0.284, 0.000}

\definecolor{tobiascyan0}         {rgb}{0.700, 0.977, 0.963}
\definecolor{tobiascyan1}         {rgb}{0.510, 0.937, 0.920}
\definecolor{tobiascyan2}         {rgb}{0.347, 0.889, 0.860}
\definecolor{tobiascyan3}         {rgb}{0.214, 0.823, 0.782}
\definecolor{tobiascyan4}         {rgb}{0.111, 0.727, 0.685}
\definecolor{tobiascyan5}         {rgb}{0.039, 0.590, 0.568}
\definecolor{tobiascyan6}         {rgb}{0.000, 0.400, 0.429}

\definecolor{tobiaslila0}         {rgb}{0.964, 0.825, 1.000}
\definecolor{tobiaslila1}         {rgb}{0.915, 0.696, 0.973}
\definecolor{tobiaslila2}         {rgb}{0.846, 0.539, 0.932}
\definecolor{tobiaslila3}         {rgb}{0.758, 0.375, 0.870}
\definecolor{tobiaslila4}         {rgb}{0.648, 0.224, 0.782}
\definecolor{tobiaslila5}         {rgb}{0.514, 0.104, 0.661}
\definecolor{tobiaslila6}         {rgb}{0.356, 0.036, 0.499}

\definecolor{black}               {rgb}{0.000, 0.000, 0.000}

\tcolor{blue}                     {tobiasblue}{3}
\tcolor{red}                      {tobiasred}{3}
\tcolor{green}                    {tobiasgreen}{3}
\tcolor{yellow}                   {tobiasyellow}{3}
\tcolor{cyan}                     {tobiascyan}{3}
\tcolor{lila}                     {tobiaslila}{3}

\tcolor{primarycolor}             {tobiasblue}{3}
\tcolor{secondarycolor}           {tobiasred}{3}
\tcolor{tertiarycolor}            {tobiasyellow}{3}
\tcolor{quaternarycolor}          {tobiasgreen}{3}

\colorlet{primarycolor-dark}      {\primarycolor{1}}
\colorlet{primarycolor}           {\primarycolor{0}}
\colorlet{primarycolor-light}     {\primarycolor{-1}}
\colorlet{primarycolor-vlight}    {\primarycolor{-3}}
\colorlet{secondarycolor-dark}    {\secondarycolor{1}}
\colorlet{secondarycolor}         {\secondarycolor{0}}
\colorlet{secondarycolor-light}   {\secondarycolor{-1}}
\colorlet{secondarycolor-vlight}  {\secondarycolor{-3}}

\colorlet{themecolor}             {\primarycolor{1}}
\colorlet{footnoteLineColor}      {\primarycolor{0}}
\colorlet{footnoteSignColor}      {\primarycolor{1}}

\tcolor{refColor}                 {tobiascyan}{3}
\tcolor{ourColor}                 {tobiaslila}{3}
\tcolor{patColor}                 {tobiasred}{3}
\tcolor{cycleColor}               {tobiasyellow}{3}
\tcolor{assignmentColor}          {tobiasblue}{3}
\tcolor{setupColor}               {tobiasgreen}{3}

\usepackage{xparse}
\usepackage{booktabs}
\usepackage{colortbl}
\usepackage{enumitem}
\usepackage{hyperref}

\usepackage{tikz}
\usepackage{pgfplots}
\usepackage{graphicx}
\usepackage{trimclip}
\usepackage{pgfplotstable}

\pgfplotsset{compat=1.16}

\usetikzlibrary{positioning}
\usetikzlibrary{shadings}
\usetikzlibrary{shapes}
\usetikzlibrary{arrows}
\usetikzlibrary{calc}
\usetikzlibrary{pgfplots.statistics}
\usetikzlibrary{decorations.text}
\usetikzlibrary{decorations.pathmorphing}
\usetikzlibrary{fit}

\usepgfplotslibrary{statistics}
\usepgfplotslibrary{colormaps}

\colorlet{nodeColor}{black!80}
\colorlet{edgeColor}{black!50}
\colorlet{axisColor}{black!80}
\colorlet{legendColor}{black!80}
\colorlet{meanColor}{black!80}

\newcommand{\gs}{\hphantom{\tiny$\cdot$}}

\newcommand{\legend}[1]{\raisebox{0.07ex}{#1}}

\pgfdeclareplotmark{markFull}{%
	{\color{white}%
		\pgfpathcircle{\pgfpointorigin}{1.5\pgfplotmarksize}%
		\pgfusepathqstroke}%
	\pgfpathcircle{\pgfpointorigin}{\pgfplotmarksize}%
	\pgfusepathqfillstroke%
}
\pgfdeclareplotmark{markCirc}{%
	{\color{white}%
		\pgfpathcircle{\pgfpointorigin}{1.5\pgfplotmarksize}%
		\pgfusepathqfillstroke}%
	\pgfpathcircle{\pgfpointorigin}{\pgfplotmarksize}%
	\pgfusepathqstroke%
}
\pgfdeclareplotmark{markBox}{%
	{\color{white}%
		\pgfpathrectangle{\pgfqpoint{-1.5\pgfplotmarksize}{-1.5\pgfplotmarksize}}{\pgfqpoint{3\pgfplotmarksize}{3\pgfplotmarksize}}%
		\pgfusepathqstroke}%
	\pgfpathrectangle{\pgfqpoint{-\pgfplotmarksize}{-\pgfplotmarksize}}{\pgfqpoint{2\pgfplotmarksize}{2\pgfplotmarksize}}%
	\pgfusepathqfillstroke%
}
\pgfdeclareplotmark{markSquare}{%
	{\color{white}%
		\pgfpathrectangle{\pgfqpoint{-1.5\pgfplotmarksize}{-1.5\pgfplotmarksize}}{\pgfqpoint{3\pgfplotmarksize}{3\pgfplotmarksize}}%
		\pgfusepathqfillstroke}%
	\pgfpathrectangle{\pgfqpoint{-\pgfplotmarksize}{-\pgfplotmarksize}}{\pgfqpoint{2\pgfplotmarksize}{2\pgfplotmarksize}}%
	\pgfusepathqstroke%
}

\tikzstyle{vertex}=[circle,line width=.5pt,minimum size=0.1pt]
\tikzstyle{routeArrow}=[->, >=stealth]
\tikzstyle{gridStyle}=[axisColor!20, line width = 0.2pt, dash pattern = on 2pt off 1pt]
\tikzstyle{axisStyle}=[axisColor, line width=0.5pt]
\tikzstyle{markSign} = [mark=o]
\tikzstyle{shortenLines} = [shorten <= 3.4pt,shorten >= 3.4pt]
\tikzstyle{pointsFull} = [mark=*, mark size=1.8pt, line width=1.2pt, only marks]
\tikzstyle{sampleFull} = [mark=*, mark size=1.8pt, line width=1.2pt, shorten <= 3.4pt,shorten >= 3.4pt]
\tikzstyle{legendFull} = [mark=markFull, mark size=1.8pt, line width=1.2pt, shorten <= -1pt,shorten >= -1pt]
\tikzstyle{pointsCirc} = [mark=o, mark size=1.8pt, line width=1.2pt, only marks]
\tikzstyle{sampleCirc} = [mark=o, mark size=1.8pt, line width=1.2pt, shorten <= 3.4pt,shorten >= 3.4pt]
\tikzstyle{legendCirc} = [mark=markCirc, mark size=1.8pt, line width=1.2pt, shorten <= -1pt,shorten >= -1pt]
\tikzstyle{sampleBox} = [mark=square*, mark size=1.5pt, line width=1.2pt, shorten <= 3.4pt,shorten >= 3.4pt]
\tikzstyle{legendBox} = [mark=markBox, mark size=1.5pt, line width=1.2pt, shorten <= -1pt,shorten >= -1pt]
\tikzstyle{sampleSquare} = [mark=square, mark size=1.5pt, line width=1.2pt, shorten <= 3.4pt,shorten >= 3.4pt]
\tikzstyle{legendSquare} = [mark=markSquare, mark size=1.5pt, line width=1.2pt, shorten <= -1pt,shorten >= -1pt]

\pgfplotsset{
	grid style={axisColor!20,line width = 0.2pt,dash pattern = on 2pt off 1pt},
	grid=both,
	axis line style={axisColor,line width = 0.4pt},
	major tick style={axisColor,line width = 0.4pt},
	minor tick style={axisColor,line width = 0.2pt},
	major tick length=3.5pt,
	minor tick length=2.0pt,
	ytick align=outside,
	xtick align=outside,
	ticklabel style = {font=\small},
	label style = {font=\small},
	boxplot/every median/.style={meanColor, line width = 1.5pt},
}

\pgfkeys{/pgf/number format/.cd, set thousands separator={\,}}

\newcommand{\boxplotLegend}[1]{%
	\raisebox{-0.1ex}{%
		\begin{tikzpicture}%
			\draw[lineColor#1,line width = \lineWidth] (0,-0.55ex) -- (0,0.55ex);%
			\draw[lineColor#1,line width = \lineWidth] (0,0) -- (0.4em,0);%
			\draw[lineColor#1,line width = \lineWidth,fill=fillColor#1] (0.4em,-0.7ex) rectangle (1.6em,0.7ex);%
			\draw[lineColor#1,line width = \lineWidth] (1.6em,0) -- (2em,0);%
			\draw[lineColor#1,line width = \lineWidth] (2em,-0.55ex) -- (2em,0.55ex);%
			\draw[meanColor,line width = 1.0pt] (1em,-0.8ex) -- (1em,0.8ex);%
		\end{tikzpicture}%
	}%
}

\usepackage{amsmath}
\usepackage{mathtools}
\usepackage{mathdots}
\usepackage{xspace}
\usepackage{bm}

\newcommand{\nonNegativeReals}{\ensuremath{\mathbb{R}^+_0}\xspace}

\newcommand{\graph}{\ensuremath{G}\xspace}
\newcommand{\augmentedGraph}{\ensuremath{\graph^+}\xspace}
\newcommand{\upwardGraph}{\ensuremath{\graph^\uparrow}\xspace}
\newcommand{\downwardGraph}{\ensuremath{\graph^\downarrow}\xspace}
\newcommand{\coreGraph}{\ensuremath{\graph^{\textsf{c}}}\xspace}
\newcommand{\shortcutGraph}{\ensuremath{\graph^{\textsf{s}}}\xspace}
\newcommand{\shortcutGraphST}{\ensuremath{\tilde{\graph}^{\textsf{s}}}\xspace}
\newcommand{\eventGraph}{\ensuremath{\graph^{\textsf{e}}}\xspace}
\newcommand{\vertices}{\ensuremath{\mathcal{V}}\xspace}
\newcommand{\coreVertices}{\ensuremath{\vertices^{\textsf{c}}}\xspace}
\newcommand{\eventVertices}{\ensuremath{\vertices^\textsf{e}}\xspace}
\newcommand{\aVertex}{\ensuremath{v}\xspace}
\newcommand{\aVertexX}{\ensuremath{v}}
\newcommand{\bVertex}{\ensuremath{w}\xspace}
\newcommand{\bVertexX}{\ensuremath{w}}
\newcommand{\cVertex}{\ensuremath{x}\xspace}
\newcommand{\cVertexX}{\ensuremath{x}}
\newcommand{\dVertex}{\ensuremath{y}\xspace}
\newcommand{\dVertexX}{\ensuremath{y}}
\newcommand{\eVertex}{\ensuremath{z}\xspace}
\newcommand{\eVertexX}{\ensuremath{z}}
\newcommand{\aSource}{\ensuremath{s}\xspace}
\newcommand{\aSourceX}{\ensuremath{s}}
\newcommand{\aTarget}{\ensuremath{t}\xspace}
\newcommand{\aTargetX}{\ensuremath{t}}
\newcommand{\targets}{\ensuremath{\vertices_\aTarget}\xspace}
\newcommand{\edges}{\ensuremath{\mathcal{E}}\xspace}
\newcommand{\coreEdges}{\ensuremath{\edges^{\textsf{c}}}\xspace}
\newcommand{\augmentedEdges}{\ensuremath{\edges^+}\xspace}
\newcommand{\upwardEdges}{\ensuremath{\edges^\uparrow}\xspace}
\newcommand{\downwardEdges}{\ensuremath{\edges^\downarrow}\xspace}
\newcommand{\shortcutEdges}{\ensuremath{\edges^\textsf{s}}\xspace}
\newcommand{\canonicalShortcuts}{\ensuremath{\edges_\text{canon}}\xspace}
\newcommand{\eventEdges}{\ensuremath{\edges^\textsf{e}}\xspace}
\newcommand{\edge}{\ensuremath{e}\xspace}
\newcommand{\edgeLength}{\ensuremath{\ell}\xspace}

\newcommand{\stops}{\ensuremath{\mathcal{S}}\xspace}
\newcommand{\aStop}{\aVertex}
\newcommand{\bStop}{\bVertex}
\newcommand{\trips}{\ensuremath{\mathcal{T}}\xspace}
\newcommand{\aTrip}{\ensuremath{T}\xspace}
\newcommand{\aTripA}{\ensuremath{\aTrip_a}\xspace}
\newcommand{\aTripB}{\ensuremath{\aTrip_b}\xspace}
\newcommand{\activeTrip}{\ensuremath{\aTrip_{\textsf{min}}}\xspace}
\newcommand{\aTripSegment}[1]{\ensuremath{\aTrip^{#1}}\xspace}
\newcommand{\aTripSegmentA}[1]{\ensuremath{\aTrip^{#1}_a}\xspace}
\newcommand{\aTripSegmentB}[1]{\ensuremath{\aTrip^{#1}_b}\xspace}
\newcommand{\aTripSegmentC}[2]{\ensuremath{\aTrip^{#1}_{#2}}\xspace}
\newcommand{\routes}{\ensuremath{\mathcal{R}}\xspace}
\newcommand{\aRoute}{\ensuremath{R}\xspace}
\newcommand{\aStopEvent}{\ensuremath{\epsilon}\xspace}

\newcommand{\aPath}{\ensuremath{P}\xspace}
\newcommand{\journeys}{\ensuremath{\mathcal{J}}\xspace}
\newcommand{\optimalJourneys}{\ensuremath{\journeys^\text{opt}}\xspace}
\newcommand{\canonicalJourneys}{\ensuremath{\journeys^\text{canon}}\xspace}
\newcommand{\candidateJourneys}{\ensuremath{\journeys^\textsf{c}}\xspace}
\newcommand{\aJourney}{\ensuremath{J}\xspace}
\newcommand{\aCandidateJourney}{\ensuremath{\aJourney^{\textsf{c}}}\xspace}
\newcommand{\aWitnessJourney}{\ensuremath{\aJourney^{\textsf{w}}}\xspace}

\newcommand{\atime}{\ensuremath{\tau}\xspace}
\newcommand{\arrivalTime}{\ensuremath{\atime_{\textsf{arr}}}\xspace}
\newcommand{\departureTime}{\ensuremath{\atime_{\textsf{dep}}}\xspace}
\newcommand{\transferTime}{\ensuremath{\atime_{\textsf{tra}}}\xspace}
\newcommand{\minTime}{\ensuremath{\atime_{\textsf{min}}}\xspace}
\newcommand{\witnessLimit}{\ensuremath{\bar{\atime}_{\textsf{wit}}}\xspace}
\newcommand{\bufferTime}{\ensuremath{\atime_\text{buf}}\xspace}
\newcommand{\departureTimes}{\ensuremath{\raisebox{-0.1pt}{\ensuremath{\mathcal{D}}}\mkern-3.2mu\raisebox{0.1pt}{\ensuremath{\mathcal{T}}}}\xspace}

\newcommand{\aQueue}{\ensuremath{Q}\xspace}
\DeclareMathOperator{\dist}{dist}
\newcommand{\searchSpace}[1]{\ensuremath{\vertices_{#1}}\xspace}

\newcommand{\labels}{\ensuremath{\mathcal{L}}\xspace}
\newcommand{\reachedIndex}{\ensuremath{r}\xspace}
\DeclareMathOperator{\Enqueue}{Enqueue}

\newcommand{\tiebreakingSequence}{\ensuremath{X}\xspace}
\newcommand{\routeIndex}{\ensuremath{\text{id}_\routes}\xspace}
\newcommand{\vertexIndex}{\ensuremath{\text{id}_\vertices}\xspace}
\newcommand{\parent}{\ensuremath{p}\xspace}
\DeclareMathOperator{\run}{run}
\newcommand{\aLabel}{\ensuremath{\ell}\xspace}

\hideLIPIcs

\begin{document}

\maketitle

\begin{abstract}
We study a multimodal journey planning scenario consisting of a public transit network and a transfer graph which represents a secondary transportation mode (e.g.,~walking, cycling, e-scooter).
The objective is to compute Pareto-optimal journeys with respect to arrival time and the number of used public transit trips.
While various existing algorithms can efficiently compute optimal journeys in either a pure public transit network or a pure transfer graph, combining the two increases running times significantly.
Existing approaches therefore typically only support limited walking between stops, either by imposing a maximum transfer distance or by requiring the transfer graph to be transitively closed.
To overcome these shortcomings, we propose a novel preprocessing technique called ULTRA~(UnLimited TRAnsfers):
Given an unlimited transfer graph, which may represent any non-schedule-based transportation mode, ULTRA computes a small number of transfer shortcuts that are provably sufficient for computing a Pareto set of optimal journeys.
These transfer shortcuts can be integrated into a variety of state-of-the-art public transit algorithms, establishing the ULTRA-Query algorithm family.
Our extensive experimental evaluation shows that ULTRA improves these algorithms from limited to unlimited transfers without sacrificing query speed.
This is true not just for walking, but also for faster transfer modes such as bicycle or car.
Compared to the state of the art for multimodal journey planning, the fastest ULTRA-based algorithm achieves a speedup of an order of magnitude.
\end{abstract}

\newpage

\section{Introduction}
\label{sec:introduction}

Research on efficient route planning algorithms has seen remarkable advances in the past two decades.
On road networks, queries can be answered in less than a millisecond with moderate preprocessing effort, even for continental-scale graphs.
Similar results are currently out of reach for public transit networks, but state-of-the-art algorithms nevertheless achieve query times of a few milliseconds on metropolitan and mid-sized country networks~\cite{Bas16b}.
Even more challenging is the multimodal journey planning problem, which combines schedule-based (i.e.,~public transit) and non-schedule-based (e.g.,~walking, cycling, driving) modes of transportation.
While this covers a greater variety of possible journeys, solving it efficiently remains difficult~\cite{Wag17}.
In this work, we consider a multimodal problem that augments public transit with a \textit{transfer graph}, which represents one arbitrary non-schedule-based transportation mode.
This transfer mode can be used at the start and end of a journey to enter and exit the public transit network, and for transferring between public transit vehicles in the middle of the journey.
Given a source and target vertex in the transfer graph and a departure time, the objective is to compute Pareto-optimal journeys with respect to arrival time and the number of used public transit trips.

\subparagraph*{Related Work.}
Journey planning algorithms for public transit networks can be divided into graph-based approaches and algorithms that operate directly on the timetable, exploiting its schedule-based structure~\cite{Bas16b}.
Graph-based approaches model the public transit network as a graph and then answer queries with Dijkstra's algorithm~\cite{Dij59}, which can be sped up by applying preprocessing techniques~\cite{Del09c,Bau11,Bas10,Bas16,Del15}.
The two main modelling approaches are the time-dependent~\cite{Pyr08,Dis08} and time-expanded~\cite{Mue07b,Pyr08} models.
The time-expanded model uses vertices to represent events in the timetable (e.g.,~a vehicle arriving at or departing from a stop) and edges to connect consecutive events.
By contrast, the time-dependent model represents stops (e.g.,~a train station) of the network as vertices and connects two stops with an edge if they are served consecutively by at least one vehicle.
Associated with each edge is a function which maps departure time to travel time.
Both models can integrate footpaths~\cite{Dis08,Bas10,Del15}, but only as direct edges between public transit stops.
This means that an unrestricted footpath network cannot be encoded efficiently, since the number of edges would be quadratic in the number of stops.
To ensure a reasonable graph size, footpaths are typically restricted to small connected components of nearby stops~\cite{Del12b}, for example by limiting the maximal duration~(e.g.,~5~minutes of walking) or distance~(e.g.,~400\,m)~\cite{Bas14,Bas16,Gia19} of footpaths.

Notable timetable-based approaches include RAPTOR~\cite{Del15b}, CSA~\cite{Dib18} and the corresponding speedup techniques, HypRAPTOR~\cite{Del17b} and ACSA~\cite{Dib18}.
Instead of exploring the public transit network with Dijkstra's algorithm, these algorithms rely on array-based scanning operations which improve cache locality.
As with the graph-based approaches, footpaths can be integrated as transfer edges between pairs of stops.
However, these are required to be \emph{one-hop transfers}, i.e.,~at most one transfer edge may be used when transferring between two public transit trips.
This removes the need for Dijkstra searches within the transfer graph, as every possible destination can be reached with a single edge.
Additionally, both RAPTOR and CSA require that the transfer graph is transitively closed, which ensures that optimal journeys never require multiple transfer edges in succession.
RAPTOR can be modified to lift this restriction~\cite{Del19}, allowing for one-hop transfers without a transitive closure.
In this case, journeys with multiple consecutive transfer edges are prohibited and the algorithm finds optimal journeys among those that remain.
This can lead to counterintuitive journeys which take detours to avoid using two transfer edges in succession.
On the other hand, computing the transitive closure significantly increases the size of the transfer graph.
As shown by Wagner and Zündorf~\cite{Wag17}, limiting the maximal transfer duration to~20~minutes before computing the transitive closure already leads to a graph that is too large for practical applications.

A special case among the timetable-based approaches is Trip-Based Routing~(TB)~\cite{Wit15}, which requires a preprocessing phase that computes transfers between pairs of trips.
This is done by enumerating all possible transfers and then applying a set of pruning rules to omit some, but not all unnecessary transfers.
TB requires a transitively closed transfer graph as input and was originally only evaluated for very sparse transfer graphs.
Because it enumerates all transfers before pruning them, the preprocessing time is highly sensitive to the size of the transfer graph.
Lehoux and Loiodice~\cite{Leh20} mitigated this by proposing an alternative transfer enumeration method which discards many unnecessary transfers before they are enumerated.
However, neither version of the TB preprocessing supports unrestricted transfer graphs.

Using a restricted transfer graph is often justified with the argument that long transfers are rarely useful.
However, experiments have shown that the availability of unrestricted walking significantly reduces travel times~\cite{Wag17,Sau18,Pha19}.
Naturally, this effect will be even stronger for faster transportation modes, such as bicycle or car.
Handling unrestricted transfer graphs (which may represent any non-schedule-based transportation mode) requires multimodal journey planning algorithms.
These algorithms typically work by interleaving an existing public transit algorithm with Dijkstra searches on the transfer graph.
Notable examples are UCCH~\cite{Dib15b} and MCR~\cite{Del13}, which are based on a time-dependent graph-based approach and RAPTOR, respectively.
Because the Dijkstra searches are expensive, these algorithms are slow compared to their pure public transit counterparts.
More recently, HL-RAPTOR and HL-CSA~\cite{Pha19} were proposed.
Here, RAPTOR and CSA are interleaved with two-hop searches based on Hub Labeling~(HL)~\cite{Abr11} instead of Dijkstra.
While this requires a moderately expensive preprocessing phase, the authors report a speedup of 1.7 over the bicriteria variant of MCR for HL-RAPTOR.

\subparagraph*{Contribution.}
Preliminary experiments~\cite{Sau18} have shown that the impact of unrestricted transfers in Pareto-optimal journeys depends heavily on their position in the journey:
\emph{Initial transfers}, which connect the source to the first public transit vehicle, and \emph{final transfers}, connecting the final vehicle to the target, are fairly common and often have a large impact on the travel time.
By contrast, \emph{intermediate transfers} between public transit trips are only occasionally relevant for optimal journeys.
This suggests that the number of unique paths in the transfer graph that occur as intermediate transfers of a Pareto-optimal journey is small.
Using this insight, we propose a new preprocessing technique called ULTRA~(UnLimited TRAnsfers), which computes a set of shortcut edges representing these paths.
The preprocessing step is carefully engineered to ensure that the number of shortcuts remains small.
Combined with efficient one-to-many searches for the initial and final transfers, these shortcuts are provably sufficient for answering all possible queries correctly.

ULTRA~shortcuts can be used without adjustment by any algorithm that requires one-hop transfers between stops.
In our experimental evaluation, we demonstrate this for RAPTOR and CSA.
The resulting multimodal algorithms have roughly the same query performance as the original restricted algorithms, regardless of the speed of the considered transfer mode.
In particular, ULTRA-CSA is the first multimodal variant of CSA.
For TB, we show that only minor changes are necessary to make ULTRA compute shortcuts between trips instead of stops.
This allows ULTRA to replace the TB preprocessing phase while enabling unlimited transfers.
We demonstrate that this significantly reduces the number of required shortcuts and the query time compared to a naive approach, i.e.,~using the output of ULTRA as input for the TB preprocessing.
Overall, ULTRA-TB outperforms the bicriteria version of MCR, which was previously the fastest multimodal algorithm, by about an order of magnitude.
This yields query times of a few milliseconds on metropolitan networks and less than 100\,ms on the much larger network of Germany.

\subparagraph*{Outline.}
The remainder of this work is structured as follows.
Section~\ref{sec:preliminaries} establishes basic notation and gives an overview of the algorithms that ULTRA builds on.
We then describe the ULTRA shortcut computation in Section~\ref{chap:ULTRA:prepro} and prove that it computes a sufficient set of transfer shortcuts.
Section~\ref{chap:ULTRA:query} explains how the transfer shortcuts can be integrated into query algorithms that require one-hop transfers.
We also present modifications to the TB query algorithm to make it more efficient in a multimodal setting.
We evaluate the performance of our preprocessing and query algorithms on real-world multimodal networks in Section~\ref{chap:ULTRA:exp}.
Finally, we summarize our results and give an outlook on potential future work in Section~\ref{chap:ULTRA:fr}.

\section{Preliminaries}
\label{sec:preliminaries}
This section establishes basic terminology and introduces foundational algorithms.

\subsection{Terminology}
\subparagraph*{Network.}
A public transit network is a 4-tuple~$(\stops,\trips,\routes,\graph)$ consisting of a set of \emph{stops} \stops, a set of \emph{trips} \trips, a set of \emph{routes} \routes, and a directed, weighted \emph{transfer graph}~$\graph=(\vertices,\edges)$.
A stop is a location in the network where passengers can board or disembark a vehicle (such as buses, trains, ferries, etc.).
A trip~$\aTrip=\langle\aStopEvent_0,\dots,\aStopEvent_k\rangle\in\trips$ is a sequence of \emph{stop events} performed by the same vehicle.
A \emph{stop event}~$\aStopEvent=(\arrivalTime(\aStopEvent), \departureTime(\aStopEvent), \aVertex(\aStopEvent))$ represents the vehicle arriving at the stop~$\aVertex(\aStopEvent)$ with the arrival time~$\arrivalTime(\aStopEvent)$ and subsequently departing from the same stop with the departure time~$\departureTime(\aStopEvent)$.
The~$i$-th stop event in~$\aTrip$ is denoted as~$\aTrip[i]$.
The length~$|\aTrip|:=k$ is the number of stop events in~$\aTrip$.
A~\emph{trip segment}~\mbox{$\aTripSegment{ij}:=\langle\aStopEvent_i,\dots,\aStopEvent_j\rangle$} is a contiguous subsequence of~\aTrip which begins at~$\aTrip[i]$ and ends at~$\aTrip[j]$.
The set of routes~\routes is a partition of~\trips such that two trips which are part of the same route visit the same sequence of stops and do not overtake each other.
A trip~$\aTripA\in\trips$ overtakes a trip~$\aTripB\in\trips$ if there exist two indices~$i<j$ such that~$\aTripA$ arrives at or departs from~$\aVertex(\aTripA[i])$ not before~$\aTripB$ but arrives at or departs from~$\aVertex(\aTripA[j])$ not after~$\aTripB$.
Given a trip~$\aTrip$, the route of~$\aTrip$ is denoted as~$\aRoute(\aTrip)$.
The length~$|\aRoute|$ of a route~$\aRoute$ is the length of any trip belonging to the route.

The transfer graph~$\graph=(\vertices,\edges)$ consists of a set of~\emph{vertices}~\vertices with~$\stops\subseteq\vertices$, and a set of~\emph{edges}~$\edges\subseteq\vertices\times\vertices$.
Traveling along an edge~$\edge=(\aVertex,\bVertex)\in\edges$ requires the \emph{transfer time}~$\transferTime(\edge)$.
The notion of transfer time carries over to paths~$\aPath=\langle\aVertex_1,\dots,\aVertex_k\rangle$ in~\graph, using the definition~$\transferTime(\aPath):=\sum_{i = 1}^{k-1}\transferTime((\aVertex_i,\aVertex_{i+1}))$.
Unlike in scenarios with limited footpaths, we impose no restrictions on~\graph.
It does not need to be transitively closed, it may be strongly connected, and transfer times may represent walking, cycling, or some other non-schedule-based mode of travel.
An example of a public transit network with an unrestricted transfer graph is shown in Figure~\ref{fig:network}.

\begin{figure}
	\centering
	\begin{tikzpicture}[scale=1]

    \node (s) at ( 0.00,  0.00) {};%
    \node (v) at ( 2.00, -2.00) {};%
    \node (w) at ( 4.00,  0.00) {};%
    \node (x) at ( 8.00, -2.00) {};%
    \node (y) at ( 8.00,  0.00) {};%
    \node (z) at (12.00,  1.50) {};%
    \node (t) at (12.00,  0.00) {};%
    
    \draw[edgeColor,line width=1pt] (s) -- (v);
    \draw[edgeColor,line width=1pt] (v) -- (w);
    \draw[edgeColor,line width=1pt] (w) -- (y);
    \draw[edgeColor,line width=1pt] (x) -- (y);
    \draw[edgeColor,line width=1pt] (y) -- (t);
    \draw[edgeColor,line width=1pt] (z) -- (t);
    
    \node[align=left,edgeColor] at ( 0.75, -1.25) {$1$};
    \node[align=left,edgeColor] at ( 3.25, -1.25) {$1$};
    \node[align=left,edgeColor] at ( 6.00,  0.30) {$4$};
    \node[align=left,edgeColor] at ( 7.75, -1.00) {$4$};
    \node[align=left,edgeColor] at (10.00,  0.30) {$4$};
    \node[align=left,edgeColor] at (12.25,  0.75) {$1$};
    
    \draw [\primarycolor{0}, line width=2.5pt, routeArrow, rounded corners = 20] (s) -- (w);
    \draw [\primarycolor{0}, line width=2.5pt, routeArrow, rounded corners = 20] (w) -- (4.00, -2.00) -- (x);
    \draw [\secondarycolor{0}, line width=2.5pt, routeArrow, rounded corners = 20] (w) -- (4.00, 1.50) -- (z);
    \draw [\tertiarycolor{0}, line width=2.5pt, routeArrow, rounded corners = 20] (x) -- (12.00, -2.00) -- (t);
  
    \node [align=left,text=\primarycolor{1}]   at ( 2.00,  0.80) {$1\rightarrow2$};%
    \node [align=left,text=\primarycolor{1}]   at ( 2.00,  0.30) {$6\rightarrow7$};%
    \node [align=left,text=\primarycolor{1}]   at ( 6.00, -1.20) {$2\rightarrow3$};%
    \node [align=left,text=\primarycolor{1}]   at ( 6.00, -1.70) {$7\rightarrow8$};%
    \node [align=left,text=\secondarycolor{1}] at ( 8.00,  1.20) {$3\rightarrow7$};%
    \node [align=left,text=\tertiarycolor{1}]  at (10.00, -0.70) {$\phantom{1}5\rightarrow7\phantom{1}$};%
    \node [align=left,text=\tertiarycolor{1}]  at (10.00, -1.20) {$\phantom{1}7\rightarrow9\phantom{1}$};%
    \node [align=left,text=\tertiarycolor{1}]  at (10.00, -1.70) {$\phantom{1}9\rightarrow11$};%

    \node at (s) [vertex,draw=nodeColor!100,fill=nodeColor!15] {\gs};%
    \node at (v) [vertex,draw=nodeColor!100,fill=nodeColor!15] {\gs};%
    \node at (w) [vertex,draw=nodeColor!100,fill=nodeColor!15] {\gs};%
    \node at (x) [vertex,draw=nodeColor!100,fill=nodeColor!15] {\gs};%
    \node at (y) [vertex,draw=nodeColor!100,fill=nodeColor!15] {\gs};%
    \node at (z) [vertex,draw=nodeColor!100,fill=nodeColor!15] {\gs};%
    \node at (t) [vertex,draw=nodeColor!100,fill=nodeColor!15] {\gs};%
    
    \node at (s) [text=nodeColor!100] {\small{$\aSource$}};%
    \node at (v) [text=nodeColor!100] {\small{$\aVertex$}};%
    \node at (w) [text=nodeColor!100] {\small{$\bVertex$}};%
    \node at (x) [text=nodeColor!100] {\small{$\cVertex$}};%
    \node at (y) [text=nodeColor!100] {\small{$\dVertex$}};%
    \node at (z) [text=nodeColor!100] {\small{$\eVertex$}};%
    \node at (t) [text=nodeColor!100] {\small{$\aTarget$}};%
\end{tikzpicture}%
	\caption[An example of a public transit network and a Pareto set of journeys.]{%
		An example of a public transit network with an unrestricted transfer graph.
		Edges in the transfer graph (gray) are labeled with their travel time.
		Routes are displayed as sequences of colored edges.
		Each edge is labeled with the departure and arrival times of the associated in trips, in the format~$\departureTime\to\arrivalTime$.
		For a query from~$\aSource$ to~$\aTarget$ with departure time~$0$, a Pareto set with respect to arrival time and number of trips consists of the journeys~$\aJourney_0=\big\langle\!\langle\aSourceX,\aVertexX,\bVertexX,\dVertexX,\aTargetX\rangle\!\big\rangle$ with arrival time~$10$, $\aJourney_1=\big\langle\!\langle\aSourceX,\aVertexX,\bVertexX\rangle,\langle\textcolor{\secondarycolor{1}}{3\rightarrow7}\rangle,\langle\eVertexX,\aTargetX\rangle\!\big\rangle$ with arrival time~$8$, and~$\aJourney_2=\big\langle\!\langle\aSourceX\rangle,\langle\textcolor{\primarycolor{1}}{1\rightarrow2},\textcolor{\primarycolor{1}}{2\rightarrow3}\rangle,\langle\cVertexX\rangle,\langle\textcolor{\tertiarycolor{1}}{5\rightarrow7}\rangle,\langle\aTargetX\rangle\!\big\rangle$ with arrival time~$7$.
	}%
	\label{fig:network}%
\end{figure}

\subparagraph*{Journeys.}
A \emph{journey} describes the movement of a passenger through the network from a source vertex~$\aSource\in\vertices$ to a target vertex~$\aTarget\in\vertices$.
Each ride of the passenger in a public transit vehicle can be described by a trip segment, whereas the transfers between the rides are represented by paths in the transfer graph.
An \emph{intermediate transfer} between two trip segments~$\aTripSegmentA{ij}$ and~$\aTripSegmentB{mn}$ is a path~\aPath in~$\graph$ such that:~(1)~the path begins with the last stop of~$\aTripSegmentA{ij}$, i.e.,~$\aVertex(\aTripA[j])$,~(2)~the path ends with the first stop of~$\aTripSegmentB{mn}$, i.e.,~$\aVertex(\aTripB[m])$, and~(3)~the transfer time of the path is sufficient to reach~$\aTripSegmentB{mn}$ after vacating~$\aTripSegmentA{ij}$.
This can be expressed formally as~$\arrivalTime(\aTripA[j])+\transferTime(\aPath)\leq\departureTime(\aTripB[m])$.
An \emph{initial transfer} before a trip segment~$\aTripSegment{ij}$ is a path in~\graph from the source~\aSource to the first stop of~$\aTripSegment{ij}$.
Correspondingly, a \emph{final transfer} after a trip segment~$\aTripSegment{ij}$ is a path in~\graph from the last stop of~$\aTripSegment{ij}$ to the target~\aTarget.

A \emph{journey}~$\aJourney=\langle\aPath_0,\aTripSegmentC{ij}{0},\dots,\aTripSegmentC{mn}{k-1},\aPath_k\rangle$ is an alternating sequence of transfers and trip segments.
Note that some or all of the transfers may be empty, i.e.,~consist of a single stop only.
Given source and target vertices~$\aSource,\aTarget\in\vertices$, we call journey~$\aJourney$ an~$\aSource$-$\aTarget$-journey if~$\aPath_0$ begins with~$\aSource$ and~$\aPath_k$ ends with~$\aTarget$.
The departure time of the journey is defined as~\mbox{$\departureTime(\aJourney):=\departureTime(\aTrip_0[i])-\transferTime(\aPath_0)$} and the arrival time as~\mbox{$\arrivalTime(\aJourney):=\arrivalTime(\aTrip_{k-1}[n])+\transferTime(\aPath_{k})$}.
The number of trips used by the journey is denoted as~$|\aJourney|:={}k$.
An important special case is a journey~$\aJourney=\langle\aPath_0\rangle$ that consists solely of a path in the transfer graph.
Since such a journey does not use any trips, it can be traversed at any time.
Thus, its departure time~$\departureTime(\aJourney)$ has to be stated separately, and its arrival time is then given by~$\arrivalTime(\aJourney):=\departureTime(\aJourney)+\transferTime(\aPath_0)$.
The \emph{vertex sequence} of~$\aJourney$ is the concatenation of its transfers: $\vertices(\aJourney)=\aPath_0\circ\aPath_1\circ\dots\circ\aPath_k$.
A \emph{subjourney} of~$\aJourney$ is a journey~$\aJourney_s=\langle\aPath'_x,\aTripSegmentC{gh}{x},\dots,\aTripSegmentC{pq}{y-1},\aPath'_y\rangle$ such that~$\langle\aTripSegmentC{gh}{x},\dots,\aTripSegmentC{pq}{y-1}\rangle$ is a contiguous subsequence of~$\aJourney$, $\aPath'_x$ is a suffix of~$\aPath_x$ and~$\aPath'_y$ is a prefix of~$\aPath_y$.
If~$x=0$ and~$\aPath'_x=\aPath_0$, we call~$\aJourney_s$ a \emph{prefix} of~$\aJourney$.
Conversely, if~$y=k$ and~$\aPath'_y=\aPath_k$, we call~$\aJourney_s$ a \emph{suffix} of~$\aJourney$.
Note that a subjourney may start or end in the middle of a transfer, but never in the middle of a trip segment.
Given two vertices~$\aVertex,\bVertex\in\vertices(\aJourney)$, the subjourney of~$\aJourney$ from~$\aVertex$ to~$\bVertex$ is denoted as~$\aJourney_{\aVertex\bVertex}$.

\subparagraph*{Problem Statement.}
To evaluate the usefulness of a journey~$\aJourney$, we mainly consider the two criteria arrival time~$\arrivalTime(\aJourney)$ and number of trips~$|\aJourney|$.
Given a set of criteria, a journey~$\aJourney$ \emph{weakly dominates} another journey~$\aJourney'$ if~$\aJourney$ is not worse than~$\aJourney'$ in any criterion.
Moreover,~$\aJourney$ \emph{strongly dominates}~$\aJourney'$ if~$\aJourney$ is strictly better than~$\aJourney'$ in at least one criterion, and not worse in the others.
Given a query consisting of source and target vertices~$\aSource,\aTarget\in\vertices$ and an earliest departure time~$\departureTime$, a journey is called \emph{feasible} if it is an~$\aSource$-$\aTarget$-journey that does not depart earlier than~$\departureTime$.
A feasible journey~\aJourney is called~\emph{Pareto-optimal} if no other feasible journey exists that strongly dominates~\aJourney.
A \emph{Pareto set} is a set~$\journeys$ containing a minimal number of Pareto-optimal journeys such that every feasible journey is weakly dominated by a journey in~$\journeys$.
For a given query, the objective is to compute a Pareto set with respect to the two criteria arrival time and number of trips.
See Figure~\ref{fig:network} for a Pareto set of journeys in the shown example network.

\subparagraph*{Departure Buffer Times.}
Many works on public transit routing (e.g.,~\cite{Pyr08,Del15}) allow a \emph{minimum change time} to be specified for each stop.
It must be observed when transferring between two trips at the same stop, but not when entering a trip after arriving via a path in the transfer graph or when entering the first trip at the start of the journey. 
The minimum change time is useful for modeling stops that represents larger stations with multiple platforms.
Here, the minimum change time represents the time needed to change between platforms.
This modeling choice is reasonable for settings with direct transfers between stops.
However, when allowing an unrestricted transfer graph, it can lead to inconsistencies.
Given a stop with minimum change time~$\atime$, if a path starting and ending at this stop with a transfer time less than~\atime exists, then taking that path would allow passengers to circumvent the minimum change time.

To prevent this, we introduce \emph{departure buffer times} as an alternative modeling approach. 
Each stop~$\aStop\in\stops$ has a non-negative departure buffer time~$\bufferTime(\aStop)$, which is the minimum amount of time that has to pass after arriving at the stop before a vehicle can be boarded.
Unlike the minimum change time, the departure buffer time always has to be observed when a trip is entered, even if the stop was reached via a transfer or if the trip is the first one in the journey.
Departure buffer times can be integrated into the network implicitly by reducing the departure times of the stop events accordingly.
For each stop event~$\aStopEvent=(\arrivalTime(\aStopEvent),\departureTime(\aStopEvent),\aVertex(\aStopEvent))$, we obtain the reduced stop event~$\aStopEvent'=(\arrivalTime(\aStopEvent),\departureTime(\aStopEvent)-\bufferTime(\aVertex(\aStopEvent)),\aVertex(\aStopEvent))$.
Note that this may cause stop events to appear as if they depart before they arrive.
However, since the departure time is only relevant when entering the trip at the current stop and not when remaining seated in the trip, this does not lead to trips that travel backwards in time.
In the following, we will not discuss departure buffer times explicitly and instead assume that they are integrated into the departure times as described here.

\subsection{Algorithms}
We now give an overview of the algorithms on which ULTRA is based.

\subparagraph*{Dijkstra's Algorithm.}
Given a graph~$\graph=(\vertices,\edges)$ with edge length function~$\edgeLength:\edges\to\nonNegativeReals$ and a source vertex~$\aSource\in\vertices$, Dijkstra's algorithm~\cite{Dij59} computes for each vertex~$\aVertex$ the length of the shortest~$\aSource$-$\aVertex$-path.
It maintains for each vertex~$\aVertex$ a~\emph{tentative distance}~$\dist[\aVertex]$, which is initialized with~$\infty$.
Additionally, it maintains a priority queue~$\aQueue$ of vertices ordered by their \emph{key}, which is the tentative distance.
Initially, $\aSource$ is inserted into~$\aQueue$ with key~$\dist[\aSource]=0$.
Then, vertices are extracted from~$\aQueue$ in increasing order of key.
Each extracted vertex~$\aVertex$ is \emph{settled} by \emph{relaxing} its outgoing edges.
An edge~$\edge=(\aVertex,\bVertex)\in\edges$ is relaxing by comparing the tentative distance~$\dist[\bVertex]$ to the distance~$\dist[\aVertex]+\edgeLength(\edge)$ which is achieved by traversing~$\edge$.
If the latter is smaller, $\dist[\bVertex]$ is updated accordingly and~$\bVertex$ is inserted into~$\aQueue$ with key~$\dist[\bVertex]$.

\subparagraph*{Contraction Hierarchies.}
To explore the transfer graph, ULTRA utilizes algorithms based on Contraction~Hierarchies~(CH)~\cite{Gei12}, a preprocessing technique originally developed to speed up one-to-one queries in road networks.
The basic building block of CH is~\emph{vertex contraction}: a vertex is contracted by removing it from the graph and inserting~\emph{shortcut edges} between its neighbors such that shortest path distances in the graph are preserved.
The CH preprocessing phase for a graph~$\graph=(\vertices,\edges)$ iteratively contracts the vertices of~$\graph$ in a heuristically determined order.
The position of a vertex in this contraction order is called its~\emph{rank}.
The output of this preprocessing phase is an~\emph{augmented graph}~$\augmentedGraph=(\vertices,\augmentedEdges)$ which contains all original edges and all inserted shortcut edges.
The augmented graph can be split into an~\emph{upward graph}~$\upwardGraph=(\vertices,\upwardEdges)$ containing only edges from lower-ranked to higher-ranked vertices, and a corresponding~\emph{downward graph}~$\downwardGraph=(\vertices,\downwardEdges)$.
Queries are answered with a bidirectional variant of Dijkstra's algorithm, where the forward search explores~$\upwardGraph$ and the backward search explores~$\downwardGraph$.

Bucket-CH~\cite{Kno07,Gei12} is an extension of CH for one-to-many queries.
It operates in three phases.
First, given the graph~$\graph=(\vertices,\edges)$, the CH precomputation is performed.
Second, given the set~$\targets\subseteq\vertices$ of targets, a \emph{bucket} containing distances to the targets is computed for every vertex.
This is done by performing a backward search on~$\downwardGraph$ from every target vertex~$\aTarget\in\targets$.
For each vertex~$\aVertex$ settled by this search with distance~$\dist(\aVertex,\aTarget)$, the entry~$(\aTarget,\dist(\aVertex,\aTarget))$ is added to the bucket of~$\aVertex$.
Finally, given a query with source vertex~$\aSource$, the algorithm performs a forward search on~$\upwardGraph$.
For each vertex~$\aVertex$ settled by this search with distance~$\dist(\aSource,\aVertex)$, the bucket of~$\aVertex$ is evaluated.
For each bucket entry~$(\aTarget,\dist(\aVertex,\aTarget))$, the shortest distance to~$\aTarget$ found so far is compared to~$\dist(\aSource,\aVertex)+\dist(\aVertex,\aTarget)$ and updated if it is improved.

Multimodal algorithms such as UCCH and MCR employ a special variant of the CH precomputation which we call~\emph{Core-CH}~\cite{Bau10, Dib15b, Del13}.
Here, the precomputation is not allowed to contract vertices that coincide with stops.
Thus, a set of \emph{core vertices}~$\coreVertices$ with~$\stops\subseteq\coreVertices\subseteq\vertices$ is left uncontracted.
In addition to the (partially) augmented graph, this yields a \emph{core graph}~$\coreGraph=(\coreVertices,\coreEdges)$, which consists of~$\coreVertices$ and all shortcuts which were inserted between core vertices.
If only stops are allowed as core vertices, the number of core edges will be quadratic in the number of stops.
This slows down both the precomputation and query algorithms to the point where they become impractical.
In practice, the contraction process is therefore stopped once the average vertex degree in the core graph surpasses a specified limit.

\subparagraph*{RAPTOR.}
To explore the public transit network, ULTRA employs algorithms from the RAPTOR family.
RAPTOR~\cite{Del15b} answers one-to-one and one-to-all queries in a public transit network with one-hop transfers.
It operates in \emph{rounds}, where the $i$-th round finds journeys with~$i$ trips by appending an additional trip to journeys found in the previous round.
For each stop~$\aStop\in\stops$ and each round~$i$, the algorithm maintains a~\emph{tentative arrival time}~$\arrivalTime(\aStop,i)$, which is the earliest arrival time among all journeys to~$\aStop$ with at most~$i$ trips found so far.
Each round consists of a~\emph{route scanning phase} followed by a~\emph{transfer relaxation phase}.
Round~$i$ assumes that every stop~$\aStop$ for which~$\arrivalTime(\aStop,i-1)$ was improved in round~$i-1$ has been~\emph{marked}.
Before either phase is performed, $\arrivalTime(\aStop,i)$ is initialized with~$\arrivalTime(\aStop,i-1)$ for each stop~$\aStop\in\stops$.
Then the route scanning phase collects all routes which visit marked stops and \emph{scans} them.
A route~$\aRoute$ is scanned by iterating across all visited stops, starting at the first marked stop.
During the scan, the algorithm maintains an~\emph{active trip}~$\activeTrip$, which is the earliest trip of~$\aRoute$ that can be entered at any of already processed stops.
Let~$\aStop$ be the~$j$-th stop of~$\aRoute$.
If~$\activeTrip$ has already been set, the algorithm checks whether exiting~$\activeTrip$ at~$\aStop$ with arrival time~$\arrivalTime(\activeTrip[j])$ improves~$\arrivalTime(\aStop,i)$.
If so, $\arrivalTime(\aStop,i)$ is updated accordingly and~$\aStop$ is marked.
Afterwards, the algorithm checks whether there is an earlier trip than~$\activeTrip$ which can be entered when arriving at~$\aStop$ with arrival time~$\arrivalTime(\aStop,i-1)$.
If so, $\activeTrip$ is updated accordingly.
After all collected routes have been scanned, the transfer relaxation phase is performed.
For every marked stop~$\aStop$, each outgoing transfer edge~$\edge=(\aStop,\bStop)\in\edges$ is relaxed.
If~$\arrivalTime(\aStop,i)+\transferTime(\edge)$ is smaller than~$\arrivalTime(\bStop,i)$, the latter is updated and~$\bStop$ is marked as well.
For a query with source stop~$\aSource\in\stops$ and departure time~$\departureTime$, the algorithm initializes~$\arrivalTime(\aSource,0)$ with~$0$ and all other arrival times in round~$0$ with~$\infty$.
Then round~$0$ is performed, which marks~$\aSource$ and relaxes its outgoing transfers.
Afterwards, new rounds are performed until no more stops have been marked.

An extension of RAPTOR called McRAPTOR~\cite{Del15b} is able to Pareto-optimize additional criteria besides arrival time and number of trips.
McRAPTOR was in turn extended to support multimodal scenarios with unlimited transfers.
The resulting algorithm, MCR~\cite{Del13}, replaces the transfer relaxation phase of (Mc)RAPTOR with a Dijkstra search on a core graph computed with Core-CH.
ULTRA employs the bicriteria variant of MCR, which was originally proposed under the name MR-$\infty$, but which we will call MR for the sake of simplicity.
MR maintains the tentative arrival time~$\arrivalTime(\aVertex,i)$ for every core vertex~$\aVertex\in\coreVertices$, not just for stops.
The transfer relaxation phase runs Dijkstra's algorithm on the core graph, using~$\arrivalTime(\cdot,i)$ as the tentative distances.
The priority queue is initialized with all marked stops, and all stops which are settled by the search are themselves marked.
Note that the Dijkstra search on the core graph can only guarantee to find shortest paths between pairs of stops.
However, the source and target vertices~$\aSource,\aTarget\in\vertices$ may not necessarily be stops.
Initial and final transfers are therefore explored with searches on the upward and downward graph produced by Core-CH, respectively.

Another RAPTOR extension, rRAPTOR~\cite{Del15b}, answers \emph{range queries}, which ask for a Pareto set of journeys for every departure time within a given interval.
rRAPTOR exploits the observation that every Pareto-optimal journey (except for a direct transfer from~$\aSource$ to~$\aTarget$) starts by entering a trip at~$\aSource$ or a stop reachable via a transfer from~$\aSource$.
This limits the number of possible departure times to a small set~$\departureTimes$ of discrete values.
For each of these departure times, rRAPTOR performs a \emph{run} of the basic RAPTOR algorithm.
The departure times are processed in descending order, and the arrival times~$\arrivalTime(\cdot,\cdot)$ are not reset between runs.
As a result, journeys found during the current run are implicitly pruned by journeys that depart later and neither arrive later nor have more trips.
This property of rRAPTOR is called \emph{self-pruning}.

\subparagraph*{Trip-Based Routing.}
A faster alternative to RAPTOR for one-to-one queries is Trip-Based Routing~(TB)~\cite{Wit15}.
It includes a preprocessing phase which computes transfers between pairs of stop events by first generating all possible transfers and then removing unnecessary ones in a ``transfer reduction'' phase.
The query algorithm resembles a breadth-first search on the set of stop events.
Instead of tentative arrival times at stops, TB maintains a~\emph{reached index}~$\reachedIndex(\aTrip)$ for each trip~$\aTrip$.
This is the index~$k$ of the first stop event~$\aTrip[k]$ which has already been reached by the search.
Initially, it is set to~$|\aTrip|$.
Like RAPTOR, TB operates in rounds, where each round scans trip segments collected in a FIFO (first-in-first-out) queue.
When the algorithm reaches a stop event~$\aTrip[j]$, it calls the~$\Enqueue$ operation:
If~$j<\reachedIndex(\aTrip)$, the trip segment~$\aTripSegment{jk}$ with~$k=\reachedIndex(\aTrip)-1$ is added to the queue of the next round.
Then the reached index is updated:
For each trip~$\aTrip'$ of the route~$\aRoute(\aTrip)$ which does not depart before~$\aTrip$, the reached index~$\reachedIndex(\aTrip')$ is set to~$\min(\reachedIndex(\aTrip'),j)$.
Initially, the algorithm processes stops which are reachable from the source stop~$\aSource$ with a transfer.
For each stop~$\aStop$ and each route~$\aRoute$ visiting~$\aStop$, the algorithm finds the earliest trip of~$\aRoute$ which can be entered at~$\aStop$ and calls the~$\Enqueue$ operation for the corresponding stop event.
Then the algorithm performs rounds until the next queue is empty.
A trip segment~$\aTripSegment{jk}$ is scanned by iterating over the stop events from~$\aTrip[j]$ to~$\aTrip[k]$.
For each stop event~$\aTrip[i]$, the outgoing precomputed transfers are relaxed.
A transfer~$(\aTrip[i],\aTrip'[i'])$ is relaxed by calling the~$\Enqueue$ operation for~$\aTrip'[i']$.
Additionally, TB maintains a Pareto set of journeys at the target stop~$\aTarget$.
If~$\aTarget$ is reachable from~$\aStop(\aTrip[i])$ via a transfer, the algorithm adds the produced journey to the Pareto set and removes dominated journeys.

\section{Shortcut Computation}
\label{chap:ULTRA:prepro}
We now present the ULTRA preprocessing phase, which computes shortcut edges that represent intermediate transfers between trips.
These shortcuts must be sufficient for answering every point-to-point query correctly.
This is achieved if every query can be answered with a Pareto set of journeys whose intermediate transfers are all represented by shortcuts.
On the other hand, the number of shortcuts should be as small as possible to allow for fast queries.

We present two variants of the ULTRA preprocessing, which differ in the granularity of the computed shortcuts:
In the \emph{stop-to-stop} variant, shortcuts connect pairs of stops.
This is sufficient for most public transit algorithms, including RAPTOR and CSA.
The \emph{event-to-event} variant computes shortcuts between stop~\emph{events}, which are required by TB.
Unlike stop-to-stop shortcuts, these also provide information about the specific trips between which a transfer is necessary.
Both variants are identical except for a few crucial details, which are discussed explicitly where appropriate.

ULTRA works by enumerating a set of journeys~\candidateJourneys with exactly two trips such that all required shortcuts occur as intermediate transfers in~\candidateJourneys.
For each enumerated journey, the intermediate transfer is unpacked and a shortcut is generated for it.
Before we describe the algorithm, we first establish a definition for~\candidateJourneys that is sufficient for answering all queries while keeping the number of shortcuts as low as possible.
We then provide a high-level overview of the ULTRA shortcut computation and prove that it enumerates~\candidateJourneys.
Afterwards, we discuss running time optimizations to make the algorithm efficient in practice.
Finally, we compare event-to-event ULTRA to the TB preprocessing and show that it is more effective at discarding unnecessary transfers.

\subsection{Enumerating a Sufficient Set of Journeys}
\label{chap:ULTRA:canonical}
Consider the subproblem where only queries between fixed source and target vertices~$\aSource,\aTarget\in\vertices$ must be answered.
Then the following naive algorithm computes a sufficient set of shortcuts:
Enumerate the set~\optimalJourneys of all~$\aSource$-$\aTarget$-journeys~$\aJourney$ which are Pareto-optimal for the departure time~$\departureTime(\aJourney)$, and generate a shortcut for every intermediate transfer that occurs in~\optimalJourneys.
This produces more shortcuts than necessary: if there are multiple Pareto-optimal journeys that are equivalent in both criteria, only one of them is required to answer a query.
The goal is therefore to find a set~$\canonicalJourneys\subseteq\optimalJourneys$ of journeys that excludes such duplicates but is still sufficient for answering all queries correctly.
We observe that every journey in~$\canonicalJourneys$ with more than two trips can be decomposed into subjourneys with two trips each.
Every shortcut that occurs in~$\canonicalJourneys$ also occurs in the much smaller set containing only these subjourneys.
To exploit this algorithmically, we require that~$\canonicalJourneys$ is closed under subjourney decomposition, i.e.,~every subjourney of a journey in~$\canonicalJourneys$ is itself contained in~$\canonicalJourneys$.

\subparagraph*{Tiebreaking Sequences.}
In order to achieve closure under subjourney decomposition, ties between equivalent journeys must be broken in a consistent manner.
For this purpose, we define total orderings on the sets of routes and vertices with a~\emph{route index} function~$\routeIndex\colon\routes\to\mathbb{N}$ and a~\emph{vertex index} function~$\vertexIndex\colon\vertices\to\mathbb{N}$.
Then ties between equivalent journeys are broken as follows:
Journeys which end with trip segments are preferred over journeys which end with (non-empty) transfers.
For journeys which end with a trip segment~$\aTripSegment{ij}$, the index of the route~$\aRoute(\aTrip)$ and the index~$i$ where the trip segment starts are used as tiebreakers, in this order.
For journeys which end with an edge~$(\bVertex,\aVertex)$, ties are broken first by considering the arrival time at~$\bVertex$, and then by considering the vertex index~$\vertexIndex(\bVertex)$.
If two journeys share a non-empty suffix, this suffix is ignored and the respective prefixes of the journeys are compared instead.
To formalize these tiebreaking rules, we associate with each $\aSource$-$\aTarget$-journey~$\aJourney$ a unique~\emph{tiebreaking sequence}.
The tiebreaking sequence~$\tiebreakingSequence(\aVertex,\aJourney)$ of a vertex~$\aVertex\in\vertices(\aJourney)$ with~$\aVertex\neq\aSource$ is defined as
\[
\tiebreakingSequence(\aVertex,\aJourney):=\begin{cases}
	\langle\arrivalTime(\aJourney_{\aSource\aVertex}),\routeIndex(\aRoute(\aTrip)),\mathmakebox[0pt][l]{i}\hphantom{\infty},\mathmakebox[0pt][l]{\infty}\hphantom{\arrivalTime(\aJourney_{\aSource\bVertex})},\mathmakebox[0pt][l]{\infty}\hphantom{\vertexIndex(\bVertex)}\rangle&\hspace{-0.15cm}\text{if }\aJourney_{\aSource\aVertex}\text{ ends with a trip segment }\aTripSegment{ij}\\
	\langle\arrivalTime(\aJourney_{\aSource\aVertex}),\mathmakebox[0pt][l]{\infty}\hphantom{\routeIndex(\aRoute(\aTrip))},\infty,\arrivalTime(\aJourney_{\aSource\bVertex}),\vertexIndex(\bVertex)\rangle&\hspace{-0.15cm}\text{if }\aJourney_{\aSource\aVertex}\text{ ends with an edge }(\bVertex,\aVertex).\\
\end{cases}
\]
The tiebreaking sequence of an~$\aSource$-$\aTarget$-journey~$\aJourney$ with vertex sequence~$\vertices(\aJourney)=\langle\aSource=\aVertex_1,\dots,\aVertex_k=\aTarget\rangle$ is defined as~$\tiebreakingSequence(\aJourney):=\tiebreakingSequence(\aVertex_k,\aJourney)\circ\dots\circ\tiebreakingSequence(\aVertex_2,\aJourney)$.
This sequence is unique among all~$\aSource$-$\aTarget$-journeys.
In particular, if two journeys~$\aJourney$ and~$\aJourney'$ end with trip segments~$\aTripSegmentA{ij}\neq\aTripSegmentB{mn}$, then their tiebreaking sequences are different.
If~$\arrivalTime(\aJourney)=\arrivalTime(\aJourney')$ and~$\aRoute(\aTripA)=\aRoute(\aTripB)$, then~$\aTripA=\aTripB$ and~$j=n$ must hold because the trips cannot overtake each other.
Then the tiebreaking sequences are different due to~$i\neq{}m$.
Sequences are ordered lexicographically: for sequences~$A=\langle{}a_1,a_2,\dots,a_k\rangle$ and~$B=\langle{}b_1,b_2,\dots,b_k\rangle$ of equal length, $A<B$ if~$a_1<b_1$, or~$a_1=b_1$ and~$\langle{}a_2,\dots,a_k\rangle<\langle{}b_2,\dots,b_k\rangle$.
For sequences of different length, the shorter one is padded with~$-\infty$ on the right side before they are compared.

\subparagraph*{Canonical Journeys.}
Because tiebreaking sequences are strictly ordered, ambiguities between equivalent journeys can be resolved by replacing the criterion arrival time with the tiebreaking sequence.
We say that an $\aSource$-$\aTarget$-journey~$\aJourney$ \emph{canonically dominates} another $\aSource$-$\aTarget$-journey~$\aJourney'$ if~$\tiebreakingSequence(\aJourney)<\tiebreakingSequence(\aJourney')$ and~$|\aJourney|\leq|\aJourney'|$.
Since the two tiebreaking sequences cannot be equal, there is no need to distinguish between strong and weak canonical dominance.
An $\aSource$-$\aTarget$-journey~$\aJourney$ is called \emph{canonical} if it is Pareto-optimal with respect to tiebreaking sequence and number of trips for the departure time~$\departureTime(\aJourney)$, i.e.,~if no other $\aSource$-$\aTarget$-journey exists which is feasible for~$\departureTime(\aJourney)$ and canonically dominates~$\aJourney$.
Since no two journeys can be equivalent in both criteria, the set which consists of all feasible canonical journeys is the only Pareto set for any given query.
We call this the \emph{canonical Pareto set}.
The set~\canonicalJourneys is the union of the canonical Pareto sets for all possible $\aSource$-$\aTarget$-queries.
This set is closed under subjourney decomposition:

\begin{lemma}\label{th:canonical_subjourney}
	For every canonical~$\aSource$-$\aTarget$-journey~$\aJourney$ and every pair~$\aVertex,\bVertex\in\vertices(\aJourney)$ of vertices visited by~$\aJourney$, the subjourney~$\aJourney_{\aVertex\bVertex}$ is canonical.
\end{lemma}
\begin{proof}
	Assume that~$\aJourney_{\aVertex\bVertex}$ is not canonical.
	Then there is a journey~$\aJourney'_{\aVertex\bVertex}$ such that~$\aJourney'_{\aVertex\bVertex}$ is feasible for~$\departureTime(\aJourney_{\aVertex\bVertex})$, $\tiebreakingSequence(\aJourney'_{\aVertex\bVertex})<\tiebreakingSequence(\aJourney_{\aVertex\bVertex})$ and~$|\aJourney'_{\aVertex\bVertex}|\leq|\aJourney_{\aVertex\bVertex}|$.
	Because~$\aJourney'_{\aVertex\bVertex}$ does not depart earlier or arrive later than~$\aJourney_{\aVertex\bVertex}$, replacing~$\aJourney_{\aVertex\bVertex}$ with~$\aJourney'_{\aVertex\bVertex}$ in~$\aJourney$ yields a feasible journey~$\aJourney'$ with~$|\aJourney'|\leq|\aJourney|$.
	Adding the prefix~$\aJourney_{\aSource\aVertex}$ to~$\aJourney'_{\aVertex\bVertex}$ and~$\aJourney_{\aVertex\bVertex}$ adds identical suffixes to both tiebreaking sequences.
	This does not change their relative order, so~$\tiebreakingSequence(\aJourney'_{\aSource\bVertex})<\tiebreakingSequence(\aJourney_{\aSource\bVertex})$.
	Similarly, adding the suffix~$\aJourney_{\bVertex\aTarget}$ to~$\aJourney'_{\aSource\bVertex}$ and~$\aJourney_{\aSource\bVertex}$ adds identical prefixes to both tiebreaking sequences, which does not change their relative order.
	Therefore, $\tiebreakingSequence(\aJourney')<\tiebreakingSequence(\aJourney)$ and~$\aJourney$ is not canonical.
\end{proof}

\subparagraph*{Candidate Journeys.}
We exploit the closure of~$\canonicalJourneys$ under subjourney decomposition by defining a suitable set of subjourneys that need to be enumerated.
A \emph{candidate} is a journey that consists of two trips connected by an intermediate transfer, but with empty initial and final transfers.
Every canonical journey that uses at least two trips can be decomposed into candidate subjourneys.
By Lemma~\ref{th:canonical_subjourney}, these subjourneys are themselves canonical.
Accordingly, every shortcut that occurs in~$\canonicalJourneys$ also occurs in the set~$\candidateJourneys\subseteq\canonicalJourneys$ of canonical candidate journeys.
A sufficient set of shortcuts can therefore be computed by enumerating~$\candidateJourneys$.

\subparagraph*{Canonical MR.}
Canonical Pareto sets can be computed by making slight modifications to MR in order to ensure proper tiebreaking:
Firstly, at the start of each round, the collected routes are sorted according to~$\routeIndex$ before they are scanned.
The second change concerns the keys of vertices in the Dijkstra priority queue.
In standard MR, the key of a vertex~$\aVertex$ in round~$i$ is the tentative arrival time~$\arrivalTime(\aVertex,i)$ at~$\aVertex$ with~$i$ trips.
This is now replaced with~$\langle\arrivalTime(\aVertex,i),\vertexIndex(\aVertex)\rangle$.
The resulting implementation of MR, which we call~\emph{canonical MR}, finds equivalent journeys in increasing order of tiebreaking sequence.
Hence, canonical journeys are found first and all other equivalent journeys are discarded because they are weakly dominated by them.
This is proven by the following lemma:
\begin{lemma}\label{th:canonical_mr}
	Canonical MR returns the canonical Pareto set for every query.
\end{lemma}
\begin{proof}
	See Appendix~\ref{app:ULTRA:canonicalMR}.
\end{proof}

The journeys returned by a straightforward (non-canonical) implementation of MR are not closed under subjourney decomposition.
An example demonstrating this is given in Appendix~\ref{app:ULTRA:nonCanonicalMR}.

\subsection{Algorithm Overview}
We now describe how~\candidateJourneys can be enumerated efficiently.
Directly applying the definition of~\candidateJourneys yields a simple but wasteful approach:
For every possible source stop and every possible departure time, a one-to-all canonical MR search restricted to the first two rounds is performed.
A candidate~\aCandidateJourney is canonical if there is no feasible journey~\aWitnessJourney with at most two trips that canonically dominates~$\aCandidateJourney$ (and is therefore found before~\aCandidateJourney by the respective canonical MR search).
If such a journey~\aWitnessJourney exists, we call it a \emph{witness} since its existence proves that~$\aCandidateJourney$ is not canonical.
Unlike candidates, witnesses may have non-empty initial or final transfers, and they may use fewer than two trips.
If there is no witness for a candidate~\aCandidateJourney, the corresponding canonical MR search will include~\aCandidateJourney in its Pareto set.
A shortcut representing the intermediate transfer of~\aCandidateJourney is then generated.

\subparagraph*{Adapting rRAPTOR.}
The reason this approach is wasteful is that it does not exploit the self-pruning property of rRAPTOR: if journeys with later departure times are explored first, they can be used to dominate worse journeys with an earlier departure time.
We therefore adapt rRAPTOR to the ULTRA setting: the RAPTOR search which is performed in each run is replaced with a canonical two-round MR search.
This version of rRAPTOR is then invoked for each possible source stop~$\aSource\in\stops$, with a departure time interval that covers the entire duration of the timetable.

We can make further improvements by carefully choosing the departure times for which runs are performed.
rRAPTOR performs an run for every possible departure time~\departureTime at~\aSource.
A departure time~\departureTime is possible if there is a stop~\aStop (which may be~\aSource itself) that is reachable from~\aSource via an initial transfer of length~$\transferTime(\aSource,\aStop)$ and a trip that departs from~\aStop at~$\departureTime+\transferTime(\aSource,\aStop)$.
If transfers are unrestricted, the number of possible departure times is very high because typically most stops in the network will be reachable from~\aSource.
Accordingly, a straightforward multimodal adaptation of rRAPTOR would perform many runs and therefore be slow.
In the context of ULTRA, however, most possible departure times require a non-empty initial transfer, which means that the corresponding runs would not find any candidates.
Since the goal is to enumerate candidates, ULTRA only performs the runs for departure events that occur directly at~\aSource.
Let~$\departureTimes=\{\departureTime^0,\dots,\departureTime^k\}$ be the set of possible departure times directly at~$\aSource$, sorted in ascending order.
The run for~$\departureTime^i$ explores candidates departing at~$\departureTime^i$ and witnesses with departure times in the interval~$[\departureTime^i,\departureTime^{i+1})$.
We define~$\departureTime^{k+1}:=\infty$ to ensure that the run for~$\departureTime^k$ explores all witnesses which depart after~$\departureTime^k$.
By integrating the witness search into the candidate runs, the algorithm skips many witnesses which would be required to answer a range query but are irrelevant for dominating candidates.
Thus, the ULTRA preprocessing is much faster than a straightforward multimodal adaptation of rRAPTOR.

\subparagraph*{Pseudocode.}
High-level pseudocode for the ULTRA shortcut computation scheme is given by Algorithm~\ref{alg:ULTRA:preprocessing}.
For each source stop~$\aSource\in\stops$, the algorithm performs the modified multimodal rRAPTOR search described above.
To avoid redundant Dijkstra searches, initial transfers to all other stops are explored only once per source stop (line~\ref{alg:ULTRA:initialTransfers}) and the results are then reused for each run in line~\ref{alg:ULTRA:collect1}.
The departure times at~\aSource for which runs need to be performed are collected in line~\ref{alg:ULTRA:collect0}.
The runs are performed in lines~\ref{alg:ULTRA:collect1}--\ref{alg:ULTRA:endloop}.
Each run consists of two canonical MR rounds, which are subdivided into three phases: collecting routes and sorting them according to~$\routeIndex$ (lines~\ref{alg:ULTRA:collect0} and~\ref{alg:ULTRA:collect1}), scanning routes (lines~\ref{alg:ULTRA:scan1} and~\ref{alg:ULTRA:scan2}), and relaxing transfers with a Dijkstra search (lines~\ref{alg:ULTRA:relax1} and~\ref{alg:ULTRA:relax2}).
After the final transfer relaxation phase in line~\ref{alg:ULTRA:relax2}, the remaining candidates which have not been dominated by witnesses are canonical, so shortcuts representing their intermediate transfers are added to the shortcut graph in line~\ref{alg:ULTRA:endloop}.

\begin{figure}[t]
	\begin{minipage}{\textwidth}
		\begin{algorithm}[H]
			\caption{ULTRA transfer shortcut computation}\label{alg:ULTRA:preprocessing}
			\Input{Public transit network~$(\stops,\trips,\routes,\graph)$, core graph~$\coreGraph=(\coreVertices,\coreEdges)$}
			\Output{Shortcut graph~$\shortcutGraph=(\stops,\shortcutEdges)$}
			\BlankLine
			\ForEach{$\aSource \in \stops$\label{alg:ULTRA:mainLoop}}{
				Clear all arrival labels and Dijkstra queues\;
				$\transferTime(\aSource,\cdot)\leftarrow{}$Compute transfer times in~$\coreGraph$ from~$\aSource$ to all stops\label{alg:ULTRA:initialTransfers}\;
				$\departureTimes\leftarrow{}$Collect departure times of trips at~$\aSource$\label{alg:ULTRA:collect0}\;
				\ForEachComment{Canonical MR run}{$\departureTime^i\in\departureTimes$ in descending order}{
					Collect and sort routes reachable within~$[\departureTime^i,\departureTime^{i+1})$\label{alg:ULTRA:collect1}\Comment[r]{first round}
					Scan routes\label{alg:ULTRA:scan1}\;
					Relax transfers\label{alg:ULTRA:relax1}\;
					Collect and sort routes serving updated stops \Comment[r]{second round}
					Scan routes\label{alg:ULTRA:scan2}\;
					$\canonicalShortcuts\leftarrow{}$Relax transfers, thereby collecting shortcuts \label{alg:ULTRA:relax2}\;
					$\shortcutEdges\leftarrow\shortcutEdges\cup\canonicalShortcuts$\label{alg:ULTRA:endloop}\;
				}
			}
		\end{algorithm}
	\end{minipage}
\end{figure}

\subparagraph*{Extracting Shortcuts.}
The final transfer relaxation phase in line~\ref{alg:ULTRA:relax2} identifies canonical candidates and extracts their shortcuts.
Whenever a stop is settled during the Dijkstra search, the algorithm checks whether the corresponding journey~\aJourney is a candidate, i.e.,~has an empty initial and final transfer.
If so, we know that~$\aJourney$ is canonical because any witness which canonically dominates it would have been found already.
Therefore, an edge representing the intermediate transfer of~$\aJourney$ is added to the shortcut graph~$\shortcutGraph$.
In order to extract the intermediate transfer, each vertex~\aVertex maintains two parent pointers~$\parent_1[\aVertex]$ and~$\parent_2[\aVertex]$, where~$\parent_k[\aVertex]$ is the parent for reaching~$\aVertex$ using~$k$~trips~(i.e.,~within the~$k$-th~MR round).
If the journey to~\aVertex ends with a trip, $\parent_k[\aVertex]$ points to the stop where this trip was entered.
If the journey ends with a transfer, it points to the stop where the transfer starts.
For a candidate ending at a stop~$\aTarget$, the shortcut representing its intermediate transfer is given by~$(\parent_1[\parent_2[\aTarget]],\parent_2[\aTarget])$.
Since intermediate transfers only need to be extracted for candidates, the parent pointer is set to a special value~$\bot$ if the corresponding journey has a non-empty initial or final transfer.
Then the final Dijkstra search in~line~\ref{alg:ULTRA:relax2} can check whether the journey ending at a stop~$\aVertex$ is a candidate or a witness by inspecting~$\parent_2[\aVertex]$.

The event-to-variant of ULTRA generates shortcuts not between stops, but between stop events.
The parent pointer definitions are changed accordingly:
If the journey to a vertex~\aVertex ends with a trip, $\parent_k[\aVertex]$ now points to the stop event where this trip was entered.
If the journey ends with a transfer, it points to the stop event where the preceding trip was exited.
Since only candidates have valid parent pointers and candidates have empty initial transfers, this preceding trip always exists.
For a candidate that ends at a stop~$\aTarget$, the corresponding shortcut is now given by~$(\parent_1[\aStop(\parent_2[\aTarget])],\parent_2[\aTarget])$.

\subparagraph*{Repairing Self-Pruning.}
Using a rRAPTOR-based approach with self-pruning allows ULTRA to discard many irrelevant candidates early on.
However, self-pruning can also cause the algorithm to discard canonical journeys.
By exploring journeys with later departure times first, rRAPTOR implicitly maximizes departure time as a third criterion.
With this additional criterion, there may be queries for which all Pareto-optimal journeys include suboptimal subjourneys.
An example of this is shown in Figure~\ref{fig:ultraWeakDominance}.
In this case, some canonical candidates are suboptimal for three criteria and therefore not found by the rRAPTOR-based scheme.
Moreover, in the depicted network, there is no Pareto set for two criteria which is closed under subjourney decomposition and only includes journeys which are Pareto-optimal for three criteria.
Hence, the problem cannot be avoided by defining~\canonicalJourneys in a different manner.
Instead, we modify the dominance criterion to ensure that canonical journeys are not discarded.

\begin{figure}
	\centering
	\begin{tikzpicture}[scale=1]

    \node (s)  at ( 0.00, 0.00) {};%
    \node (a)  at ( 2.00, 0.00) {};%
    \node (b)  at ( 4.00, 0.00) {};%
    \node (c)  at ( 6.00, 0.00) {};%
    \node (d)  at ( 8.00, 0.00) {};%
    \node (e)  at (10.00, 0.00) {};%
    \node (t)  at (12.00, 0.00) {};%
    \node (bb) at ( 6.00, 1.50) {};%
    
    \draw [\primarycolor{0}, line width=2.5pt, routeArrow, rounded corners = 20] (s) -- (a);
    \draw [\secondarycolor{0}, line width=2.5pt, routeArrow, rounded corners = 20] (b) -- (c);
    \draw [\tertiarycolor{0}, line width=2.5pt, routeArrow, rounded corners = 20] (e) -- (t);
    \draw [\quaternarycolor{0}, line width=2.5pt, routeArrow, rounded corners = 20] (bb) -- (8.0, 1.5) -- (d);

    \node [align=left,text=\primarycolor{1}] at ( 1.00, 0.30) {$0\rightarrow1$};%
    \node [align=left,text=\secondarycolor{1}] at ( 5.00, 0.30) {$2\rightarrow3$};%
    \node [align=left,text=\quaternarycolor{1}] at ( 7.00, 1.80) {$4\rightarrow5$};%
    \node [align=left,text=\tertiarycolor{1}] at (11.00, 0.30) {$7\rightarrow8$};%

    \draw [edgeColor, line width=1pt]  (a) -- (b) node [midway, anchor=south] {$1$};
    \draw [edgeColor, line width=1pt]  (c) -- (d) node [midway, anchor=south] {$1$};
    \draw [edgeColor, line width=1pt]  (d) -- (e) node [midway, anchor=south] {$1$};
    \draw [edgeColor, line width=1pt, rounded corners = 20]  (b) -- (4.0, 1.5) -- (bb) node [midway, anchor=south east] {$1$};

    \node at (s) [vertex,draw=nodeColor!100,fill=nodeColor!15] {\gs};%
    \node at (a) [vertex,draw=nodeColor!100,fill=nodeColor!15] {\gs};%
    \node at (b) [vertex,draw=nodeColor!100,fill=nodeColor!15] {\gs};%
    \node at (c) [vertex,draw=nodeColor!100,fill=nodeColor!15] {\gs};%
    \node at (d) [vertex,draw=nodeColor!100,fill=nodeColor!15] {\gs};%
    \node at (e) [vertex,draw=nodeColor!100,fill=nodeColor!15] {\gs};%
    \node at (t) [vertex,draw=nodeColor!100,fill=nodeColor!15] {\gs};%
    \node at (bb) [vertex,draw=nodeColor!100,fill=nodeColor!15] {\gs};%

    \node at (s) [text=nodeColor!100] {\small{$\aSource$}};%
    \node at (a) [text=nodeColor!100] {\small{$a$}};%
    \node at (b) [text=nodeColor!100] {\small{$b$}};%
    \node at (c) [text=nodeColor!100] {\small{$c$}};%
    \node at (d) [text=nodeColor!100] {\small{$d$}};%
    \node at (e) [text=nodeColor!100] {\small{$e$}};%
    \node at (t) [text=nodeColor!100] {\small{$\aTarget$}};%
    \node at (bb) [text=nodeColor!100] {\small{$c^\prime$}};%

\end{tikzpicture}%
	\caption[An example network where every journey that is Pareto-optimal with respect to the three criteria arrival time, number of trips, and departure time includes a suboptimal subjourney.]{%
		An example network where every $\aSource$-$\aTarget$-journey that is Pareto-optimal with respect to the three criteria arrival time, number of trips, and departure time includes a suboptimal subjourney.
		Transfer edges (gray) are labeled with their travel time, while trips (colored) are labeled with~$\departureTime\to\arrivalTime$.
		The two Pareto-optimal journeys are~$\aJourney=\big\langle\!
			\langle\aSourceX\rangle,
			\langle\textcolor{\primarycolor{1}}{0\rightarrow1}\rangle,
			\langle{}a,b\rangle,
			\langle\textcolor{\secondarycolor{1}}{2\rightarrow3}\rangle,
			\langle{}c,d,e\rangle,
			\langle\textcolor{\tertiarycolor{1}}{7\rightarrow8}\rangle,
			\langle\aTargetX\rangle
			\!\big\rangle$ and~$\aJourney'=\big\langle\!
			\langle\aSourceX\rangle,
			\langle\textcolor{\primarycolor{1}}{0\rightarrow1}\rangle,
			\langle{}a,b,c'\rangle,
			\langle\textcolor{\quaternarycolor{1}}{4\rightarrow5}\rangle,
			\langle{}d,e\rangle,
			\langle\textcolor{\tertiarycolor{1}}{7\rightarrow8}\rangle,
			\langle\aTargetX\rangle
			\!\big\rangle$.
		The subjourney~$\aJourney_{b\aTarget}$ of~$\aJourney$ is not Pareto-optimal because it has an earlier departure time than~$\aJourney'_{b\aTarget}$ and is otherwise equivalent.
		Likewise, the subjourney~$\aJourney'_{\aSource{}d}$ is suboptimal because it has a later arrival time than~$\aJourney_{\aSource{}d}$.
	}%
	\label{fig:ultraWeakDominance}%
\end{figure}

For a journey~$\aJourney$, let~$\run(\aJourney)$ be the highest~$i$ with~$\departureTime^i\in\departureTimes$ such that~$\departureTime(\aJourney)\geq\departureTime^i$.
This is the run in which our modified rRAPTOR finds~$\aJourney$.
For each vertex~$\aVertex$ and round~$i$, the algorithm maintains a label~$\aLabel(\aVertex,i)=\left(\arrivalTime(\aVertex,i),\parent_i[\aVertex],\run(\aVertex,i)\right)$, where~$\arrivalTime(\aVertex,i)$ is the tentative arrival time, $\parent_i[\aVertex]$ is the parent pointer, and~$\run(\aVertex,i)$ is the run of the journey corresponding to this label, which we denote as~$\aJourney(\aVertex,i)$.
Let~$\aLabel=(\arrivalTime,\parent,j)$ be the label of a new journey~$\aJourney$ which is found by the algorithm at~$\aVertex$ in round~$i$.
Normally, rRAPTOR discards~$\aJourney$ if it is weakly dominated by~$\aJourney(\aVertex,i)$, i.e.,~$\arrivalTime(\aVertex,i)\leq\arrivalTime$.
Otherwise, it replaces~$\aLabel(\aVertex,i)$ with~$\aLabel$.
Our modified algorithm only discards~$\aJourney$ if one of the following conditions is fulfilled: (1)~$\aJourney$~is not a prefix of a candidate, i.e.,~$\parent=\bot$, (2)~$\aJourney$~is strongly dominated by~$\aJourney(\aVertex,i)$, i.e.,~$\arrivalTime(\aVertex,i)<\arrivalTime$ or~$\arrivalTime(\aVertex,i-1)\leq\arrivalTime$, or~(3),~$\aJourney(\aVertex,i)$ was found in the current run, i.e.,~$\run(\aVertex,i)=j$.
With this modified dominance condition, we can prove that the ULTRA preprocessing computes a sufficient shortcut graph:
\begin{theorem}\label{th:correctness}
	For every canonical journey~$\aJourney=\langle\aPath_0,\aTripSegmentC{ij}{0},\dots,\aTripSegmentC{mn}{k-1},\aPath_k\rangle$, every intermediate transfer in~$\aJourney$ is represented by an edge in the shortcut graph computed by ULTRA.
\end{theorem}
\begin{proof}
	Consider an intermediate transfer~$\aPath_{x+1}$ of~$\aJourney$ and the corresponding candidate subjourney~\smash{\mbox{$\aCandidateJourney=\langle\aTripSegmentC{gh}{x},\aPath_{x+1},\aTripSegmentC{pq}{x+1}\rangle$}}.
	We show that the modified rRAPTOR search for the source stop~$\aStop(\aTrip_x[g])$ finds this candidate in the run for~$\departureTime(\aCandidateJourney)$ and inserts a shortcut for it.
	Assume~$\aCandidateJourney$ is not found.
	Then some prefix~$\aJourney'$ of~$\aCandidateJourney$ is discarded by the search in favor of a witness~$\aWitnessJourney$.
	By Lemma~\ref{th:canonical_subjourney}, ~$\aJourney'$ is canonical and therefore not strongly dominated by~$\aWitnessJourney$.
	Then by our modified dominance criterion, $\aWitnessJourney$ must have been found in the same canonical MR run as~$\aJourney'$.
	However, by Lemma~\ref{th:canonical_mr}, canonical MR discards~$\aWitnessJourney$ in favor of~$\aJourney'$, a contradiction.
\end{proof}

\subsection{Optimizations}
We now discuss running time optimizations which are not mentioned in the high-level overview given by Algorithm~\ref{alg:ULTRA:preprocessing}.
These optimizations are crucial for achieving fast preprocessing times.

\subparagraph*{Initial Route Collection.}
An rRAPTOR run with departure time~$\departureTime^i$ explores journeys that depart at~$\aSource$ within the interval~$[\departureTime^i,\departureTime^{i+1})$.
Line~\ref{alg:ULTRA:collect1} collects the set~$\routes(\departureTime^i)$ of routes which must be scanned in the first round of this run.
This set consists of all routes~\aRoute for which there is a stop~\aStop visited by~\aRoute and a trip~\aTrip of~\aRoute such that~$\departureTime(\aTrip,\aStop)-\transferTime(\aSource,\aStop)\in[\departureTime^i,\departureTime^{i+1})$.
In order to speed up this step, the set~$\routes(\departureTime^i)$ is precomputed when~$\departureTime^i$ is added to the set~$\departureTimes$ of candidate departure times in line~\ref{alg:ULTRA:collect0}.
This leads to the following procedure for calculating~$\departureTimes$ and~$\routes(\cdot)$:
First, the algorithm collects all departure triplets~$(\aVertex,\departureTime,\aRoute)$ of departure stop~$\aVertex$, departure time~$\departureTime$, and route~$\aRoute$ which occur in the network.
They are then sorted by their departure time at~$\aSource$, which is~$\departureTime-\transferTime(\aSource,\aVertex)$, and processed in descending order.
The algorithm maintains a tentative set~$\routes'$ of routes for the next candidate departure time that is added to~$\departureTimes$.
For each triplet~$(\aVertex,\departureTime,\aRoute)$, the algorithm checks whether~$\aVertex=\aSource$.
If~$\aVertex\neq\aSource$, $\aRoute$ is added to~$\routes'$.
Otherwise, $\departureTime$ is a candidate departure time.
If~$\departureTime$ is already contained in~$\departureTimes$, the algorithm already found another route departing from~$\aSource$ at~$\departureTime$, so~$\aRoute$ is added to~$\routes(\departureTime)$.
Otherwise, $\departureTime$ is added to~$\departureTimes$, $\routes(\departureTime)$ is set to $\routes'\cup\{\aRoute\}$, and~$\routes'$ is cleared.

\subparagraph*{Limited Dijkstra Searches.}
The algorithm can be sped up by introducing a stopping criterion to the Dijkstra search for final transfers in line~\ref{alg:ULTRA:relax2}.
For this purpose, the preceding route scanning phase in line~\ref{alg:ULTRA:scan2} counts the number of stops which are marked because their tentative arrival time is improved by a candidate.
Whenever such a stop is settled in line~\ref{alg:ULTRA:relax2}, the counter is decreased.
Once the counter reaches zero, we know that the Dijkstra search has processed all candidates which have been found in this run, so it is stopped.

A similar stopping criterion is applied to the intermediate Dijkstra search in line~\ref{alg:ULTRA:relax1}.
Here, the first route scanning phase in line~\ref{alg:ULTRA:scan1} counts the stops whose tentative arrival time is improved by a candidate prefix, i.e.,~a journey with an empty initial transfer.
As in line~\ref{alg:ULTRA:relax2}, the Dijkstra search is stopped as soon as no such stops are left in the queue.
This does not affect the correctness of the computed shortcut graph~$\shortcutGraph$, since all candidates are still processed.
However, some of the witnesses that are pruned might be required to dominate non-canonical candidates.
In this case, superfluous shortcuts will be added to~$\shortcutGraph$.
This can be counteracted by continuing the Dijkstra search for some time after the last candidate prefix has been extracted.
We introduce a parameter~\witnessLimit called the \emph{witness limit} which determines how long the search continues.
Let~$\arrivalTime$ be the arrival time of the last extracted candidate prefix.
Instead of stopping the Dijkstra search immediately, it continues until the smallest element in the queue has an arrival time greater than~$\arrivalTime+\witnessLimit$.

Once a Dijkstra search is stopped, the remaining witness labels are kept in the queue since they may dominate candidates in later runs.
This requires that the two Dijkstra searches in lines~\ref{alg:ULTRA:relax1} and~\ref{alg:ULTRA:relax2} use separate queues, so that labels from the final Dijkstra search of a previous run do not interfere with the intermediate Dijkstra search of the current run.
As a consequence, if a label is discarded because it is dominated, it must be explicitly removed from any queues that still contain it.
Moreover, the run in which a label is settled may no longer be the same one in which it was enqueued.
Accordingly, the run in which a journey~$\aJourney$ is found may no longer equal~$\run(\aJourney)$.
To ensure that the dominance condition is applied correctly, the run of a newly created label is carried over from its parent label, rather than setting it to the currently performed run.

With these changes, the only remaining part of the algorithm that performs an unlimited Dijkstra search on the core graph is the initial transfer relaxation in line~\ref{alg:ULTRA:initialTransfers}.
Unlike the searches for the intermediate and final transfers, this search is only performed once for every source stop instead of once per run, so its impact on the overall running time is small.

\subparagraph*{Pruning with Found Shortcuts.}
Once a shortcut has been found and added to the shortcut graph~$\shortcutGraph$, it is no longer necessary to find candidates which produce the same shortcut.
We exploit this by further restricting the definition of candidates: a journey is only classified as a candidate if its intermediate transfer is not contained in the set of already computed shortcuts.
Since this reduces the number of candidates, the stopping criterion for the Dijkstra searches in lines~\ref{alg:ULTRA:relax1} and~\ref{alg:ULTRA:relax2} may be applied earlier, further saving preprocessing time.

Whenever a potential candidate is found during the second route scanning phase in line~\ref{alg:ULTRA:scan2}, the stop-to-stop variant of ULTRA checks if the corresponding shortcut is already contained in~$\shortcutGraph$.
If so, the journey is classified as a witness by setting its parent pointer to~$\bot$.
In the event-to-event variant, this check is more expensive since the number of shortcuts is much larger.
Furthermore, since an event-to-event shortcut typically occurs in much fewer candidate journeys than its stop-to-stop counterpart, it is much less likely that the shortcut is already contained in~$\shortcutGraph$.
Our preliminary experiments showed that the benefit of potentially dismissing a candidate no longer outweighs the work required to look up the shortcut.
Therefore, the check is skipped in the event-to-event variant.

When a candidate is extracted from the Dijkstra queue in line~\ref{alg:ULTRA:relax2} and a shortcut is inserted for it, there may be other candidates remaining in the queue that use the same intermediate transfer.
These must be turned into witnesses by setting the respective parent pointers to~$\bot$.
This requires keeping track of all candidates belonging to a particular shortcut.
Within a single canonical MR run, the search can find at most one intermediate transfer ending at a particular stop or stop event.
In stop-to-stop ULTRA, each stop~\aStop therefore maintains a list of all non-dominated candidates whose intermediate transfer ends at~\aStop.
The event-to-event variant does the same for each stop event.
When a shortcut is inserted, all candidates in the corresponding list are turned into witnesses.

\subparagraph*{Transfer Graph Contraction.}
As with~MCR~\cite{Del13}, the Dijkstra searches are performed on a core graph, which is constructed with Core-CH in advance.
Since~ULTRA~only needs to compute journeys between pairs of stops, rather than arbitrary vertices in the transfer graph, only transfers that start and end at stops are relevant.
Accordingly, the initial and final transfer searches which MCR performs on the upward and downward CH graphs can be omitted.

Another type of contraction is performed for cliques of stops which have a pairwise distance of~0 in the transfer graph.
These cliques typically occur when different platforms of a larger station are modeled as individual stops.
Each such clique is contracted into a single stop in order to decrease the number of canonical MR runs that need to be performed.
The number of runs for a source stop~\aSource is equal to the number of unique departure times at~\aSource.
If a departure time occurs at multiple stops within a clique with transfer distance~0, then the algorithm will perform one run for this departure time at each stop.
The journeys found by these runs will be identical, save for initial transfers of length~0.
By contracting the clique into a single stop, these redundant runs are merged into one.
This does not affect the correctness of the algorithm since it is conceptually equivalent to allowing candidates to begin with an initial transfer of length~0.

\subparagraph*{Parallelization.}
Finally, we observe that~ULTRA~allows for trivial parallelization.
The preprocessing algorithm searches for candidates once for every possible source stop~(line~\ref{alg:ULTRA:mainLoop} of Algorithm~\ref{alg:ULTRA:preprocessing}).
As these searches are mostly independent of each other, they can be distributed to parallel threads and the results are then combined in a final sequential step.
The only aspect of the algorithm that introduces a dependency between the searches for different source stops is the restricted candidate definition: a journey is only considered a candidate if no shortcut has yet been added for its intermediate transfer.
If a shortcut was added by a different thread, the algorithm will not notice this.
However, since this is merely a performance optimization, the algorithm remains correct if only shortcuts added by the current thread are considered.

\subsection{Integration with Trip-Based Routing}
Unlike other public transit algorithms, TB on its own already requires a preprocessing step, even when used without~ULTRA.
One possible approach for enabling unlimited transfers in TB is with a \emph{sequential} three-phase algorithm:
First, shortcuts between stops are generated with the stop-to-stop variant of the~ULTRA preprocessing.
These are then used as input for the TB preprocessing, which generates event-to-event shortcuts that can be used by the~ULTRA-TB query.
However, we will show that an \emph{integrated} two-phase approach is superior.
Here, the TB preprocessing is replaced entirely by the event-to-event variant of the~ULTRA preprocessing.
The resulting shortcuts between stop events are then used as input for the~ULTRA-TB query.
The advantage of the integrated approach is that it produces fewer shortcuts because ULTRA applies stricter pruning rules than the TB preprocessing.
Both algorithms enumerate journeys with at most two trips in order to find witnesses which prove that a potential shortcut is not necessary.
The TB preprocessing does this in a ``transfer reduction'' step, after all potential shortcuts have been generated.
Since the latter is no longer feasible with unlimited transfers, ULTRA interleaves the generation and pruning of shortcuts.
Furthermore, ULTRA examines a larger set of witnesses.
In the TB preprocessing, witnesses must start with the same stop event as the candidate, whereas ULTRA also considers witnesses that start with an non-empty initial transfer or a different initial trip.
Furthermore, because the TB preprocessing explores intermediate transfers by iterating along the stop sequence of the initial trip in reverse, a candidate cannot be pruned by witnesses that exit the initial trip before the candidate.
Overall,~ULTRA has more options for pruning candidates and thus produces fewer shortcuts.

\section{Query Algorithms}
\label{chap:ULTRA:query}
ULTRA shortcuts can be combined with any public transit query algorithm that normally requires one-hop transfers.
The idea is to replace the original transfer graph with the precomputed shortcuts and run the algorithm on the resulting network.
Some algorithms, including RAPTOR, CSA and TB, normally require that the transfer graph is transitively closed.
While this is not the case for the ULTRA shortcut graph, this is not a problem:
Theorem~\ref{th:correctness} proves that journeys with two consecutive shortcut edges are never required to answer a query correctly.
Accordingly, if a transitive edge is missing in the shortcut graph, we know that it is never required as part of an optimal journey.

While the shortcut graph covers intermediate transfers between two trips, it does not provide any information for transferring from the source to the first trip or for transferring from the last trip to the target.
In this section we describe how initial and final transfers can be integrated into the query algorithms efficiently.
Additionally, we describe optimizations for the TB query algorithm that make it more efficient in a scenario with unlimited transfers.

\subsection{Query Algorithm Framework}
The ULTRA query algorithm exploits the fact that for initial and final transfers, one endpoint of the transfer is fixed.
All initial transfers start at the source vertex~$\aSource$ of the query, whereas all final transfers end at the target vertex~$\aTarget$.
Therefore, initial and final transfers can be explored with two additional one-to-many queries on the original transfer graph: a forward query to compute distances from~$\aSource$ to all stops, and a backward query for the distances from all stops to~$\aTarget$.
ULTRA uses Bucket-CH for this task, as it is one of the fastest known one-to-many algorithms and allows for optimization of local queries.
Thus, ULTRA requires three preprocessing steps in total:
First, a core graph is constructed with Core-CH.
This is then used as input for the transfer shortcut computation outlined in Section~\ref{chap:ULTRA:prepro}.
The third step is the Bucket-CH preprocessing for the original transfer graph~$\graph$.
The query algorithm then takes as input the public transit network, the transfer shortcut graph, and the Bucket-CH data.
Pseudocode for the query algorithm is shown in Algorithm~\ref{alg:ULTRA:query}.

\begin{figure}[t]
	\begin{minipage}{\textwidth}
		\begin{algorithm}[H]
			\caption{ULTRA query algorithm framework}\label{alg:ULTRA:query}
			\Input{%
				Public transit network~$(\stops,\trips,\routes,\graph)$,\linebreak
				transfer shortcut graph~$\shortcutGraph=(\stops,\shortcutEdges)$, Bucket-CH data for~$\graph$,\linebreak
				source vertex~$\aSource$, departure time~$\departureTime$, and target vertex~$\aTarget$
			}
			\Output{Pareto set~$\journeys$ of~$\aSource$-$\aTarget$-journeys for departure time~$\departureTime$}
			\BlankLine
			$(\transferTime(\aSource,\aTarget),\searchSpace{\aSource},\searchSpace{\aTarget})\leftarrow{}$Run a CH query from~$\aSource$ to~$\aTarget$ with departure time~$\departureTime$\;\label{alg:ULTRA:ch}
			$\transferTime(\aSource,\cdot)\leftarrow{}$Evaluate the vertex-to-stop buckets for vertices in~$\searchSpace{\aSource}$\;\label{alg:ULTRA:bch:forward}
			$\transferTime(\cdot,\aTarget)\leftarrow{}$Evaluate the stop-to-vertex buckets for vertices in~$\searchSpace{\aTarget}$\;\label{alg:ULTRA:bch:backward}
			\BlankLine
			$\shortcutGraphST\leftarrow(\stops\cup\{\aSource,\aTarget\},\shortcutEdges)$\label{alg:ULTRA:copyGraph}\;
			Add edge~$(\aSource,\aTarget)$ with travel time~$\transferTime(\aSource,\aTarget)$\label{alg:ULTRA:directEdge}\;
			\ForEach{$\aVertex \in \stops\setminus\{\aSource,\aTarget\}$ with~$\transferTime(\aSource,\aVertex)<\transferTime(\aSource,\aTarget)$}{
				Add edge~$(\aSource,\aVertex)$ to~$\shortcutGraphST$ with travel time~$\transferTime(\aSource,\aVertex)$\label{alg:ULTRA:forwardEdges}\;
			}
			\ForEach{$\aVertex \in \stops\setminus\{\aSource,\aTarget\}$ with~$\transferTime(\aVertex,\aTarget)<\transferTime(\aSource,\aTarget)$}{
				Add edge~$(\aVertex,\aTarget)$ to~$\shortcutGraphST$ with travel time~$\transferTime(\aVertex,\aTarget)$\label{alg:ULTRA:backwardEdges}\;
			}
			\BlankLine
			Run black-box public transit algorithm on~$(\stops\cup\{\aSource,\aTarget\},\trips,\routes,\shortcutGraphST)$\label{alg:ULTRA:blackBox}\;
		\end{algorithm}
	\end{minipage}
\end{figure}

A query begins with a bidirectional CH search from~$\aSource$ to~$\aTarget$ in line~\ref{alg:ULTRA:ch}.
This yields the travel time~$\transferTime(\aSource,\aTarget)$ for a direct transfer from~$\aSource$ to~$\aTarget$ (which may be~$\infty$ if no direct transfer is possible).
A naive approach would then perform a forward Bucket-CH query from~$\aSource$ and a reverse Bucket-CH query from~$\aTarget$, yielding for every stop~$\aStop$ the initial transfer distance~$\transferTime(\aSource,\aStop)$ and the final transfer distance~$\transferTime(\aStop,\aTarget)$.
However, not all of these distances are actually needed.
An initial transfer to a stop~$\aStop$ cannot be part of an optimal journey if~$\transferTime(\aSource,\aStop)\geq\transferTime(\aSource,\aTarget)$, since any journey containing the initial transfer will be dominated by the direct transfer from~$\aSource$ to~$\aTarget$.
Likewise, no optimal journey can include a final transfer to a stop~$\aStop$ with~$\transferTime(\aStop,\aTarget)\geq\transferTime(\aSource,\aTarget)$.
The algorithm exploits this by using the forward and backward search spaces~$\searchSpace{\aSource}$ and~$\searchSpace{\aTarget}$ of the bidirectional CH query.
Because the CH search is stopped once the shortest~$\aSource$-$\aTarget$-path has been found, these contain no vertices whose distance from~$\aSource$ and to~$\aTarget$, respectively, is greater than~$\transferTime(\aSource,\aTarget)$.
Therefore, it is sufficient to scan the forward buckets of all vertices in~$\searchSpace{\aSource}$ (line~\ref{alg:ULTRA:bch:forward}) and the backward buckets of all vertices in~$\searchSpace{\aTarget}$ (line~\ref{alg:ULTRA:bch:backward}).
Additional query time can be saved by sorting the entries of each bucket in ascending order of distance during the preprocessing phase.
Then the scan for the forward bucket of a vertex~$\aVertex$ can be stopped once it reaches a stop~$\bVertex$ within the bucket with~$\transferTime(\aSource,\aVertex)+\transferTime(\aVertex,\bVertex)\geq\transferTime(\aSource,\aTarget)$ (and analogously for backward buckets).
Doing so drastically improves local queries, as they do not need to evaluate all stops, but only stops that are close to the source or target.

After the distances for the initial and final transfers have been computed, the algorithm creates a temporary copy~$\shortcutGraphST$ of the shortcut graph~$\shortcutGraph$, which contains~$\aSource$ and~$\aTarget$ as additional vertices.
In lines~\ref{alg:ULTRA:directEdge}--\ref{alg:ULTRA:backwardEdges}, this temporary graph is complemented with edges for the initial and final transfers, and the direct transfer from~$\aSource$ and~$\aTarget$, using the distances obtained from the Bucket-CH queries.
Finally, a public transit algorithm is invoked as a black box on the public transit network with the temporary graph~$\shortcutGraphST$ in line~\ref{alg:ULTRA:blackBox}.
The temporary graph is sufficient for obtaining correct results, as it contains edges for all necessary initial, intermediate and final transfers, and an edge for a direct transfer from source to target.
Since there are no additional requirements on the black-box public transit algorithm, it is easy to see that any existing algorithm can be used with ULTRA shortcuts.

If the public transit algorithm is not treated as a black box, the performance can be improved further by omitting the construction of~$\shortcutGraphST$.
Most public transit algorithms, including RAPTOR and CSA, maintain a tentative arrival time at each stop, which is improved as new journeys are found.
Instead of adding an edge from~$\aSource$ to a stop~$\aStop$, the tentative arrival time of~$\aStop$ can be initialized with~$\departureTime+\transferTime(\aSource,\aVertex)$.
To incorporate final transfers, whenever the tentative arrival time at a stop~$\aStop$ is set to some value~$\atime$, the algorithm can try to improve the tentative arrival time at~$\aTarget$ with~$\atime+\transferTime(\aStop,\aTarget)$.

\subsection{Improved TB Query}
Unlike most algorithms, TB already distinguishes between initial/final and intermediate transfers, exploring different graphs for both.
The original transfer graph~$\graph$ is only used for the initial and final transfers, while intermediate transfers are explored using the precomputed event-to-event transfers.
In the context of ULTRA, this requires a modification to the query framework shown in Algorithm~\ref{alg:ULTRA:query}:
The temporary graph~$\shortcutGraphST$ now only contains the edges added for the initial and final transfers, but not the ULTRA shortcuts.
The query then uses~$\shortcutGraphST$ for the initial and final transfers, and the unmodified event-to-event shortcut graph~$\eventGraph=(\eventVertices,\eventEdges)$ for the intermediate transfers.

Additionally, the TB query algorithm can be optimized further for networks with unlimited transfers.
The original query, as introduced by Witt~\cite{Wit15}, is optimized for a use case where only a few stops are reachable with an initial or final transfer.
However, with unlimited transfers, it is typical for almost all stops to be reachable.
Therefore, we restructure the query to allow the huge number of possible initial and final transfers to be processed more efficiently.
Pseudocode for the modified query is given by Algorithm~\ref{alg:ULTRA:trip:query}.
In the following, we describe this algorithm in detail.

\begin{figure}
	\begin{minipage}{\textwidth}
		\begin{algorithm}[H]
			\caption{ULTRA-Trip-Based query}\label{alg:ULTRA:trip:query}
			\Input{%
				Public transit network~$(\stops,\trips,\routes,\graph)$,\\ transfer~shortcut~graph~$\eventGraph=(\eventVertices,\eventEdges)$,
				Bucket-CH data for~$\graph$,\\
				source vertex~$\aSource$, departure time~$\departureTime$, and target vertex~$\aTarget$
			}
			\Output{Labels~$\labels$ representing Pareto set of~$\aSource$-$\aTarget$-journeys for departure time~$\departureTime$}
			\BlankLine
			$(\transferTime(\aSource,\aTarget),\searchSpace{\aSource},\searchSpace{\aTarget})\leftarrow{}$Run a CH query from~$\aSource$ to~$\aTarget$ with departure time~$\departureTime$\;\label{alg:ULTRA:trip:bch:begin}
			$\transferTime(\aSource,\cdot)\leftarrow{}$Evaluate the vertex-to-stop buckets for vertices in~$\searchSpace{\aSource}$\;
			$\transferTime(\cdot,\aTarget)\leftarrow{}$Evaluate the stop-to-vertex buckets for vertices in~$\searchSpace{\aTarget}$\;\label{alg:ULTRA:trip:bch:end}
			\BlankLine
			$\minTime\leftarrow\departureTime+\transferTime(\aSource,\aTarget)$\;
			\lIf{$\minTime<\infty$}{$\labels\leftarrow\{(\minTime,0)\}$}\label{alg:ULTRA:trip:bch:journey}
			\lForEach{$\aVertex\in\stops$}{
				$\arrivalTime(\aSource,\aVertex)\leftarrow{}\departureTime+\transferTime(\aSource,\aVertex)$
			}
			\BlankLine
			$\routes',\aQueue_1\leftarrow\emptyset$\;
			\ForEach{$\aVertex\in\stops$ with~$\transferTime(\aSource,\aVertex)<\transferTime(\aSource,\aTarget)$\label{alg:ULTRA:trip:it:begin}}{
				$\routes'\leftarrow\routes'\cup\{\text{Routes from~\routes that contain~\aVertex}\}$\;\label{alg:ULTRA:trip:it:collect}
			}
			\ForEach{$\aRoute\in\routes'$}{
				$\activeTrip\leftarrow\bot$\;\label{alg:ULTRA:trip:it:mintrip}
				\For{$i$ from $0$ to $|\aRoute|-1$}{
					$\aVertex\leftarrow{}i\text{-th stop of }\aRoute$\;
					\lIf{$\arrivalTime(\aSource,\aVertex)\geq\minTime$}{\Continue}
					$\activeTrip'\leftarrow{}$earliest~$\aTrip\in\aRoute$ departing from~$\aVertex$\;\label{alg:ULTRA:trip:it:act}
					\If{$\activeTrip'$ is earlier than~$\activeTrip$}{
						$\activeTrip\leftarrow\activeTrip'$\;
						$\Enqueue(\activeTrip[i],\aQueue_1)$\;\label{alg:ULTRA:trip:it:enqueue}
						\lIf{$\activeTrip$ is the first trip in~$\aRoute$}{\Break}\label{alg:ULTRA:trip:it:end}
					}
				}
			}
			\BlankLine
			$n\leftarrow1$\;\label{alg:ULTRA:trip:ts:begin}
			\While{$\aQueue_n$ is not empty}{
				\ForEach{$\aTripSegment{jk}\in\aQueue_n$}{
					\For{$i$ from $j$ to $k$}{
						\lIf{$\arrivalTime(\aTrip[i])\geq\minTime$\label{alg:ULTRA:trip:ts:final}}{\Break}
						\If{$\arrivalTime(\aTrip[i])+\transferTime(\aVertex(\aTrip[i]),\aTarget)<\minTime$}{
							$\minTime\leftarrow\arrivalTime(\aTrip[i])+\transferTime(\aVertex(\aTrip[i]),\aTarget)$\;
							$\labels\leftarrow{}\labels\cup\{(\minTime,n)\}$, removing dominated labels\;
						}
					}
				}
				$\aQueue_{n+1}\leftarrow\emptyset$\;
				\ForEach{$\aTripSegment{jk}\in\aQueue_n$}{
					\For{$i$ from $j$ to $k$}{
						\lIf{$\arrivalTime(\aTrip[i])\geq\minTime$\label{alg:ULTRA:trip:ts:prune}}{\Break}
						\ForEach{\smash{$(\aTrip[i],\aTrip'[i'])\in\eventEdges$}}{
							$\Enqueue(\aTrip'[i'],\aQueue_{n+1})$\label{alg:ULTRA:trip:ts:enqueue}
						}
					}
				}
				$n\leftarrow{}n+1$\;\label{alg:ULTRA:trip:ts:end}
			}
		\end{algorithm}
	\end{minipage}
\end{figure}

\subparagraph*{Initial Transfer Evaluation.}
As in the generic ULTRA query, the algorithm begins with the Bucket-CH search~(lines~\ref{alg:ULTRA:trip:bch:begin}--\ref{alg:ULTRA:trip:bch:end}).
This yields a minimum arrival time~$\arrivalTime(\aSource,\aStop)$ for every reached stop~$\aStop$ as well as the minimum arrival time~$\minTime$ at~$\aTarget$ via a direct transfer.
If~$\minTime<\infty$, a label representing the~$\aSource$-$\aTarget$-journey with zero trips is added to the result set~$\labels$ in line~\ref{alg:ULTRA:trip:bch:journey}.
The algorithm then identifies trips which are reachable via an initial transfer~(lines~\ref{alg:ULTRA:trip:it:begin}--\ref{alg:ULTRA:trip:it:end}).
In the original TB query~\cite{Wit15}, this is done by iterating over all stops that are reachable via an initial transfer.
For each such stop~$\aVertex$ and each route~$\aRoute$ visiting~$\aVertex$, the algorithm identifies the earliest trip of~$\aRoute$ that can be entered at~$\aVertex$ after taking the initial transfer.
This approach is efficient as long as the number of stops reachable via an initial transfer is small.
However, in a scenario with unlimited transfers, where almost all stops are reachable, consecutive stops of a route often share the same earliest reachable trip.
This can cause the same trip to be found multiple times, leading to redundant work.
To avoid this, we propose a new approach for evaluating the initial transfers, which is based on two steps of the~RAPTOR algorithm: collecting updated routes and scanning routes.

Lines~\ref{alg:ULTRA:trip:it:begin} and~\ref{alg:ULTRA:trip:it:collect} collect all routes which visit a stop that is reachable via an initial transfer.
This is analogous to collecting routes that visit marked stops at the beginning of a~RAPTOR round.
Then, all collected routes are scanned.
As in RAPTOR, a route~$\aRoute$ is scanned by processing its stops in the order in which they are visited by~$\aRoute$.
The algorithm maintains an active trip~$\activeTrip$, which is the earliest trip of~$\aRoute$ that is reachable from any of the already processed stops.
Initially, $\activeTrip$ is set to a dummy value~$\bot$ (line~\ref{alg:ULTRA:trip:it:mintrip}).
Let~$\aStop$ be the next stop to be processed while scanning~$\aRoute$.
To check if~$\activeTrip$ can be improved, the algorithm finds the earliest trip~$\activeTrip'$ of~$\aRoute$ that can be boarded when arriving at~$\aStop$ with the arrival time~$\arrivalTime(\aSource,\aStop)$.
If no reachable trip has been found for any of the previous stops in~$\aRoute$~(i.e.,~$\activeTrip=\bot$), then~$\activeTrip'$ is found with a binary search.
Otherwise, the algorithm starts a linear search from~$\activeTrip$ and looks backward for earlier trips.
Since~$\activeTrip'$ is often not much earlier than~$\activeTrip$, this is faster than a binary search in practice.
Note that~$\activeTrip'$ will not be found if it is later than~$\activeTrip$, but in this case entering~$\activeTrip'$ at~$\aStop$ does not produce an optimal journey, so it can be discarded.
If~$\activeTrip'$ is earlier than~$\activeTrip$, then~$\activeTrip$ is updated and the~$\Enqueue$ operation is called for the corresponding stop event in line~\ref{alg:ULTRA:trip:it:enqueue}.
The~$\Enqueue$ operation itself is unchanged from the original TB query.
If~$\activeTrip'$ is the earliest trip in~$\aRoute$, the remainder of the route scan can be skipped.

\subparagraph*{Trip Scanning.}
The trip scanning phase~(lines~\ref{alg:ULTRA:trip:ts:begin}--\ref{alg:ULTRA:trip:ts:end}) is identical to the original TB query algorithm except for the evaluation of final transfers.
It is organized in rounds, where the~$n$-th round scans the trip segments that were previously collected in the FIFO queue~$\aQueue_n$.
A trip segment~$\aTripSegment{jk}$ is scanned by iterating over all stop events from~$\aTrip[j]$ to~$\aTrip[k]$.
While scanning a stop event~$\aTrip[i]$, the algorithm checks whether a final transfer from the~$i$-th stop of the trip~$\aTrip$ to the target exists in line~\ref{alg:ULTRA:trip:ts:final}.
If such a transfer exists and improves the earliest known arrival time~$\minTime$ at the target, then the algorithm has found a new Pareto-optimal journey.
In this case,~$\minTime$ is updated and a label representing the newly found journey is added to the result set~$\labels$.
If~$\labels$ already contains a label with~$n$ trips~(note that a Pareto set can only contain one such label), this label is replaced.
After the final transfers have been evaluated, the algorithm relaxes the outgoing shortcuts from~$\aTrip[i]$.
For each shortcut~$(\aTrip[i], \aTrip'[i'])\in\eventEdges$, the~$\Enqueue$ operation is called for~$\aTrip'[i']$.
This adds the relevant segment of~$\aTrip'$ to the queue~$\aQueue_{n+1}$ of trips which are scanned in the next round.

Note that the trips in~$\aQueue_n$ are scanned twice, once to evaluate the final transfers and then again to relax transfer shortcuts.
This is done for two reasons:
First, separating the two scans improves memory locality, as~$\transferTime(\cdot,\aTarget)$ is only accessed by the first scan, and~$\eventEdges$ is only accessed by the second scan.
Secondly, $\minTime$ is improved throughout the first scan, which enables stricter pruning of trips that cannot contribute to Pareto-optimal journeys in line~\ref{alg:ULTRA:trip:ts:prune} of the second scan.

\subparagraph*{Data Structures and Memory Layout.}
In order to achieve optimal performance, the query algorithm needs to use a streamlined memory layout.
To this end, the~FIFO queues~$\aQueue_n$ are implemented using dynamic arrays.
This enables an efficient~$\Enqueue$ operation and efficient scanning of the entries in~$\aQueue_n$.
The shortcuts~$\eventEdges$ are stored in an array such that all outgoing shortcuts of a stop event~$\aTrip[i]$ are consecutive in memory and the outgoing shortcuts of the next stop event~$\aTrip[i+1]$ follow directly afterwards.
Finally, note that the trip scanning step only needs access to the arrival time~$\arrivalTime(\aTrip[i])$ and the stop~$\aVertex(\aTrip[i])$ of a stop event~$\aTrip[i]$.
Therefore, these values are stored separately from the departure time~$\departureTime(\aTrip[i])$ of the stop event, which improves memory locality.

\section{Experiments}
\label{chap:ULTRA:exp}

All algorithms were implemented in C\raisebox{0.15ex}{\small++}17 and compiled with GCC version 10.3.0 and optimization flag -O3.
Experiments were performed on the following machines:

\begin{description}[leftmargin=28pt]
	\item[\parbox{23pt}{Xeon}] A machine with two 8-core Intel Xeon Skylake SP Gold 6144 CPUs clocked at~3.50\,GHz, with a boost frequency of~4.2\,GHz,~192\,GiB of~\mbox{DDR4-2666}~RAM, and~24.75\,MiB of L3 cache.
	\item[\parbox{23pt}{Epyc}] A machine with two 64-core AMD Epyc Rome 7742 CPUs clocked at~2.25\,GHz, with a boost frequency of~3.4\,GHz,~1024\,GiB of~\mbox{DDR4-3200} RAM, and~256\,MiB of L3 cache.
\end{description}

\subsection{Networks}
\begin{table}[t]%
	\caption[Sizes of public transit networks.]{%
		Sizes of the public transit networks and the accompanying transfer graphs.
		Also reported is the number of edges in the transitively closed transfer graph used to compare ULTRA to uni-modal algorithms.
	}%
	\label{tbl:networks}%
	\begin{tabular*}{\textwidth}{@{\,}l@{\extracolsep{\fill}}r@{\extracolsep{\fill}}r@{\extracolsep{\fill}}r@{\extracolsep{\fill}}r@{\,\,}}
		\toprule
		& \hspace{20mm}\llap{Stuttgart} & \hspace{20mm}\llap{London} & \hspace{21mm}\llap{Switzerland} & \hspace{20mm}\llap{Germany} \\
		\midrule
		Stops                                 &                       13\,584 &                    19\,682 &                         25\,125 &                    243\,167 \\
		Routes                                &                       12\,351 &                     1\,955 &                         13\,786 &                    230\,255 \\
		Trips                                 &                       91\,304 &                   114\,508 &                        350\,006 &                 2\,381\,394 \\
		Stop events                           &                   1\,561\,972 &                4\,508\,644 &                     4\,686\,865 &                48\,380\,936 \\[5pt]
		Transfer graph vertices \hspace{-3mm} &                   1\,166\,604 &                   181\,642 &                        603\,691 &                 6\,870\,496 \\
		Transfer graph edges    \hspace{-3mm} &                   3\,682\,232 &                   575\,364 &                     1\,853\,260 &                21\,367\,044 \\[5pt]
		Transitive graph edges                &                   1\,369\,928 &                3\,212\,206 &                     2\,639\,402 &                22\,571\,280 \\
		\bottomrule
	\end{tabular*}
\end{table}

We evaluated our algorithms on the transportation networks of Stuttgart, London, Switzerland, and Germany.
The Stuttgart network represents the greater region of Stuttgart and comprises two identical business days.
It was previously used in~\cite{Mal13} and~\cite{Bri17} and is not publicly available.
The public transit timetable of London was obtained from Transport for London\footnote{\url{https://data.london.gov.uk}} and covers a single Tuesday in the periodic summer schedule of 2011.
It was previously used to evaluate RAPTOR~\cite{Del15b}, MCR~\cite{Del13} and TB~\cite{Wit15}.
The Switzerland network was extracted from a publicly available GTFS feed\footnote{\url{http://gtfs.geops.ch/}} and consists of two successive business days (30th and 31st of May 2017).
Lastly, the Germany network was provided by Deutsche Bahn for research purposes and is not publicly available.
It is based on data from \url{bahn.de} for Winter 2011/2012, comprising two successive identical days, and was previously used to evaluate CSA~\cite{Dib18} and TB~\cite{Wit15}.
Both the Switzerland and Germany networks were previously used in~\cite{Wag17}.

We constructed unrestricted transfer graphs by extracting road graphs, including pedestrian zones and staircases, from OpenStreetMap\footnote{\url{https://download.geofabrik.de/}}.
Unless stated otherwise, we used walking as the transfer mode, assuming a constant speed of 4.5 km/h.
The transfer graph was connected to the public transit network using the procedure outlined in~\cite{Wag17}.
For each stop~$\aVertex\in\stops$, we located its (geographically) nearest neighbor~$\bVertex\in\vertices$ in the transfer graph.
If~$\aVertex$ and~$\bVertex$ were less than 5 meters apart and~$\aVertex$ was also the nearest neighbor of~$\bVertex$, we identified~$\aVertex$ with~$\bVertex$.
Otherwise, we added a new vertex for~$\aVertex$ and connected it to~$\bVertex$ if their distance was less than 100 meters.
Afterwards, vertices with degree one and two were contracted unless they coincided with stops.
Remote and isolated parts of the networks were removed by applying a bounding box and removing everything except the largest connected component.

To obtain transitively closed transfer graphs (for comparison with standard RAPTOR, CSA and TB), we inserted edges between all stops whose distance in the transfer graph lies below a certain threshold~(9~minutes for Stuttgart and Switzerland,~8~minutes for Germany,~4~minutes for London), and then computed the transitive closure.
Following~\cite{Wag17}, the thresholds were chosen so that the resulting graph has an average vertex degree of about 100.
An overview of the networks is given in Table~\ref{tbl:networks}.

\subsection{Preprocessing}

In this section we evaluate the performance of the~ULTRA~preprocessing phase, which includes the Core-CH transfer graph contraction, the shortcut computation, and the Bucket-CH computation.
We analyze the effects of the parameters core degree, witness limit, and transfer speed in detail for the Switzerland network, and then discuss more general results for all four networks.

\subparagraph*{Core Degree and Witness Limit.}
\begin{figure*}
	\newcommand{\plotHeight}{6.5cm}
\newcommand{\plotShift}{\plotHeight+0.5cm}
\newcommand{\plotWidth}{0.48\textwidth}

\begin{tikzpicture}

\colorlet{plotColor1}{\green{0}}
\colorlet{plotColor2}{\cyan{1}}
\colorlet{plotColor3}{\blue{0}}
\colorlet{plotColor4}{\lila{0}}
\colorlet{plotColor5}{\red{0}}
\colorlet{plotColor6}{\yellow{0}}

\begin{scope}


\begin{scope}[shift={(1.20, 1.95)}]

   \begin{axis}[
      title={Stop-to-stop shortcuts},
      title style={anchor=north,yshift=10pt},
      height=\plotHeight,
      width=\plotWidth,
      xmin=7.5,
      xmax=20.5,
      ymin=450,
      ymax=1410,
      xlabel={\small{}Core degree},
      ylabel={\small{}Preprocessing time [min]},
      ylabel style = {yshift=-1pt},
      xtick={8, 10, 12, 14, 16, 18, 20},
      xticklabel=\pgfmathparse{\tick}${\pgfmathprintnumber{\pgfmathresult}}$,
      ytick={480, 660, 840, 1020, 1200, 1380},
      yticklabel=\pgfmathparse{(\tick / 60)}${\pgfmathprintnumber{\pgfmathresult}}$\!,
      xtick pos=left,
      ytick pos=left,
      minor y tick num={2},
      grid=major,
      at={(0.04\linewidth,\plotShift)}]
      
      \addplot[sampleFull,color=plotColor1] table[x=CoreDegree,y=Time0,col sep=tab]{figures/coreDegreeWitnessLimit_stop.dat};
      \addplot[sampleFull,color=plotColor2] table[x=CoreDegree,y=Time1800,col sep=tab]{figures/coreDegreeWitnessLimit_stop.dat};
      \addplot[sampleFull,color=plotColor3] table[x=CoreDegree,y=Time3600,col sep=tab]{figures/coreDegreeWitnessLimit_stop.dat};
      \addplot[sampleFull,color=plotColor4] table[x=CoreDegree,y=Time7200,col sep=tab]{figures/coreDegreeWitnessLimit_stop.dat};
      \addplot[sampleFull,color=plotColor5] table[x=CoreDegree,y=Time14400,col sep=tab]{figures/coreDegreeWitnessLimit_stop.dat};
      \addplot[sampleFull,color=plotColor6] table[x=CoreDegree,y=Time172800,col sep=tab]{figures/coreDegreeWitnessLimit_stop.dat};

   \end{axis}

    \begin{axis}[
    title={Event-to-event shortcuts},
    title style={anchor=north,yshift=10pt},
    height=\plotHeight,
    width=\plotWidth,
    xmin=7.5,
    xmax=20.5,
    ymin=450,
    ymax=1410,
    xlabel={\small{}Core degree},
    ylabel={\small{}Preprocessing time [min]},
    ylabel style = {yshift=-1pt},
    xtick={8, 10, 12, 14, 16, 18, 20},
    xticklabel=\pgfmathparse{\tick}${\pgfmathprintnumber{\pgfmathresult}}$,
    ytick={480, 660, 840, 1020, 1200, 1380},
    yticklabel=\pgfmathparse{(\tick / 60)}${\pgfmathprintnumber{\pgfmathresult}}$\!,
    xtick pos=left,
    ytick pos=left,
    minor y tick num={2},
    grid=major,
    at={(0.54\linewidth,\plotShift)}]
    
    \addlegendimage{legendFull,plotColor1}\label{legend:cd:0}
    \addlegendimage{legendFull,plotColor2}\label{legend:cd:30}
    \addlegendimage{legendFull,plotColor3}\label{legend:cd:60}
    \addlegendimage{legendFull,plotColor4}\label{legend:cd:120}
    \addlegendimage{legendFull,plotColor5}\label{legend:cd:240}
    \addlegendimage{legendFull,plotColor6}\label{legend:cd:max}

    \addplot[sampleFull,color=plotColor1] table[x=CoreDegree,y=Time0,col sep=tab]{figures/coreDegreeWitnessLimit_event.dat};
    \addplot[sampleFull,color=plotColor2] table[x=CoreDegree,y=Time1800,col sep=tab]{figures/coreDegreeWitnessLimit_event.dat};
    \addplot[sampleFull,color=plotColor3] table[x=CoreDegree,y=Time3600,col sep=tab]{figures/coreDegreeWitnessLimit_event.dat};
    \addplot[sampleFull,color=plotColor4] table[x=CoreDegree,y=Time7200,col sep=tab]{figures/coreDegreeWitnessLimit_event.dat};
    \addplot[sampleFull,color=plotColor5] table[x=CoreDegree,y=Time14400,col sep=tab]{figures/coreDegreeWitnessLimit_event.dat};
    \addplot[sampleFull,color=plotColor6] table[x=CoreDegree,y=Time172800,col sep=tab]{figures/coreDegreeWitnessLimit_event.dat};
    \end{axis}
    
   \begin{axis}[
      title={Stop-to-stop shortcuts},
      title style={anchor=north,yshift=10pt},
      scaled y ticks = false,
      y tick label style={/pgf/number format/fixed},
      height=\plotHeight,
      width=\plotWidth,
      xmin=7.5,
      xmax=20.5,
      ymin=170.6675,
      ymax=170.7225,
      xlabel={\small{}Core degree},
      ylabel={\small{}Shortcuts \raisebox{0.8pt}{[}k\raisebox{0.8pt}{]}},
      ylabel style = {yshift=-3pt},
      xtick={8, 10, 12, 14, 16, 18, 20},
      xticklabel=\pgfmathparse{\tick}${\pgfmathprintnumber{\pgfmathresult}}$,
      ytick={170.67, 170.68, 170.69, 170.70, 170.71, 170.72},
      yticklabel=\pgfmathparse{\tick}${\pgfmathprintnumber{\pgfmathresult}}$\!,
      xtick pos=left,
      ytick pos=left,
      at={(0.04\linewidth,0)}]

      \addplot[sampleFull,color=plotColor1] table[x=CoreDegree,y=Shortcuts0,col sep=tab]{figures/coreDegreeWitnessLimit_stop.dat};
      \addplot[sampleFull,color=plotColor2] table[x=CoreDegree,y=Shortcuts1800,col sep=tab]{figures/coreDegreeWitnessLimit_stop.dat};
      \addplot[sampleFull,color=plotColor3] table[x=CoreDegree,y=Shortcuts3600,col sep=tab]{figures/coreDegreeWitnessLimit_stop.dat};
      \addplot[sampleFull,color=plotColor4] table[x=CoreDegree,y=Shortcuts7200,col sep=tab]{figures/coreDegreeWitnessLimit_stop.dat};
      \addplot[sampleFull,color=plotColor5] table[x=CoreDegree,y=Shortcuts14400,col sep=tab]{figures/coreDegreeWitnessLimit_stop.dat};
      \addplot[sampleFull,color=plotColor6] table[x=CoreDegree,y=Shortcuts172800,col sep=tab]{figures/coreDegreeWitnessLimit_stop.dat};

   \end{axis}
   
   \begin{axis}[
   title={Event-to-event shortcuts},
   title style={anchor=north,yshift=10pt},
   scaled y ticks = false,
   y tick label style={/pgf/number format/fixed},
   height=\plotHeight,
   width=\plotWidth,
   xmin=7.5,
   xmax=20.5,
   ymin=6935.75,
   ymax=6938.5,
   xlabel={\small{}Core degree},
   ylabel={\small{}Shortcuts \raisebox{0.8pt}{[}k\raisebox{0.8pt}{]}},
   ylabel style = {yshift=-3pt},
   xtick={8, 10, 12, 14, 16, 18, 20},
   xticklabel=\pgfmathparse{\tick}${\pgfmathprintnumber{\pgfmathresult}}$,
   ytick={6936, 6937, 6938},
   yticklabel=\pgfmathparse{\tick}${\pgfmathprintnumber{\pgfmathresult}}$\!,
   xtick pos=left,
   ytick pos=left,
   at={(0.54\linewidth,0)}]
   
   \addplot[sampleFull,color=plotColor1] table[x=CoreDegree,y=Shortcuts0,col sep=tab]{figures/coreDegreeWitnessLimit_event.dat};
   \addplot[sampleFull,color=plotColor2] table[x=CoreDegree,y=Shortcuts1800,col sep=tab]{figures/coreDegreeWitnessLimit_event.dat};
   \addplot[sampleFull,color=plotColor3] table[x=CoreDegree,y=Shortcuts3600,col sep=tab]{figures/coreDegreeWitnessLimit_event.dat};
   \addplot[sampleFull,color=plotColor4] table[x=CoreDegree,y=Shortcuts7200,col sep=tab]{figures/coreDegreeWitnessLimit_event.dat};
   \addplot[sampleFull,color=plotColor5] table[x=CoreDegree,y=Shortcuts14400,col sep=tab]{figures/coreDegreeWitnessLimit_event.dat};
   \addplot[sampleFull,color=plotColor6] table[x=CoreDegree,y=Shortcuts172800,col sep=tab]{figures/coreDegreeWitnessLimit_event.dat};
   
   \end{axis}

\end{scope}

   \arrayrulecolor{legendColor}
   \node[inner sep=0pt,outer sep=0pt] (legend) at (0,0.05) {};
   \node[inner sep=0pt,outer sep=0pt,anchor=south west] at (legend) {
      \small
      \begin{tabular*}{\textwidth}{|@{~~}l@{~~~~}r@{\extracolsep{\fill}}r@{\extracolsep{\fill}}r@{\extracolsep{\fill}}r@{\extracolsep{\fill}}r@{\extracolsep{\fill}}r@{~~}|}
         \hline
                                                &                                  &                                  &
                                                &                                  &                                  &                                                          \\[-7pt]
         Witness limit ($\witnessLimit$) [min]: & \legend{\ref{legend:cd:0}}~0     & \legend{\ref{legend:cd:30}}~30   & 
         \legend{\ref{legend:cd:60}}\,60        & \legend{\ref{legend:cd:120}}~120 & \legend{\ref{legend:cd:240}}~240 & \legend{\ref{legend:cd:max}}~\raisebox{0.15ex}{$\infty$} \\[2pt]
         \hline
      \end{tabular*}
   };
   \arrayrulecolor{black}

\end{scope}

\end{tikzpicture}%
	\caption[Impact of core degree and witness limit on the ULTRA preprocessing.]{%
		Impact of the core degree and the witness limit on the running time of the ULTRA preprocessing and the number of computed shortcuts, measured for the Switzerland network on the Xeon machine.
		Preprocessing time includes both contracting the transfer graph and computing the shortcuts.
		The time required for the Bucket-CH computation, which is independent of both parameters, is excluded.
	}
	\label{fig:coreDegreeWitnessLimit}
\end{figure*}

The two main parameters influencing the performance of the~ULTRA~preprocessing are the average vertex degree of the contracted core graph and the witness limit~$\witnessLimit$.
Figure \ref{fig:coreDegreeWitnessLimit} shows the impact of these parameters on the Switzerland network.
The lowest preprocessing times are achieved with a core degree of~14.
While the actual shortcut computation is slightly faster for higher core degrees, this is offset by the increased time required to contract the transfer graph.
The witness limit~$\witnessLimit$ has a larger impact on the preprocessing time.
Choosing a witness limit of~0 instead of~$\infty$ nearly cuts the preprocessing time in half.
Regardless of core degree or witness limit, the event-to-event variant takes about one minute longer than the stop-to-stop variant.
Both parameters have a negligible effect on the number of computed shortcuts.
For all following experiments, we therefore choose a core degree of~14 and a witness limit of~0 to minimize the preprocessing time.
The only exception is the Germany network, where we use a core degree of~20.
This is because the share of the Core-CH computation in the overall running time is significantly lower for this network, due to its much larger size.
Preprocessing results for the stop-to-stop variant on all four networks are listed in Table~\ref{tbl:ULTRA:preprocessing}.

\begin{table}
	\center
	\caption[Stop-to-stop ULTRA preprocessing results.]{%
		Overview of the stop-to-stop ULTRA preprocessing results.
		All running times were measured on the Xeon machine and are displayed as (hh:)mm:ss.
		The Core-CH and Bucket-CH computations were run sequentially, whereas the shortcut computation used all 16 cores.
	}%
	\label{tbl:ULTRA:preprocessing}%
	\begin{tabular*}{\textwidth}{@{\,}l@{\hspace{10pt}}r@{\extracolsep{\fill}}r@{\extracolsep{\fill}}r@{\extracolsep{\fill}}r@{\,}}
		\toprule
		&        Stuttgart & \hspace{21mm}\llap{London} & \hspace{22mm}\llap{Switzerland} & \hspace{22mm}\llap{Germany} \\
		\midrule
		Core-CH time              &             1:45 &                       0:19 &                            1:09 &                       20:16 \\[1pt]
		Number of core vertices   &          25\,631 &                    23\,860 &                         33\,219 &                    313\,351 \\[1pt]
		Number of core edges      &         358\,842 &                   334\,112 &                        465\,067 &                 6\,267\,050 \\[5pt]
		Shortcut computation time &             4:27 &                      18:01 &                            8:54 &                     8:01:25 \\[1pt]
		Number of shortcuts       &          83\,086 &                   190\,388 &                        170\,713 &                 2\,907\,691 \\[5pt]
		Bucket-CH time            &             2:13 &                       0:11 &                            0:43 &                       14:49 \\
		\bottomrule
	\end{tabular*}
\end{table}

\subparagraph*{ULTRA-TB Preprocessing.}

\begin{table}[t]
	\center
	\caption[TB preprocessing results.]{%
		Number of shortcuts and preprocessing times for different~TB preprocessing variants.
		``Transitive'' refers to the original TB preprocessing on the transitively closed transfer graph.
		``Sequential'' uses stop-to-stop ULTRA shortcuts as input for the TB preprocessing, whereas ``integrated'' uses event-to-event ULTRA shortcuts directly.
		``Optimized'' refers to the improved TB preprocessing algorithm of Lehoux and Loiodice~\cite{Leh20}.
		Running times were measured on the Xeon machine with 16 cores and are displayed as (hh:)mm:ss.
	}%
	\label{tbl:ULTRA:tripBasedPreprocessing}%
	\begin{tabular*}{\textwidth}{@{\,}l@{\hspace{10pt}}r@{\extracolsep{\fill}}r@{\extracolsep{\fill}}r@{\extracolsep{\fill}}r@{\,}}
		\toprule
		&        Stuttgart & \hspace{21mm}\llap{London} & \hspace{22mm}\llap{Switzerland} & \hspace{22mm}\llap{Germany} \\
		\midrule
		Shortcuts (transitive)            &             7\,387\,445 &               50\,242\,519 &                    31\,507\,264 & 458\,826\,534 \\
		Shortcuts (transitive, optimized) &             7\,387\,586 &               50\,240\,558 &                    31\,507\,543 & 458\,763\,050 \\
		Shortcuts (sequential)            &            19\,361\,708 &               53\,179\,082 &                    65\,485\,696 &            1\,195\,573\,925 \\
		Shortcuts (sequential, optimized)            &            19\,361\,129 &               53\,181\,238 &                    65\,484\,976 &            1\,195\,509\,797 \\
		Shortcuts (integrated)            &             1\,973\,321 &                8\,576\,120 &                     6\,938\,012 &               77\,515\,291 \\[5pt]
		Time (transitive)                 &                    9:30 &                    1:42:35 &                         1:01:54 & 73:43:07 \\
		Time (transitive, optimized)      &                    0:37 &                      13:12 &                            4:41 & 2:55:06 \\
		Time (sequential)                 &                    4:41 &                      18:43 &                            9:40 &                     8:57:46 \\
		Time (sequential, optimized)                 &                    4:37 &                      18:28 &                            9:24 &                     8:22:37 \\
		Time (integrated)                 &                    4:42 &                      20:43 &                            9:40 &                     8:37:49 \\
		\bottomrule
	\end{tabular*}
\end{table}

To evaluate the effectiveness of the event-to-event ULTRA shortcut computation, we compare it to the original TB preprocessing, using the transitively closed transfer graphs as input, and to a naive sequential approach, i.e.,~using stop-to-stop ULTRA shortcuts as input for the TB preprocessing.
An overview of the results is given in Table~\ref{tbl:ULTRA:tripBasedPreprocessing}.
The integrated ULTRA preprocessing drastically reduces the amount of shortcuts compared to the sequential approach.
This reduction ranges from a factor of~6 for the London network to over~15 for Germany.
Regarding computation time, the sequential approach using the optimized TB preprocessing proposed by~\cite{Leh20} is only marginally faster than the integrated approach.
Overall, the integrated preprocessing is clearly preferable since it produces much fewer shortcuts with only a minor overhead in running time.

Remarkably, event-to-event ULTRA significantly outperforms the original TB preprocessing in both number of shortcuts and computation time, despite operating on an unrestricted transfer graph instead of a transitively closed one.
This underscores that the original TB preprocessing was only designed for very limited transfer graphs and confirms the findings of Lehoux and Loiodice~\cite{Leh20} that it does not scale well for larger graphs.
Compared to the optimized TB preprocessing, ULTRA is slower by a factor of about~2--3 on most networks.
On the Stuttgart network, the slowdown is about~8.
The difference is explained by the fact that Stuttgart is the only network where the transitively closed transfer graph has fewer edges than the full transfer graph.
Overall, the preprocessing results show that ULTRA is much more effective than the TB preprocessing at identifying necessary transfers, at the cost of a somewhat higher preprocessing time.

\subparagraph*{Parallelization.}

\begin{table}[t]
	\center
	\caption[Impact of parallelization on the stop-to-stop ULTRA shortcut computation.]{%
		Impact of parallelization on the running time of the stop-to-stop ULTRA shortcut computation.
		Running times are displayed as (hh:)mm:ss.
	}%
	\label{tbl:ULTRA:preprocessingtime}%
	\begin{tabular*}{\textwidth}{@{\,}l@{\hspace{10pt}}r@{\hspace{10pt}}r@{\extracolsep{\fill}}r@{\extracolsep{\fill}}r@{\extracolsep{\fill}}r@{\,}}
		\toprule
		Machine & Cores &        Stuttgart & \hspace{21mm}\llap{London} & \hspace{22mm}\llap{Switzerland} & \hspace{22mm}\llap{Germany} \\
		\midrule
		\multirow{5}{*}{Xeon}
		& 1   &            59:28 &                    4:00:31 &                         2:00:29 &                   100:02:46 \\[0.5pt]
		& 2   &            30:42 &                    2:05:06 &                         1:02:24 &                    54:12:12 \\[0.5pt]
		& 4   &            15:49 &                    1:06:24 &                           32:17 &                    29:02:18 \\[0.5pt]
		& 8   &             8:28 &                      34:52 &                           17:13 &                    15:26:13 \\[0.5pt]
		& 16  &             4:27 &                      18:01 &                            8:54 &                     8:01:25 \\[5pt]
		\multirow{8}{*}{Epyc}
		& 1   &          1:14:37 &                    4:53:01 &                         2:25:26 &                   122:35:42 \\[0.5pt]
		& 2   &            40:38 &                    2:43:33 &                         1:21:57 &                    72:42:27 \\[0.5pt]
		& 4   &            20:10 &                    1:19:21 &                           40:39 &                    37:56:49 \\[0.5pt]
		& 8   &            10:03 &                      39:54 &                           20:23 &                    19:11:35 \\[0.5pt]
		& 16  &             5:05 &                      19:54 &                           10:08 &                     9:49:56 \\[0.5pt]
		& 32  &             2:37 &                      10:08 &                            5:11 &                     4:57:06 \\[0.5pt]
		& 64  &             1:29 &                       5:52 &                            2:55 &                     2:56:49 \\[0.5pt]
		& 128 &             0:54 &                       3:44 &                            1:57 &                     2:53:57 \\
		\bottomrule
	\end{tabular*}
\end{table}

The previous experiments used all~16~cores of the Xeon machine for the shortcut computation.
To assess the impact of parallelization on the preprocessing time, we evaluate the running time of the stop-to-stop shortcut computation for different numbers of threads.
Additionally, we compare running times of the Epyc machine, which has worse single-core performance but contains more cores.
Running times on both machines are listed in Table~\ref{tbl:ULTRA:preprocessingtime}.
Overall, the parallelized shortcut computation achieves good speedups for all networks on both machines.
For the Switzerland network, the maximal speedup is~13.5 on the Xeon machine and~74.6 on the Epyc machine.
The speedup for the entire preprocessing phase, including the sequential Core-CH and Bucket-CH computation times on the Xeon machine, drops to~11.4 and~38.6, respectively.
Independently of the network, we observe the smallest speedup when switching from~64~threads to~128~threads on the Epyc machine.
In this case the speedup is most likely limited by the memory bandwidth.

The results are similar for the event-to-event variant.
On the Switzerland network, the single-threaded performance on the Xeon machine is~2:07:00 for the sequential approach and~2:10:10 for the integrated approach.
This corresponds to speedup factors of~13.1 and~13.5, respectively, which matches the speedups observed for the stop-to-stop variant and the TB preprocessing.

\subparagraph*{Transfer Speed.}
To test the impact of the transfer mode on the shortcut computation, we changed the transfer speed in the Switzerland network from 4.5\,km/h to values between 1\,km/h and 140\,km/h.
We considered two ways of applying the transfer speed:
In the first version, the speed on an edge is not allowed to exceed the speed limit given in the road network.
This models fast transfer modes such as cars fairly realistically.
In the second version, speed limits are ignored and the same constant speed is assumed for every edge.
This allows us to analyze to which extent the effects observed in the first version are caused by the speed limit data.
Figure~\ref{fig:transferSpeedPreprocessing} reports the preprocessing times and number of shortcuts (both stop-to-stop and event-to-event) measured for each configuration.
In all measurements, the preprocessing time remained below 15 minutes.
The number of stop-to-stop shortcuts initially increases with the transfer speed until it peaks at about~300\,000 between~10 and~20\,km/h (roughly the speed of a bicycle).
In the event-to-event variant, the behavior is the opposite: the number of shortcuts is highest for~1\,km/h and decreases from there.
Above 20~km/h, both variants exhibit a slight increase in the number of shortcuts, which is more pronounced if speed limits are obeyed.
Overall, the results show that ULTRA is practical for all transfer speeds, both in terms of preprocessing time and the number of shortcuts.

\begin{figure*}
	\input{figures/transferSpeedPreprocessing}%
	\caption[Impact of the transfer speed on the ULTRA preprocessing.]{%
		Impact of transfer speed on preprocessing time and number of shortcuts, measured on the Switzerland network with a core degree of~14 and a witness limit of~0.
		Speed limits were obeyed for the red lines and ignored for the blue lines.
		For the two lines at the bottom of the right plots, shortcuts were only added if the source and target of the candidate journey are connected by a path in the transfer graph.
	}
	\label{fig:transferSpeedPreprocessing}
\end{figure*}

To explain the difference in behavior between the two variants, consider how the transfer speed affects Pareto-optimal journeys.
As the transfer mode becomes faster, it becomes increasingly feasible to cover large distances in the transfer graph quickly.
This has two effects:
On the one hand, more witnesses which require long initial or final transfers become feasible and start dominating slower candidates.
Accordingly, the number of canonical candidates decreases, from~409 million for~1\,km/h to~114 million for~10\,km/h.
This explains the decrease in the number of event-to-event shortcuts.
On the other hand, longer intermediate transfers between trips also become feasible.
This means that although there are fewer canonical candidates for higher transfer speeds, the shortcuts that occur in them tend to cover larger distances in the transfer graph.
The number of stop pairs within a certain distance of each other grows roughly quadratically with the distance.
This explains why the number of stop-to-stop shortcuts rises for higher transfer speeds even as the number of event-to-event shortcuts declines.

Once the transfer speed becomes faster than public transit, the direct transfer from source to target will dominate all other journeys, including all candidates.
Accordingly, we should expect the number of shortcuts to eventually reach~0 for very high transfer speeds.
The reason why this is not observed in our measurements is that not all stops in our network instances are reachable from each other in the transfer graph.
Consider what happens in the shortcut computation for journeys between stops~$\aSource$~and~$\aTarget$~that are isolated in the transfer graph.
In this case, a direct transfer is not possible, regardless of the transfer speed.
In fact, unless there is a route that serves both~$\aSource$~and~$\aTarget$, all $\aSource$-$\aTarget$-journeys with at most two trips are candidates and the shortcut computation will add shortcuts for the canonical ones.
In our Switzerland network,~624 stops are isolated in the transfer graph, usually as a result of incomplete or imperfect data.
If we omit shortcuts for candidates whose source and target stop are not connected in the transfer graph, the number of shortcuts behaves as expected:
If speed limits are obeyed, a few shortcuts remain even for the highest transfer speed.
If they are ignored, a direct transfer is always the fastest option and thus no shortcuts are required.

\subparagraph*{Shortcut Graph Structure.}

\begin{figure*}
	\newcommand{\plotHeight}{6.35cm}
\newcommand{\plotWidth}{0.48\textwidth}

\begin{tikzpicture}

\tcolor{histogramColorA}{tobiasblue}{4}
\tcolor{histogramColorB}{tobiasblue}{4}

\begin{scope}

\clip (0,0) rectangle (\textwidth,6.78);

\begin{scope}[shift={(1.2, 1.88)}]

   \begin{axis}[
      ybar stacked,
      height=\plotHeight,
      width=\plotWidth,
      xmin=0,
      xmax=18,
      ymin=0,
      ymax=30,
      xlabel={Transfer time [s]},
      ylabel={\small{}ULTRA shortcuts \raisebox{0.8pt}{[}k\raisebox{0.8pt}{]}},
      ylabel style = {yshift=-4pt},
      ytick={0, 5, 10, 15, 20, 25, 30},
      xtick={0, 3, 6, 9, 12, 15, 18},
      xticklabel=\pgfmathparse{\tick}$2\smash{{}^{\pgfmathprintnumber{\pgfmathresult}}}$,
      yticklabel=\pgfmathparse{\tick}{${\pgfmathprintnumber{\pgfmathresult}}$},
      xtick pos=left,
      ytick pos=left,
      grid=major,
      minor x tick num={2},
      at={(0.0\linewidth,0)}]
   \end{axis}

   \begin{axis}[
      ybar stacked,
      height=\plotHeight,
      width=\plotWidth,
      xmin=-0.5,
      xmax=17.5,
      ymin=0,
      ymax=30,
      ticks=none,
      grid=none,
      at={(0.0\linewidth,0)}]

      \addplot[draw=none,fill=\histogramColorA{0},color=\histogramColorA{0},bar width=5pt,area legend] table[x=EdgeLength,y=ShortcutsFiltered,col sep=tab]{figures/shortcutDistributionStop.dat};\label{legend:tb:nis}
      \addplot[draw=none,fill=\histogramColorA{-2},color=\histogramColorA{-2},bar width=5pt,area legend] table[x=EdgeLength,y=Diff,col sep=tab]{figures/shortcutDistributionStop.dat};\label{legend:tb:wis}

   \end{axis}

   \begin{axis}[
      height=\plotHeight,
      width=\plotWidth,
      xmin=0,
      xmax=1,
      ymin=0,
      ymax=1,
      ticks=none,
      grid=none,
      at={(0.0\linewidth,0)}]
   \end{axis}

   \begin{axis}[
      ybar stacked,
      height=\plotHeight,
      width=\plotWidth,
      xmin=0,
      xmax=18,
      ymin=0,
      ymax=2400,
      xlabel={Transfer time [s]},
      ylabel={\small{}ULTRA-TB shortcuts \raisebox{0.8pt}{[}k\raisebox{0.8pt}{]}},
      ylabel style = {yshift=-3pt},
      ytick={0, 400, 800, 1200, 1600, 2000, 2400},
      xtick={0, 3, 6, 9, 12, 15, 18},
      xticklabel=\pgfmathparse{\tick}$2\smash{{}^{\pgfmathprintnumber{\pgfmathresult}}}$,
      yticklabel=\pgfmathparse{\tick}{${\pgfmathprintnumber{\pgfmathresult}}$},
      xtick pos=left,
      ytick pos=left,
      grid=major,
      minor x tick num={2},
      at={(0.53\linewidth,0)}]
   \end{axis}

   \begin{axis}[
      ybar stacked,
      height=\plotHeight,
      width=\plotWidth,
      xmin=-0.5,
      xmax=17.5,
      ymin=0,
      ymax=2400,
      ticks=none,
      grid=none,
      at={(0.53\linewidth,0)}]

      \addplot[draw=none,fill=\histogramColorA{0},color=\histogramColorA{0},bar width=5pt,area legend] table[x=EdgeLength,y=ShortcutsFiltered,col sep=tab]{figures/shortcutDistributionEventLength.dat};
      \addplot[draw=none,fill=\histogramColorA{-2},color=\histogramColorA{-2},bar width=5pt,area legend] table[x=EdgeLength,y=Diff,col sep=tab]{figures/shortcutDistributionEventLength.dat};

   \end{axis}

   \begin{axis}[
      height=\plotHeight,
      width=\plotWidth,
      xmin=0,
      xmax=1,
      ymin=0,
      ymax=1,
      ticks=none,
      grid=none,
      at={(0.53\linewidth,0)}]
   \end{axis}

\end{scope}

   \arrayrulecolor{legendColor}
   \node[inner sep=0pt,outer sep=0pt] (legend) at (0,0.05) {};
   \node[inner sep=0pt,outer sep=0pt,anchor=south west] at (legend) {
      \footnotesize
      \begin{tabular*}{\textwidth}{|@{~~}l@{\extracolsep{\fill}}r@{~~}|}
         \hline
	                                                                            &                                                                       \\[-6pt]
         \legend{\ref{legend:tb:nis}} Shortcuts of connected candidate journeys & \legend{\ref{legend:tb:wis}} Shortcuts of isolated candidate journeys \\[2pt]
         \hline
      \end{tabular*}
   };
   \node[inner sep=0pt,outer sep=0pt,xshift=\textwidth] (legendTwo) at (legend) {};

\end{scope}

\end{tikzpicture}%
	\caption[Transfer time distribution of ULTRA shortcuts.]{%
		Distribution of the ULTRA shortcuts with respect to their transfer time for the Switzerland network.
		The bar between~$2^i$ and $2^{i-1}$ corresponds to the number of shortcuts with a transfer time in the interval~$[2^i , 2^{i-1})$.
		An exception is the first bar, which also contains shortcuts with a transfer time of less than a second.
		The dark blue portion of each bar represents shortcuts where the source and the target of the corresponding candidate journey are connected by a path in the transfer graph.
		\textit{Left:}~Shortcuts between stops as computed by the ULTRA preprocessing.
		\textit{Right:}~Shortcuts between stop events as computed by the ULTRA-TB preprocessing.
	}
	\label{fig:shortcutDistributionTwo}
\end{figure*}

The stop-to-stop shortcut graph computed by ULTRA for Switzerland is structurally very different from the transitively closed transfer graph we created for comparison with pure public transit algorithms.
This is already evidenced by the fact that the shortcut graph is much less dense, containing only~6\% as many edges as the transitively closed graph.
Furthermore, the transitive graph consists of many small fully connected components, with the largest one containing only~1\,004 vertices.
By contrast, the largest strongly connected component in the shortcut graph contains~10\,891 vertices, which corresponds to~43\% of all stops.
Accordingly, a transitive closure of the shortcut graph would contain more than~100~million edges.

As Wagner and Zündorf~\cite{Wag17} observed when constructing a transitively closed transfer graph, preserving all transfers with a duration of up to a few minutes already leads to an average vertex degree of more than~100.
This means that algorithms which require a transitively closed transfer graph cannot be efficient and at the same time guarantee that long transfers are found.
Figure~\ref{fig:shortcutDistributionTwo}~(left side) shows the distribution of travel times for the ULTRA shortcuts.
Note that the high number of shortcuts with travel time~0 is caused by cases where multiple stops model the same physical location.
Most of the shortcuts have a travel time of more than~9\,minutes~($\approx2^{9}$\,seconds) and are therefore not contained in the transitive transfer graph.
In fact, only~26\,826 edges are shared between the two graphs, which constitute~1.0\% of all transitive edges and~15.7\% of all shortcuts.
Altogether, this shows that the transitively closed graph fails to represent most of the relevant intermediate transfers, at the expense of many superfluous ones.

As with the transfer speed experiment, Figure~\ref{fig:shortcutDistributionTwo} distinguishes between shortcuts generated by candidates whose source and target stop are connected in the transfer graph~(dark blue) and shortcuts where source and target are isolated~(light blue).
We observe that most of the very long shortcuts are produced by candidates with isolated stops.
To analyze how often longer shortcuts are required, we examine the distribution of the event-to-event shortcuts in Figure~\ref{fig:shortcutDistributionTwo}~(right side).
Since stop events occur at a fixed point time, a stop-to-stop shortcut that is required at several times throughout the day corresponds to multiple event-to-event shortcuts.
Thus, the number of event-to-event shortcuts with a certain travel time reflects more accurately how frequently these shortcuts are required.
Approximately one third of all event-to-event shortcuts have a travel time of~0.
Most of these connect pairs of trips at the same stop and therefore have no stop-to-stop counterpart.
Among the remaining shortcuts, most have a travel time between~1~minute~($\approx2^6$\,s) and~34~minutes~($\approx2^{11}$\,s).
This is in contrast to the stop-to-stop shortcuts, most of which have a travel time of more than one hour~($\approx2^{12}$\,s).
This shows that very long shortcuts are only rarely required.
Furthermore, the fraction of shortcuts that are generated by candidates with isolated source and target is much lower in the event-to-event variant than in the stop-to-stop variant.

\subsection{Queries}
To evaluate the impact of ULTRA on the query performance, we test three public transit algorithms: CSA, RAPTOR and TB.
For CSA and RAPTOR, we compare our new ULTRA variant to the original algorithm on a transitively closed transfer graph and a multimodal variant with Dijkstra searches.
For TB, no multimodal variants have been proposed thus far.
We therefore compare the original TB algorithm on a transitively closed transfer graph to ULTRA-TB with sequential and integrated preprocessing.
Since we do not consider parallelized query algorithms, we use the Xeon machine~(which has better single-core performance) for all following experiments.

Additional experiments evaluating the impact of the query distance on the running times can be found in Appendix~\ref{app:ULTRA:queryDistance}.
Furthermore, a comparison to the HL-based approaches proposed by Phan and Viennot~\cite{Pha19} can be found in Appendix~\ref{app:ULTRA:hub-labeling}.
Since the original evaluation of the HL-based algorithms was based on a comparison of running times measured on different machines, we reimplemented all query algorithms and evaluated them on the same machine.
In these experiments, we were only able to observe a marginal speedup of HL-RAPTOR compared to MR.

\subparagraph*{CSA Queries.}

\begin{table}[t]
	\center
	\caption[Query performance of ULTRA-CSA.]{%
		Query performance for CSA, MCSA, and ULTRA-CSA.
		Query times are divided into two phases: initialization including initial transfers (Init.), and connection scans including intermediate transfers (Scan).
		All results are averaged over 10\,000 random queries.
		Note that CSA~(marked with $^\ast$) only supports stop-to-stop queries with transitive transfers.
		The other two algorithms have been evaluated for vertex-to-vertex queries on the full graph.
	}%
	\label{tbl:queryCSA}%
	\begin{tabular*}{\textwidth}{@{\,}l@{\extracolsep{\fill}}l@{\extracolsep{\fill}}c@{\extracolsep{\fill}}r@{\extracolsep{\fill}}r@{\extracolsep{\fill}}r@{\extracolsep{\fill}}r@{\extracolsep{\fill}}r@{\,}}
		\toprule
		\multirow{2}{*}{Network} & \multirow{2}{*}{Algorithm} & \multirow{2}{*}{\parbox{26pt}{\shortstack[c]{\vspace{0.04cm}\\Full \vspace{0.15cm} \\ graph \vspace{-0.16cm}}}} & \multicolumn{2}{c}{Scans [k]} & \multicolumn{3}{c}{Time [ms]} \\
		\cmidrule(){4-5} \cmidrule(){6-8}
		&                        &           & Connection   & Edge     & Init.  & Scan    & Total\\
		\midrule
		\multirow{3}{*}{Stuttgart}
		& CSA$^\ast$ & $\circ$   &      52.6 &    281 &   0.0 &   1.4 &   1.4\\[1pt]
		& MCSA       & $\bullet$ &     113.7 &    238 &  10.1 &   6.4 &  16.5\\[1pt]
		& ULTRA-CSA  & $\bullet$ &     113.4 &     42 &   1.2 &   1.7 &   2.9\\[5pt]
		\multirow{3}{*}{London}
		& CSA$^\ast$ & $\circ$   &      83.9 &    663 &   0.0 &   3.0 &   3.0\\[1pt]
		& MCSA       & $\bullet$ &      58.2 &    182 &   4.6 &   4.5 &   9.1\\[1pt]
		& ULTRA-CSA  & $\bullet$ &      57.7 &     53 &   0.8 &   1.9 &   2.7\\[5pt]
		\multirow{3}{*}{Switzerland}
		& CSA$^\ast$ & $\circ$   &     135.2 &    787 &   0.1 &   4.9 &   4.9\\[1pt]
		& MCSA       & $\bullet$ &      88.2 &    241 &   8.4 &   8.1 &  16.4\\[1pt]
		& ULTRA-CSA  & $\bullet$ &      87.6 &     59 &   1.1 &   2.9 &   4.0\\[5pt]
		\multirow{3}{*}{Germany}
		& CSA$^\ast$ & $\circ$   &  2\,587.8 & 6\,351 &   1.3 & 144.3 & 145.5\\[1pt]
		& MCSA       & $\bullet$ &  1\,662.1 & 3\,191 & 142.8 & 195.2 & 338.0\\[1pt]
		& ULTRA-CSA  & $\bullet$ &  1\,657.3 &    877 &  22.4 & 107.4 & 129.8\\[-1pt]
		\bottomrule
	\end{tabular*}
\end{table}

Unlike the other algorithms we evaluate, CSA only supports optimizing arrival time as the sole criterion.
While Profile CSA, a CSA variant for range queries, also supports optimizing the number of trips as a second criterion, no bicriteria variant of basic CSA has been published thus far.
We conducted preliminary experiments which showed that a bicriteria variant of CSA is outperformed by RAPTOR.
Therefore, we only consider single-criterion optimization for CSA.
Unlike RAPTOR, no Dijkstra-based multimodal variant of CSA has been proposed thus far.
We therefore implemented a naive multimodal version of CSA, which we call MCSA~(multimodal CSA), as a baseline for our comparison.
This algorithm alternates connection scans with Dijkstra searches on the contracted core graph, in a similar manner to MCR.
Query times for all three CSA variants are reported in Table~\ref{tbl:queryCSA}.

\begin{table}[t]
	\center
	\caption[Query performance of ULTRA-RAPTOR.]{%
		Query performance for RAPTOR, MR, and ULTRA-RAPTOR.
		Query times are divided into phases:~initialization, including scanning initial transfers~(Init.), collecting routes~(Coll.), scanning routes~(Scan), and relaxing transfers~(Relax).
		All results are averaged over 10\,000 random queries.
		Note that RAPTOR~(marked with~$^\ast$) only supports stop-to-stop queries with transitive transfers, whereas the other three algorithms support vertex-to-vertex queries on the full graph and have been evaluated accordingly.
	}%
	\label{tbl:queryRAPTOR}%
	\begin{tabular*}{\textwidth}{@{\,}l@{\extracolsep{\fill}}l@{\hspace{-5pt}}c@{}r@{\extracolsep{\fill}}r@{\extracolsep{\fill}}r@{\extracolsep{\fill}}r@{\extracolsep{\fill}}r@{}r@{\extracolsep{\fill}}r@{\,}}
		\toprule
		\multirow{2}{*}{Network} & \multirow{2}{*}{Algorithm} & \multirow{2}{*}{\!\!\parbox{26pt}{\shortstack[c]{\vspace{0.04cm}\\Full \vspace{0.15cm} \\ graph \vspace{-0.16cm}}}\!} & \multicolumn{2}{c}{Scans [k]} & \multicolumn{5}{c}{Time [ms]} \\
		\cmidrule(){4-5}  \cmidrule(){6-10}
		&                          &            & Route  & Edge      & Init.  &   Coll.  & Scan   & Relax  & Total\\
		\midrule
		\multirow{4}{*}{Stuttgart}
		& RAPTOR$^\ast$ & $\circ$    &    19.8 &     756 &   0.2 &   1.6 &   2.1 &   2.1 &   5.9\\[1pt]
		& MR            & $\bullet$  &    35.6 &     687 &  12.3 &   5.2 &   5.2 &  11.1 &  33.5\\[1pt]
		& ULTRA-RAPTOR  & $\bullet$  &    37.9 &     105 &   1.4 &   3.5 &   3.5 &   1.0 &   9.6\\[5pt]
		\multirow{4}{*}{London}
		& RAPTOR$^\ast$ & $\circ$    &     4.4 &  2\,573 &   0.3 &   1.1 &   2.2 &   5.4 &   8.9\\[1pt]
		& MR            & $\bullet$  &     5.0 &     500 &   6.4 &   1.9 &   2.7 &   7.0 &  18.0\\[1pt]
		& ULTRA-RAPTOR  & $\bullet$  &     5.4 &     179 &   1.2 &   1.5 &   2.3 &   1.2 &   6.2\\[5pt]
		\multirow{4}{*}{Switzerland}
		& RAPTOR$^\ast$ & $\circ$    &    26.2 &  2\,115 &   0.4 &   2.4 &   5.0 &   5.0 &  12.8\\[1pt]
		& MR            & $\bullet$  &    33.0 &     731 &  10.6 &   4.8 &   7.2 &  11.7 &  34.1\\[1pt]
		& ULTRA-RAPTOR  & $\bullet$  &    35.9 &     177 &   1.6 &   3.3 &   6.2 &   1.4 &  12.5\\[5pt]
		\multirow{4}{*}{Germany}
		& RAPTOR$^\ast$ & $\circ$    & \!472.9 & 26\,420 &   7.0 & 102.6 & 120.4 &  74.2 & 304.2\\[1pt]
		& MR            & $\bullet$  & \!541.4 & 12\,359 & 154.2 & 187.5 & 153.5 & 236.2 & 731.4\\[1pt]
		& ULTRA-RAPTOR  & $\bullet$  & \!599.7 &  3\,165 &  33.0 & 144.0 & 151.7 &  33.3 & 362.1\\[-1pt]
		\bottomrule
	\end{tabular*}
\end{table}

On all networks, ULTRA-CSA has a similar running time to CSA with transitively closed transfers.
Caution has to be taken when comparing these running times since CSA does not support fully multimodal vertex-to-vertex queries and was therefore evaluated on a different set of stop-to-stop queries.
Nonetheless, our experiments demonstrate that ULTRA enables CSA to use unrestricted transfers without performance loss.
Compared to MCSA, the ULTRA approach is faster by about a factor of 3--4 on most networks and even more on the Stuttgart network, which has a particularly large transfer graph.
By replacing the Core-CH search of MCSA with a Bucket-CH query, ULTRA speeds up the exploration of initial and final transfers by a factor of 6--8.
The time required for the exploration of intermediate transfers is difficult to measure directly because it is interleaved with the individual connection scans.
Nevertheless, we observe that using ULTRA shortcuts speeds up the connection scanning phase in its entirety by a factor of 2--4 compared to MCSA.

On all networks except Stuttgart, the multimodal variants scan significantly fewer connections than CSA on the transitively closed transfer graph.
This is a direct result of the fact that fully multimodal journeys usually have a shorter travel time~\cite{Wag17}.
Since CSA scans connections in chronological order, the number of scanned connections correlates directly with the earliest arrival time of the query.
The Stuttgart network exhibits the opposite behavior because the transfer graph covers a much larger geographical area than the public transit network.
Therefore, if the source and target are picked among all vertices instead of only stops, the average query distance increases and the search space becomes larger.

\subparagraph*{RAPTOR Queries.}
To evaluate RAPTOR, we used the MR variant of MCR as the multimodal baseline algorithm.
The results of our comparison are shown in Table~\ref{tbl:queryRAPTOR}.
The share of the overall running time spent exploring the transfer graph~(i.e.,~the \emph{Init} and \emph{Relax} phases) is reduced from~50--75\% for MR to~20--40\% for ULTRA-RAPTOR.
The \emph{Init} phase exhibits the same speedup that was already observed for CSA.
Since RAPTOR explores intermediate transfers in a separate phase, the impact of using ULTRA shortcuts can now be measured directly.
Compared to the Dijkstra searches on the core graph performed by MR, exploring the transfer shortcuts is up to an order of magnitude faster.
Overall, ULTRA-RAPTOR is 2--3 times as fast as MR and has a similar running time to RAPTOR with transitive transfers.

\subparagraph*{Trip-Based Queries.}

\begin{table}[t]
	\center
	\caption[Query performance of ULTRA-TB.]{%
		Query performance for TB and ULTRA-TB (sequential and integrated).
		Query times are divided into phases:~the Bucket-CH query~\mbox{(B-CH)}, the initial transfer evaluation~(Initial), and the scanning of trips~(Scan).
		All results are averaged over~10\,000 random queries.
		Note that TB~(marked with~$^\ast$) only supports stop-to-stop queries with transitive transfers, whereas the other two algorithms support vertex-to-vertex queries on the full graph.
	}%
	\label{tbl:queryTB}%
	\begin{tabular*}{\textwidth}{@{\,}l@{\extracolsep{\fill}}l@{\hspace{-5pt}}c@{\extracolsep{\fill}}r@{\extracolsep{\fill}}r@{\extracolsep{\fill}}r@{\extracolsep{\fill}\!}r@{\extracolsep{\fill}}r@{\extracolsep{\fill}}r@{\hspace{1pt}}}
		\toprule
		\multirow{2}{*}{Network} & \multirow{2}{*}{Algorithm} & \multirow{2}{*}{\parbox{26pt}{\shortstack[c]{\vspace{0.04cm}\\Full \vspace{0.15cm} \\ graph \vspace{-0.16cm}}}} & \multicolumn{2}{c}{Scans [k]} & \multicolumn{4}{c}{Time [ms]} \\
		\cmidrule(){4-5}  \cmidrule(){6-9}
		&                                            &            &     Trip &     Shortcut &   B-CH &  Initial &    Scan &   Total\\
		\midrule
		\multirow{3}{*}{Stuttgart}
		& TB$^\ast$              & $\circ$    &     10.9 &        223 &   0.0 &     0.0 &    1.5 &    1.6\\[1pt]
		& ULTRA-TB (seq.)        & $\bullet$  &     25.1 &     1\,417 &   1.2 &     1.0 &    5.8 &    7.9\\[1pt]
		& ULTRA-TB (int.)        & $\bullet$  &     15.3 &        112 &   1.1 &     0.8 &    1.7 &    3.6\\[5pt]
		\multirow{3}{*}{London}
		& TB$^\ast$              & $\circ$    &     15.3 &        830 &   0.0 &     0.0 &    3.7 &    3.7\\[1pt]
		& ULTRA-TB (seq.)        & $\bullet$  &     23.5 &     1\,021 &   0.8 &     0.7 &    5.1 &    6.6\\[1pt]
		& ULTRA-TB (int.)        & $\bullet$  &     14.5 &        153 &   0.8 &     0.6 &    1.9 &    3.3\\[5pt]
		\multirow{3}{*}{Switzerland}
		& TB$^\ast$              & $\circ$    &     23.4 &        662 &   0.0 &     0.0 &    4.5 &    4.5\\[1pt]
		& ULTRA-TB (seq.)        & $\bullet$  &     34.9 &     1\,620 &   1.0 &     1.2 &    7.1 &    9.3\\[1pt]
		& ULTRA-TB (int.)        & $\bullet$  &     19.5 &        138 &   1.0 &     1.0 &    2.2 &    4.3\\[5pt]
		\multirow{3}{*}{Germany}
		& TB$^\ast$              & $\circ$    &  \!389.1 &    16\,331 &   0.0 &     0.0 &  106.6 &  106.9\\[1pt]
		& ULTRA-TB (seq.)        & $\bullet$  &  \!467.5 &    43\,219 &  19.9 &    19.3 &  162.6 &  202.0\\[1pt]
		& ULTRA-TB (int.)        & $\bullet$  &  \!196.5 &     2\,057 &  19.6 &    19.3 &   37.9 &   77.0\\[-1pt]
		\bottomrule
	\end{tabular*}
\end{table}

We continue with evaluating our improved ULTRA-TB query algorithm.
Table~\ref{tbl:queryTB} compares the query performance for ULTRA-TB with sequential and integrated preprocessing, as well as the original TB query algorithm on the transitively closed transfer graph.
ULTRA-TB with integrated preprocessing achieves significantly lower query times than the state of the art.
Depending on the network, it has a speedup of~2--5 over~ULTRA-RAPTOR and~5--10 over MR, which was previously the fastest multimodal journey planning algorithm~(cf.~Table~\ref{tbl:queryRAPTOR}).
As with RAPTOR and CSA, ULTRA-TB is able to match the query performance of the original TB algorithm despite solving a harder multimodal problem.
Furthermore, ULTRA-TB achieves a similar performance to ULTRA-CSA, despite optimizing an additional criterion.

While ULTRA-TB with sequential preprocessing still outperforms other algorithms, it is slower than the integrated version by a factor of~2.
This is because the integrated preprocessing reduces the number of relaxed shortcuts by around an order of magnitude.
This in turn reduces the overall search space and thereby the number of scanned trips.
Overall, the trip scanning phase is sped up by a factor of~3--4 and only takes up around half of the overall query time.
The remaining half is spent performing the Bucket-CH searches and evaluating initial trips, both of which are unaffected by the number of transfer shortcuts.

\subparagraph*{Impact of Transfer Speed.}

\begin{figure*}
	\input{figures/transferSpeedQuery}%
	\caption[Impact of transfer speed on query times and travel times.]{%
		Impact of transfer speed on query times and travel times, measured on the Switzerland network with a core degree of 14 and a witness limit of~0.
		All results were averaged over 10\,000 random queries.
		\textit{Left:} Query performance of~MR, ULTRA-RAPTOR and~ULTRA-TB.
		Speed limits were obeyed during the construction of the transfer graph.
		For MR and ULTRA-RAPTOR, query times are divided into route collecting/scanning, transfer relaxation, and remaining time.
		\textit{Right:} Total travel time and time spent on initial/final and intermediate transfers for the journey with minimal arrival time.
		The time required for a direct transfer from source to target is shown for reference.
		To allow for this comparison, we only chose random queries where the source and target vertex are connected in the transfer graph.
	}
	\label{fig:transferSpeedQuery}
\end{figure*}

In addition to overall query performance, we also measured how the query times of MR, ULTRA-RAPTOR and ULTRA-TB are impacted by the transfer speed.
Results are shown in Figure~\ref{fig:transferSpeedQuery}~(left side).
The performance gains for ULTRA-RAPTOR compared to MR are similar for all transfer speeds, and in fact slightly better for higher speeds.
To explain this, observe that the time required for the route scanning phase decreases as the transfer speed increases.
This is because the total number of rounds and thus the number of scanned routes decreases for higher transfer speeds.
ULTRA-RAPTOR benefits more from this since the share of the route scanning phase in the overall running time is greater for ULTRA-RAPTOR than for MR.
In all cases, the entire query time for ULTRA-RAPTOR is similar to or lower than the time that MR takes for the route scanning phases only.
ULTRA-TB achieves its highest speedup over the other two algorithms for medium transfer speeds, where the number of event-to-event shortcuts is lowest.
For very high transfer speeds, the Bucket-CH search for the initial and final transfers starts to dominate the overall running time of both ULTRA-based algorithms.
Accordingly, the speedup of ULTRA-TB over ULTRA-RAPTOR decreases.

The impact of the transfer speed on the travel time of the fastest journey is shown in Figure~\ref{fig:transferSpeedQuery}~(right side).
As the transfer speed increases, the overall travel time decreases.
The time that is spent on an initial or final transfer also decreases at first, but its share in the overall travel time becomes larger.
From~10\,km/h onward, transferring directly from source to target starts becoming the best option for more queries, and consequently the time spent on initial and final transfers starts increasing.
For very high transfer speeds, a direct transfer is almost always the fastest option.
This matches our observation that intermediate transfers become useless for high transfer speeds unless the source and target are isolated from each other in the transfer graph.
In contrast to initial and final transfers, intermediate transfers have a very small impact on the overall travel time, further demonstrating that long intermediate transfers are rarely needed.

\section{Conclusion}
\label{chap:ULTRA:fr}

We proposed ULTRA, a technique which accelerates the computation of Pareto-optimal journeys in a public transit network with an unrestricted transfer graph.
The centerpiece of ULTRA is a preprocessing step which computes shortcuts that provably represent all necessary intermediate transfers.
With parallelization, this step takes only a few minutes for metropolitan and mid-sized country networks, and about~3 hours for Germany.
The number of computed shortcuts is low, regardless of the speed of the transfer mode.
ULTRA shortcuts can be used without adjustments by any public transit algorithm that requires one-hop transfers.
This enables the computation of unrestricted multimodal journeys without incurring the performance losses of existing multimodal algorithms.
In particular, combining ULTRA with CSA yields the first efficient multimodal variant of CSA.
To combine ULTRA with TB, we developed tailored versions of the ULTRA preprocessing and the TB query.
The resulting ULTRA-TB algorithm outperforms MR, the fastest previously known multimodal algorithm for bicriteria optimization, by an order of magnitude.

Future work could involve extending ULTRA to support more optimization criteria, such as walking distance or cost, and multiple non-schedule-based transportation modes.
Furthermore, it would be interesting to adapt our approach to scenarios where public transit vehicles can be delayed.
Without adaptation, ULTRA can no longer guarantee optimal results in such a setting, since journeys with delayed vehicles might require additional intermediate transfers which are not covered by the shortcut set.
We suspect, however, that the underlying assumption of ULTRA~(i.e.,~the set of required intermediate transfers is small) is still valid in a scenario with delays.

\newpage

\bibliography{references}

\newpage
\appendix
 \section{Correctness of Canonical MR}
\label{app:ULTRA:canonicalMR}

The correctness of canonical MR, which is described in Section~\ref{chap:ULTRA:canonical}, is proven by the following lemma.
\begin{lemma}
	Canonical MR returns the canonical Pareto set for every query.
\end{lemma}
\begin{proof}
	It follows from the correctness of MR that canonical MR returns a valid Pareto set.
	We show that this is the canonical Pareto set.
	Consider a query with source and target vertices~$\aSource,\aTarget\in\vertices$ and departure time~$\departureTime$.
	Let~$\aJourney$ be a journey in the canonical Pareto set for this query and~$\aJourney'$ another journey which is feasible for~$\departureTime$, with~$\arrivalTime(\aJourney)=\arrivalTime(\aJourney')$, $|\aJourney|=|\aJourney'|$ and~$\tiebreakingSequence(\aJourney)<\tiebreakingSequence(\aJourney')$.    
	Let~$\aVertex$ be the vertex where the longest shared suffix of~$\aJourney$ and~$\aJourney'$ starts, and let~$\aJourney_{\aSource\aVertex}$ and~$\aJourney'_{\aSource\aVertex}$ be the corresponding prefixes.
	We show that canonical MR discards~$\aJourney'_{\aSource\aVertex}$ in favor of~$\aJourney_{\aSource\aVertex}$.
	
	By the definition of the tiebreaking sequence, $\tiebreakingSequence(\aVertex,\aJourney_{\aSource\aVertex})<\tiebreakingSequence(\aVertex,\aJourney'_{\aSource\aVertex})$.
	If~$\arrivalTime(\aJourney_{\aSource\aVertex})<\arrivalTime(\aJourney'_{\aSource\aVertex})$, then it follows from the correctness of MR that canonical MR discards~$\aJourney'_{\aSource\aVertex}$ in favor of~$\aJourney_{\aSource\aVertex}$.
	Assume therefore that~$\arrivalTime(\aJourney_{\aSource\aVertex})=\arrivalTime(\aJourney'_{\aSource\aVertex})$.
	Then the first entries of~$\tiebreakingSequence(\aVertex,\aJourney_{\aSource\aVertex})$ and~$\tiebreakingSequence(\aVertex,\aJourney'_{\aSource\aVertex})$ are identical.
	The comparison depends on whether the journeys end with a trip segment or an edge:
	\begin{itemize}
		\item Case 1a: $\aJourney_{\aSource\aVertex}$ ends with a trip segment~$\aTripSegmentA{ij}$ and~$\aJourney'_{\aSource\aVertex}$ ends with an edge.
		Then~$\aJourney_{\aSource\aVertex}$ is found in the route scanning phase of round~$|\aJourney_{\aSource\aVertex}|$ and~$\aJourney'_{\aSource\aVertex}$ is discarded when it is found in the subsequent transfer relaxation phase.
		\item Case 1b: $\aJourney_{\aSource\aVertex}$ ends with a trip segment~$\aTripSegmentA{ij}$ and~$\aJourney'_{\aSource\aVertex}$ ends with a trip segment~$\aTripSegmentB{mn}$.
		If~$\aRoute(\aTripA)\neq\aRoute(\aTripB)$, it follows that~$\routeIndex(\aRoute(\aTripA))<\routeIndex(\aRoute(\aTripB))$.
		Then the route scanning phase of round~$|\aJourney_{\aSource\aVertex}|$ scans~$\aRoute(\aTripA)$ before~$\aRoute(\aTripB)$, finds~$\aJourney_{\aSource\aVertex}$ first and discards~$\aJourney'_{\aSource\aVertex}$
		If both journeys use the same route, then it follows that~$\aTripA=\aTripB$ and~$j=n$.
		Since~$\tiebreakingSequence(\aVertex,\aJourney_{\aSource\aVertex})<\tiebreakingSequence(\aVertex,\aJourney'_{\aSource\aVertex})$, it follows that~$i<j$.
		Then the scan of route~$\aRoute(\aTripA)$ finds~$\aJourney_{\aSource\aVertex}$ when it enters at the~$i$-th stop of~$\aRoute(\aTripA)$ and discards~$\aJourney'_{\aSource\aVertex}$ because it does not improve the active trip at the~$j$-th stop.
		\item Case 2: $\aJourney_{\aSource\aVertex}$ ends with an edge~$(\bVertex,\aVertex)$.
		Then the second and third entries of~$\tiebreakingSequence(\aVertex,\aJourney_{\aSource\aVertex})$ are~$\infty$, so~$\aJourney'_{\aSource\aVertex}$ must also end with an edge~$(\cVertex,\aVertex)$.
		Since the longest shared suffix of~$\aJourney_{\aSource\aVertex}$ and~$\aJourney'_{\aSource\aVertex}$ is empty, $\bVertex\neq\cVertex$ must hold.
		It follows from~$\tiebreakingSequence(\aVertex,\aJourney_{\aSource\aVertex})<\tiebreakingSequence(\aVertex,\aJourney'_{\aSource\aVertex})$ that~$\langle\arrivalTime(\aJourney_{\aSource\bVertex}),\vertexIndex(\bVertex)\rangle<\langle\arrivalTime(\aJourney'_{\aSource\cVertex}),\vertexIndex(\cVertex)\rangle$.
		Then the Dijkstra search in round~$|\aJourney_{\aSource\aVertex}|$ extracts~$\bVertex$ from the priority queue before~$\cVertex$.
		Thus, the edge~$(\bVertex,\aVertex)$ is relaxed before~$(\cVertex,\aVertex)$, $\aJourney_{\aSource\aVertex}$ is found first and~$\aJourney'_{\aSource\aVertex}$ discarded.
	\end{itemize}
	In all cases, canonical MR discards~$\aJourney'_{\aSource\aVertex}$ and therefore also~$\aJourney'$.
	Since this is the case for every journey~$\aJourney'$ which is equivalent to~$\aJourney$ but has a higher tiebreaking sequence, it follows that canonical MR finds the canonical journey~$\aJourney$.
\end{proof}

\section{Subjourney Decomposition for MR}
\label{app:ULTRA:nonCanonicalMR}
Enumerating the set~\canonicalJourneys of canonical journeys defined in Section~\ref{chap:ULTRA:canonical} requires modifications to MR.
These modifications are necessary in order to obtain closure under subjourney decomposition, which is required to restrict the search to candidates.
To demonstrate this, consider the following alternative definition for~\canonicalJourneys: an $\aSource$-$\aTarget$-journey~$\aJourney$ is canonical if it is computed by a (non-canonical) MR search from~$\aSource$ to~$\aTarget$ for departure time~$\departureTime(\aJourney)$.
Using this definition, \canonicalJourneys is not closed under subjourney decomposition.
This is because the order in which two equivalent journeys are explored by MR can flip if the same prefix is added to both journeys.
An example of this is shown in Figure~\ref{fig:nonCanonicalMR}.
The two $\aSource$-$\aTarget$-journeys~$\aJourney$ and~$\aJourney'$ are equivalent and differ only in the route that is used for the second trip.
The order in which the journeys are found depends on the order in which these routes are scanned.
At the start of each round, RAPTOR iterates over all stops which were marked in the previous round and collects all routes which visit them.
The order in which these routes are then scanned is not specified in the original description of RAPTOR~\cite{Del15b}, but it is natural to scan them in the order in which they were collected.
Routes visiting the same stop are collected in the order defined by~$\routeIndex$.
However, if multiple stops were updated, the order in which the routes are collected and scanned depends on the order in which the stops were reached in the previous round.

In Figure~\ref{fig:nonCanonicalMR}, if the search is started from~\aSource, the stop~$x$ is reached before~$b$.
Therefore, the red route is collected before the yellow route, which does not visit~$x$, and~$\aJourney'$ is explored before~$\aJourney$.
However, if the shared prefix~$\aJourney_{\aSource{}b}$ is omitted and the search is started at~$b$, the yellow route is preferred and~$\aJourney_{b\aTarget}$ is found.
Thus, if the ULTRA preprocessing used non-canonical MR, it would generate event-to-event shortcuts for the intermediate transfers of~$\aJourney'_{\aSource{}c}$ and~$\aJourney_{b\aTarget}$, but not for those of~$\aJourney_{\aSource{}c}$ and~$\aJourney'_{b\aTarget}$.
As a result, neither~$\aJourney$ nor~$\aJourney'$ could be reconstructed from these shortcuts.

\begin{figure}
	\centering
	\begin{tikzpicture}[scale=1]

    \node (s)  at ( 0.00, 1.00) {};%
    \node (a)  at ( 2.00, 1.00) {};%
    \node (b1) at ( 4.00, 1.75) {};%
    \node (b2)  at ( 4.00, 1.00) {};%
    \node (b3) at ( 4.75, 0.25) {};%
    \node (b_label) at (4.375, 2.25) {};%
    \node (c1) at ( 7.00, 1.75) {};%
    \node (c2)  at ( 7.00, 1.00) {};%
    \node (c3) at ( 7.00, 0.25) {};%
    \node (c_label) at (7.00, 2.25) {};%
    \node (d)  at ( 9.00, 1.00) {};%
    \node (t)  at (11.00, 1.00) {};%
    \node (x1) at ( 4.00, -2.00) {};%
    \node (x2) at ( 4.75, -2.50) {};%
    \node (x_label) at (4.375, -3.00) {};%
    
    \node [fit=(b1)(b2)(b3)(b_label),line width=.5pt, draw=nodeColor!100,fill=nodeColor!15,rounded corners=0.1cm] {};%
    \node [fit=(x1)(x2)(x_label),line width=.5pt, draw=nodeColor!100,fill=nodeColor!15,rounded corners=0.1cm] {};%
    \node [fit=(c1)(c2)(c3)(c_label),line width=.5pt, draw=nodeColor!100,fill=nodeColor!15,rounded corners=0.1cm] {};%

    \draw [edgeColor, line width=1pt, rounded corners = 20]  (a) -- (2.0, 1.75) -- (b1);
    \draw [edgeColor, line width=1pt]  (a) -- (b2);
    \draw [edgeColor, line width=1pt, rounded corners = 20]  (a) -- (2.25, 0.25) -- (b3);
    \draw [edgeColor, line width=1pt, rounded corners = 20]  (a) -- (2.375, -2.0) -- (x1);
    \draw [edgeColor, line width=1pt, rounded corners = 20]  (a) -- (2.0, -2.5) -- (x2);
    \draw [edgeColor, line width=1pt, rounded corners = 20]  (c1) -- (9.0, 1.75) -- (d);
    \draw [edgeColor, line width=1pt]  (c2) -- (d);
    \draw [edgeColor, line width=1pt, rounded corners = 20]  (c3) -- (9.0, 0.25) -- (d);
    
    \node[align=left,edgeColor] at (3.25, 2.05) {$2$};
    \node[align=left,edgeColor] at (3.25, 1.30) {$2$};
    \node[align=left,edgeColor] at (3.25, 0.55) {$2$};
    \node[align=left,edgeColor] at (3.25, -1.75) {$1$};
    \node[align=left,edgeColor] at (3.25, -2.25) {$1$};
    \node[align=left,edgeColor] at (8.00, 2.05) {$1$};
    \node[align=left,edgeColor] at (8.00, 1.30) {$1$};
    \node[align=left,edgeColor] at (8.00, 0.55) {$1$};
    
    \draw [\primarycolor{0}, line width=2.5pt, routeArrow, rounded corners = 20] (s) -- (a);
    \draw [\secondarycolor{0}, line width=2.5pt, routeArrow, rounded corners = 20] (x1) -- (b2);
    \draw [\secondarycolor{0}, line width=2.5pt, routeArrow, rounded corners = 20] (x2) -- (b3);
    \draw [\tertiarycolor{0}, line width=2.5pt, routeArrow, rounded corners = 20] (b1) -- (c1);
    \draw [\secondarycolor{0}, line width=2.5pt, routeArrow, rounded corners = 20] (b2) -- (c2);
    \draw [\secondarycolor{0}, line width=2.5pt, routeArrow, rounded corners = 20] (b3) -- (c3);
    \draw [\quaternarycolor{0}, line width=2.5pt, routeArrow, rounded corners = 20] (d) -- (t);
    
    \node [align=left,text=\primarycolor{1}] at ( 1.00, 1.30) {$0\rightarrow1$};%
    \node [align=left,label={[rotate=90,text=\secondarycolor{1}]$1\rightarrow3$}] at ( 3.95, -1.00) {};%
    \node [align=left,label={[rotate=90,text=\secondarycolor{1}]$5\rightarrow7$}] at ( 4.70, -1.00) {};%
    \node [align=left,text=\tertiarycolor{1}] at ( 5.75, 2.05) {$3\rightarrow4$};%
    \node [align=left,text=\secondarycolor{1}] at ( 5.75, 1.30) {$3\rightarrow4$};%
    \node [align=left,text=\secondarycolor{1}] at ( 5.75, 0.55) {$7\rightarrow8$};%
    \node [align=left,text=\quaternarycolor{1}] at (10.00, 1.30) {$5\rightarrow6$};%
    
    \node at (s) [vertex,draw=nodeColor!100,fill=nodeColor!15] {\gs};%
    \node at (a) [vertex,draw=nodeColor!100,fill=nodeColor!15] {\gs};%
    \node at (b1) [vertex,draw=\tertiarycolor{1}!100,fill=\tertiarycolor{1}!15] {\gs};%
    \node at (b2) [vertex,draw=\secondarycolor{1}!100,fill=\secondarycolor{1}!15] {\gs};%
    \node at (b3) [vertex,draw=\secondarycolor{1}!100,fill=\secondarycolor{1}!15] {\gs};%
    \node at (c1) [vertex,draw=\tertiarycolor{1}!100,fill=\tertiarycolor{1}!15] {\gs};%
    \node at (c2) [vertex,draw=\secondarycolor{1}!100,fill=\secondarycolor{1}!15] {\gs};%
    \node at (c3) [vertex,draw=\secondarycolor{1}!100,fill=\secondarycolor{1}!15] {\gs};%
    \node at (d) [vertex,draw=nodeColor!100,fill=nodeColor!15] {\gs};%
    \node at (t) [vertex,draw=nodeColor!100,fill=nodeColor!15] {\gs};%
    \node at (x1) [vertex,draw=\secondarycolor{1}!100,fill=\secondarycolor{1}!15] {\gs};%
    \node at (x2) [vertex,draw=\secondarycolor{1}!100,fill=\secondarycolor{1}!15] {\gs};%

    \node at (s) [text=nodeColor!100] {\small{$\aSource$}};%
    \node at (a) [text=nodeColor!100] {\small{$a$}};%
    \node at (b1) [text=nodeColor!100] {\small{$b_1$}};%
    \node at (b2) [text=nodeColor!100] {\small{$b_2$}};%
    \node at (b3) [text=nodeColor!100] {\small{$b_3$}};%
    \node at (b_label) [text=nodeColor!100] {\small{$b$}};%
    \node at (c1) [text=nodeColor!100] {\small{$c_1$}};%
    \node at (c2) [text=nodeColor!100] {\small{$c_2$}};%
    \node at (c3) [text=nodeColor!100] {\small{$c_3$}};%
    \node at (c_label) [text=nodeColor!100] {\small{$c$}};%
    \node at (d) [text=nodeColor!100] {\small{$d$}};%
    \node at (t) [text=nodeColor!100] {\small{$\aTarget$}};%
    \node at (x1) [text=nodeColor!100] {\small{$x_1$}};%
    \node at (x2) [text=nodeColor!100] {\small{$x_2$}};%
    \node at (x_label) [text=nodeColor!100] {\small{$x$}};%
\end{tikzpicture}%
	\caption[An example network showing that the set of journeys computed by MR is not closed under subjourney decomposition.]{%
		An example network where the set of journeys computed by MR is not closed under subjourney decomposition.
		Transfer edges (gray) are labeled with their travel time, while trips (colored) are labeled with~$\departureTime\to\arrivalTime$.
		Trips with the same color belong to the same route.
		Stops (gray) which are visited by multiple trips are subdivided into stop events (colored).
		There are two equivalent Pareto-optimal~$\aSource$-$\aTarget$-journeys:
		$\aJourney=\big\langle\!
			\langle{}\aSourceX\rangle,
			\langle\textcolor{\primarycolor{1}}{0\rightarrow1}\rangle,
			\langle{}a,b\rangle,
			\langle\textcolor{\tertiarycolor{1}}{3\rightarrow4}\rangle,
			\langle{}c,d\rangle,
			\langle\textcolor{\quaternarycolor{1}}{5\rightarrow6}\rangle,
			\langle{}\aTargetX\rangle
			\!\big\rangle$ and~$\aJourney'=\big\langle\!
			\langle{}\aSourceX\rangle,
			\langle\textcolor{\primarycolor{1}}{0\rightarrow1}\rangle,
			\langle{}a,b\rangle,
			\langle\textcolor{\secondarycolor{1}}{3\rightarrow4}\rangle,
			\langle{}c,d\rangle,
			\langle\textcolor{\quaternarycolor{1}}{5\rightarrow6}\rangle,
			\langle{}\aTargetX\rangle
			\!\big\rangle$.
		Let~$\aRoute_Y$ denote the yellow route and~$\aRoute_R$ the red route.
		Assume that~$\routeIndex(\aRoute_Y)<\routeIndex(\aRoute_R)$.
		Then an MR query from~$b$ for the departure time~$3$ scans~$\aRoute_Y$ first and consequently finds~$\aJourney_{b\aTarget}=\big\langle\!
			\langle{}b\rangle,
			\langle\textcolor{\tertiarycolor{1}}{3\rightarrow4}\rangle,
			\langle{}c,d\rangle,
			\langle\textcolor{\quaternarycolor{1}}{5\rightarrow6}\rangle,
			\langle{}\aTargetX\rangle
			\!\big\rangle$.
		However, a query from~\aSource reaches~$x$ and collects~$\aRoute_R$ there before it reaches~$b$ and collects~$\aRoute_Y$.
		Note that~$\big\langle\!
			\langle{}\aSourceX\rangle,
			\langle\textcolor{\primarycolor{1}}{0\rightarrow1}\rangle,
			\langle{}a,x\rangle,
			\langle\textcolor{\secondarycolor{1}}{1\rightarrow3,3\rightarrow4}\rangle,
			\langle{}c,d\rangle,
			\langle\textcolor{\quaternarycolor{1}}{5\rightarrow6}\rangle,
			\langle{}\aTargetX\rangle
			\!\big\rangle$ is not a valid journey because~$\aRoute_R$ departs too early at~$x$ to be entered.
		However, $\aRoute_R$ is still collected at~$x$ because there is a later trip departing at~$5$, which can be entered.
		Thus, a query from~$\aSource$ scans~$\aRoute_R$ before~$\aRoute_Y$ and therefore finds~$\aJourney'$ and its subjourney~$\aJourney'_{\aSource{}c}=\big\langle\!
			\langle{}\aSourceX\rangle,
			\langle\textcolor{\primarycolor{1}}{0\rightarrow1}\rangle,
			\langle{}a,b\rangle,
			\langle\textcolor{\secondarycolor{1}}{3\rightarrow4}\rangle,
			\langle{}c\rangle
			\!\big\rangle$.
		Overall, the set of journeys output by MR includes~$\aJourney'$ and~$\aJourney_{b\aTarget}$, but not~$\aJourney$ and~$\aJourney'_{b\aTarget}$, so it is not closed under subjourney decomposition.
	}%
	\label{fig:nonCanonicalMR}%
\end{figure}

\section{Impact of Query Distance}
\label{app:ULTRA:queryDistance}
Figure~\ref{fig:ultraGeoRankResultsPlot} compares the running times of the three fastest bicriteria algorithms~(MR, ULTRA-RAPTOR, and~ULTRA-TB) depending on the query distance, using geo-rank as a measurement for distance.
Geo-rank queries are generated by picking a source vertex~\aSource uniformly at random and sorting all vertices by their geographical distance to~\aSource.
The~$i$-th vertex in this order is then the target of the geo-rank query for rank~$i$.
For our comparison in Figure~\ref{fig:ultraGeoRankResultsPlot}, we generated and evaluated~10\,000 of these queries for the Germany network.
For all geo-ranks, ULTRA-TB is an order of magnitude faster than MR.
ULTRA-RAPTOR lies between these two algorithms and is closer to ULTRA-TB for local queries, and closer to MR for long-range queries.
Furthermore, we observe that many short-range queries can be solved in less than one millisecond by ULTRA-TB with integrated preprocessing.

A geo-rank-based evaluation on the Germany network was also performed for the original TB algorithm by Witt~\cite{Wit15}.
While the results are similar to ours, they contain significantly more outliers, especially for low ranks.
Across all geo-ranks, the evaluation for the original TB algorithm shows a considerable number of queries that take more than~10 milliseconds.
These can be attributed to queries where the source vertex is located in particularly sparse parts of the public transit network.
In these regions, the correlation between geo-rank and actual distance is poor, and thus a query can be a long-range query despite having a low geo-rank.
Adding an unrestricted transfer graph improves the correlation between geo-rank and query complexity, which explains why we observe fewer outliers in comparison.

\begin{figure}
	\centering
	\input{figures/ultraGeoRankResultsPlotRAPTOR}%
	\caption[Comparison of query times depending on the geo-rank.]{%
		Comparison of query times depending on the geo-rank for the Germany network.
		For each geo-rank, we evaluated~10\,000 random vertex-to-vertex queries.
		We compare the previously fastest multimodal algorithm~(MR) to the two bicriteria ULTRA query algorithms: ULTRA-RAPTOR and ULTRA-TB.
	}
	\label{fig:ultraGeoRankResultsPlot}
\end{figure}

\section{Comparison to HL-Based Approach}
\label{app:ULTRA:hub-labeling}
An alternative to ULTRA is the HL-based approach proposed by~\cite{Pha19}.
Instead of using precomputed one-hop transfers, this approach explores the transfer graph with two-hop searches based on Hub Labeling~\cite{Abr11}.
For HL-RAPTOR in the bicriteria setting, the authors report a speedup of~1.7 for HL-RAPTOR over MR.
However, this figure is based on a comparison to the MR query times reported by Delling et al.~\cite{Del13}, which were measured on an older machine and likely for a different set of queries.
To obtain a fair comparison, we therefore re-implemented the HL-RAPTOR query, building on the same RAPTOR code that we also used for MR and ULTRA-RAPTOR.

\begin{table}[t]
	\center
	\caption[HL preprocessing results.]{%
		Overview of the~HL preprocessing results.
		Running times were measured on the Xeon machine with a single core and are displayed as (hh:)mm:ss.
		Only the outgoing hub edges are reported.
		Since all evaluated transfer graphs are symmetrical, the number of incoming hub edges is identical.
	}%
	\label{tbl:HL:preprocessing}%
	\begin{tabular*}{\textwidth}{@{\,}l@{\hspace{10pt}}r@{\extracolsep{\fill}}r@{\extracolsep{\fill}}r@{\extracolsep{\fill}}r@{\,}}
		\toprule
		&        Stuttgart & \hspace{21mm}\llap{London} & \hspace{22mm}\llap{Switzerland} & \hspace{22mm}\llap{Germany} \\
		\midrule
		Preprocessing time             &       1:07:14 &         3:56 &        21:14 &         52:48:23 \\[1pt]
		Outgoing hub edges of vertices & 153\,323\,291 & 13\,314\,082 & 53\,744\,836 & 1\,320\,767\,674 \\[1pt]
		Outgoing hub edges of stops    &   1\,898\,414 &  1\,363\,960 &  1\,952\,586 &     45\,075\,714 \\[1pt]
		Average vertex out-degree      &         131.4 &         73.3 &         89.0 &            192.2 \\
		\bottomrule
	\end{tabular*}
\end{table}

\subparagraph{Preprocessing.}
For the HL preprocessing, we used the HL implementation provided by the authors\footnote{\url{https://github.com/lviennot/hub-labeling/}}.
Results for the preprocessing phase are reported in Table~\ref{tbl:HL:preprocessing}.
Note that Phan and Viennot~\cite{Pha19} only computed hubs between pairs of stops, since they only evaluated their algorithms for stop-to-stop queries.
In order to support vertex-to-vertex queries on the full transfer graph, we computed hubs between all pairs of vertices.
Unlike ULTRA, the HL preprocessing is not easily parallelizable and was therefore performed on a single core.
While the HL preprocessing has a lower single-core preprocessing time, ULTRA becomes significantly faster with parallelization.
An exception to this is the London network, which has a particularly small transfer graph but a large and complex public transit network.
Since the HL preprocessing only considers the transfer graph while ULTRA has to consider both, ULTRA is outperformed here even when parallelized.
Regarding memory consumption, ULTRA clearly outperforms HL since the number of shortcuts is much smaller than the number of hub edges.

\subparagraph{Queries.}
\begin{table}[t]
	\center
	\caption[Query performance of HL-RAPTOR.]{%
		Query performance for MR, HL-RAPTOR, and ULTRA-RAPTOR.
		Query times are divided into phases:~initialization, including scanning initial transfers~(Init.), collecting routes~(Coll.), scanning routes~(Scan), and relaxing transfers~(Relax).
		All results are averaged over 10\,000 random queries.
	}%
	\label{tbl:queryHL}%
	\begin{tabular*}{\textwidth}{@{\,}l@{\extracolsep{\fill}}l@{\extracolsep{\fill}}r@{\extracolsep{\fill}}r@{\extracolsep{\fill}}r@{\extracolsep{\fill}}r@{\extracolsep{\fill}}r@{}r@{\extracolsep{\fill}}r@{\,}}
		\toprule
		\multirow{2}{*}{Network} & \multirow{2}{*}{Algorithm} & \multicolumn{2}{c}{Scans [k]} & \multicolumn{5}{c}{Time [ms]} \\
		\cmidrule(){3-4}  \cmidrule(){5-9}
		&               & Route  & Edge      & Init.  &   Coll.  & Scan   & Relax  & Total\\
		\midrule
		\multirow{4}{*}{Stuttgart}
		& MR            &    35.6 &     687 &  12.3 &   5.2 &   5.2 &  11.1 &  33.5\\[1pt]
		& HL-RAPTOR     &    39.3 &  3\,068 &   1.9 &   5.3 &   5.1 &  18.2 &  30.7\\[1pt]
		& ULTRA-RAPTOR  &    37.9 &     105 &   1.4 &   3.5 &   3.5 &   1.0 &   9.6\\[5pt]
		\multirow{4}{*}{London}
		& MR            &     5.0 &     500 &   6.4 &   1.9 &   2.7 &   7.0 &  18.0\\[1pt]
		& HL-RAPTOR     &     5.6 &  1\,599 &   0.9 &   1.8 &   2.7 &   8.0 &  13.4\\[1pt]
		& ULTRA-RAPTOR  &     5.4 &     179 &   1.2 &   1.5 &   2.3 &   1.2 &   6.2\\[5pt]
		\multirow{4}{*}{Switzerland}
		& MR            &    33.0 &     731 &  10.6 &   4.8 &   7.2 &  11.7 &  34.1\\[1pt]
		& HL-RAPTOR     &    38.4 &  2\,337 &   1.6 &   5.0 &   7.8 &  17.8 &  32.1\\[1pt]
		& ULTRA-RAPTOR  &    35.9 &     177 &   1.6 &   3.3 &   6.2 &   1.4 &  12.5\\[5pt]
		\multirow{4}{*}{Germany}
		& MR            & \!541.4 & 12\,359 & 154.2 & 187.5 & 153.5 & 236.2 & 731.4\\[1pt]
		& HL-RAPTOR     & \!629.4 & 41\,773 &  28.8 & 214.8 & 183.4 & 381.4 & 808.3\\[1pt]
		& ULTRA-RAPTOR  & \!599.7 &  3\,165 &  33.0 & 144.0 & 151.7 &  33.3 & 362.1\\[-1pt]
		\bottomrule
	\end{tabular*}
\end{table}
Table~\ref{tbl:queryHL} compares the query performance of HL-RAPTOR to MR and ULTRA-RAPTOR.
We chose not to evaluate HL-CSA because the results reported by~Phan and Viennot~\cite{Pha19} indicate that it is only marginally faster than HL-RAPTOR, and therefore slower than MCSA.
This was confirmed by our preliminary experiments.
We observe that HL-RAPTOR only slightly outperforms MR on the three smaller networks, and is slower on Germany.
Its speedup comes entirely from the initial transfer exploration.
In the intermediate transfer phase, using the precomputed hubs is actually slower than the Dijkstra search performed by MR.
This is due to the very high number of edges in the hub graph.
The first few rounds of a query typically reach most stops in the network, so most hub edges which are adjacent to a stop need to be relaxed.
This causes HL-RAPTOR to relax more than three times as many edges as MR.
HL-RAPTOR performs best on the London network, achieving a speedup of 1.3 over MR.
By comparison, ULTRA-RAPTOR is faster than the HL-based approach by a factor of more than~2 on all networks.

\end{document}